\documentclass[a4paper]{JHEP3}

\usepackage{caption}
\usepackage{subcaption}

\usepackage{graphicx}
\usepackage{bm}        
\usepackage{amssymb}   
\usepackage{amsmath}
\usepackage{cite}
\usepackage{pifont}
\usepackage{slashed}
\usepackage{multirow}
\usepackage{array}
\usepackage{feynarts}
\usepackage{color}

\usepackage{IEEEtrantools}

\usepackage{listings}
\let\normalcolor\relax

\definecolor{shaded}{RGB}{210,210,210}
\definecolor{bianco}{RGB}{255,255,255}
\definecolor{rosso}{RGB}{210,0,0}
\definecolor{blu}{RGB}{0,0,210}
\definecolor{nero}{RGB}{0,0,0}

\newcommand{\ulult}{$\tilde u_L \tilde u_L^\ast$-type}%
\newcommand{\ulurt}{$\tilde u_L \tilde u_R^\ast$-type}%
\newcommand{\ulclt}{$\tilde u_L \tilde c_L^\ast$-type}%
\newcommand{\ulslt}{$\tilde u_L \tilde s_L^\ast$-type}%
\newcommand{\uldlt}{$\tilde u_L \tilde d_L^\ast$-type}%
\newcommand{\ulurtulslt}{$\tilde u_L \tilde u_R^\ast ~\text{and}\\ \tilde u_L \tilde s_L^\ast$-type}%
\newcommand{\ord}{\mathcal{O}}

\allowdisplaybreaks

\newcommand{\Order}[2]{\mathcal{O}(\alpha_s^{#1} \alpha^{#2})}
\newcommand{\psec}[3]{\sigma^{{#1,#2}}_{#3}}
\newcommand{\csec}[3]{\bar \sigma^{#1,#2}_{#3}}

\title{
Electroweak corrections to 
squark--antisquark   production at the LHC
}

\author{Wolfgang Hollik  \\
Max-Planck-Institut f\"ur Physik, F\"ohringer Ring 6, D-80805 M\"unchen, Germany; \\
Email: \email{hollik@mpp.mpg.de}}

\author{Jonas M. Lindert \\ 
Physik-Institut, Universit\"at Z\"urich,
Wintherturerstrasse 190, CH-8057 Z\"urich,
Switzerland; 
E-mail: \email{lindert@physik.uzh.ch}
}    

\author{Edoardo Mirabella\thanks{Now at Fleuchaus \& Gallo Partnerschaft mbB, Patent- und Rechtsanw\"alte} \\
Max-Planck-Institut f\"ur Physik, F\"ohringer Ring 6, D-80805 M\"unchen, Germany; \\
E-mail: \email{mirabell@mpp.mpg.de}
}    

\author{Davide Pagani \\ 
Centre for Cosmology, Particle Physics and Phenomenology (CP3), Universit\'e  Catholique de Louvain, B-1348 
Louvain-la-Neuve, Belgium; \\
E-mail: \email{davide.pagani@uclouvain.be}
}

\preprint{{\small CP3-15-15, \;  MPP-2015-106,  \;  ZU-TH 13/15}}

\abstract{We present the calculation of the electroweak corrections for squark--antisquark pair production at the LHC within the Minimal Supersymmetric Standard Model. Taking into account all possible chirality and light-flavor configurations, we evaluate the NLO EW corrections, which are of $\ord(\alpha_s^2 \alpha)$, as well as the subleading tree-level contributions of $\ord(\alpha_s \alpha)$ and $\ord(\alpha^2)$.
Numerical results are presented for several scans in the SUSY parameter space and relevant differential distributions are investigated. The impact of the electroweak corrections is nonnegligible and strongly depends on the chirality configuration of the produced squarks. Our analysis includes a discussion of photon--gluon initiated processes with a focus on the impact of the corresponding large PDF uncertainties. 
}

\begin{document}

\section{Introduction}
Supersymmetry (SUSY)~\cite{Wess:1974tw}  is one of the theoretically most appealing beyond -the-Standard-Model (BSM) scenarios. In particular, the minimal supersymmetric extension of the 
Standard Model (MSSM)~\cite{Nilles:1983ge,Haber:1984rc,Barbieri:1987xf} predicts a  light scalar Higgs boson that is compatible with the resonance
observed at the LHC experiments~\cite{Aad:2012tfa,Chatrchyan:2012ufa}.
Moreover, the MSSM naturally stabilizes the electroweak vacuum and, assuming R-parity conservation,  provides also a dark matter candidate.

These  features  give to  SUSY  a central role in the BSM searches at the LHC.  Within the MSSM the analyses of the first-run LHC  data  have investigated 
 several final states and signatures ~\cite{AtlasPage,CmsPage}. All the analyses found no significant deviation from SM predictions, setting limits
 on the SUSY spectrum. Still, the second run of the LHC, at higher energies and higher luminosity, will probe larger regions of the MSSM parameter space 
 that are not excluded at the moment by experimental searches. 
 
At the LHC, the production of a pair of colored SUSY particles ($\tilde q \tilde q',\tilde q \tilde q'^{*},\tilde g \tilde g, \tilde g \tilde q^{(*)}$) plays a crucial role in the context of direct searches; their cross sections are the largest among all the possible production mechanisms.   
The phenomenological relevance of this class of processes has triggered an extensive effort in improving the precision of their 
theoretical predictions. Not only the leading-order (LO) contributions~\cite{Kane:1982hw,Harrison:1982yi,Reya:1984yz,Dawson:1983fw,Baer:1985xz}, but also the complete set of next-to-leading
 order (NLO) QCD corrections~\cite{Beenakker:1994an,Beenakker:1995fp,Beenakker:1996ch, Beenakker:1997ut,Hollik:2012rc,GoncalvesNetto:2012yt,Goncalves:2014axa} and  most of the NLO electroweak (EW) corrections~\cite{Bozzi:2005sy,Alan:2007rp,Hollik:2007wf,Bornhauser:2007bf,Beccaria:2008mi,Hollik:2008yi,Hollik:2008vm,Bornhauser:2009ru,Mirabella:2009ap,Arhrib:2009sb,Germer:2010vn,Germer:2011an,Germer:2014jpa} 
  are available in the literature. Approximate next-to-next-to-leading  order (NNLO) QCD  corrections
 to squark--antisquark production  have been computed as well~\cite{Langenfeld:2009eg,Langenfeld:2010vu, Langenfeld:2012ti,Broggio:2013uba}.  
 
 Besides fixed order calculations, the large corrections in the  threshold region
 have been computed at the next-to-leading (NLL)~\cite{Kulesza:2008jb,Kulesza:2009kq,Beenakker:2009ha,Beenakker:2010nq,Beenakker:2011fu}  and next-to-next-to-leading  
 (NNLL)~\cite{Beenakker:2011sf,  Broggio:2013cia,Pfoh:2013iia, Beenakker:2013mva, Beenakker:2014sma} logarithmic 
 accuracy.  Effective field theory approaches allowed  also for the resummation of  soft-gluon and Coulomb corrections~\cite{Beneke:2009nr,Beneke:2010da,Falgari:2012hx,Falgari:2012sq}. Moreover,
 gluino and stop  bound states have been studied~\cite{Kauth:2009ud,Kim:2014yaa} and the bound-state  effects in  gluino--gluino and squark--gluino
 production have also been analyzed~\cite{Hagiwara:2009hq,Kauth:2011vg,Kauth:2011bz}. 
 
 Also the squark and  gluino decay rate is  known at  the NLO accuracy, including both QCD~\cite{Beenakker:1996dw, Djouadi:1996wt,Beenakker:1996de} 
 and EW~\cite{Guasch:2001kz,Guasch:2002ez} corrections. Phenomenological studies systematically including the NLO QCD corrections to the combined production and decays of squark--squark and stop--antistop pairs have been respectively performed in ~\cite{Hollik:2012rc,Hollik:2013xwa} and \cite{Boughezal:2012zb,Boughezal:2013pja}. In~\cite{Gavin:2013kga,Gavin:2014yga}, the NLO QCD 
 corrections  to squark--squark  and squark--antisquark  production have been  matched  to parton showers.
 
From the experimental side,  the lack of a signal from direct searches has increased the lower bounds on the masses of the supersymmetric particles, especially for those that are colored.  
In particular, the most stringent bounds have been set on light-flavor 
 squarks and gluinos.  For instance, assuming a simplified MSSM scenario with only the gluino and degenerate light-flavor
 squarks production with decay into a massless neutralino,  the gluino (light-flavor squark) mass 
 should be  heavier  than $1.4$~TeV ($1.6$~TeV). Under the same hypotheses, degenerate light-flavor 
 squarks  and gluinos should be heavier than $1.8$~TeV~\cite{Aad:2014wea}.  
 
 In these experimental analyses the simulation of the signal
takes into account only part of the aforementioned higher-order corrections. In particular  signal 
 events are obtained by following~\cite{Kramer:2012bx}, i.e., they 
 are produced at LO accuracy and afterwards they are globally reweighted via the $K$-factors from 
 the inclusive NLO+NLL QCD  total cross section that is obtained from  the codes {\tt Prospino}~\cite{Beenakker:1996ed} 
 and {\tt  NLL-fast}~\cite{Beenakker:2011fu}.   The theoretical uncertainty affecting the signal is evaluated by   
 varying  both the parton distribution function (PDF) sets  and the renormalization/factorization scale. 
 It is worth to notice  that this procedure does not account for the  kinematical dependence of the higher-order 
contributions and for the significant impact of the NLO corrections  to the decays of the produced 
particles~\cite{Boughezal:2012zb,Boughezal:2013pja,Hollik:2012rc,Hollik:2013xwa}.  Moreover  corrections of  EW origin are completely ignored both in the evaluation of  signal  and in the estimation of the 
theoretical uncertainty. Since EW corrections naturally involve most of the MSSM particles via loop corrections, they formally depend on the specific scenario considered. For instance, a large mass gap between colored and uncolored particles may in principle enhance their effect beyond the percent level. In order to have under control the impact of higher-order corrections, it is necessary to explicitly check these effects   and identify for which parameters EW corrections possibly show sizable dependencies. 

In this paper we perform this type of analyses,  focussing on the hadronic associate production of a light-flavor squark 
(a squark of the first two generations) together with a light-flavor  antisquark,   
\begin{align}
PP \to  \tilde q_\alpha \tilde q'^\ast_\beta  \, ,
\label{Eq:Processes}
\end{align}
where $q, q' =u,d,c,s$, while $\alpha$ and $\beta$ denote the chirality of the corresponding squark, $\alpha, \beta =L,R$.
We compute  the EW corrections to the processes~(\ref{Eq:Processes})  extending and completing the analysis  of~\cite{Hollik:2008yi},
which focused only on same-flavor, same-chirality squark--antisquark production,  $PP \to  \tilde q_\alpha \tilde q^\ast_\alpha$.  Moreover, we present for this process  the first 
phenomenological study combining NLO QCD and NLO EW corrections. Our study mainly focuses on the NLO corrections to the 
inclusive squark--antisquark production cross section, and it can be directly used in experimental analyses to estimate
 the theoretical uncertainty related to the missing EW corrections. Additionally, the impact of the EW corrections on kinematic distributions for undecayed squarks is considered as well. 

The outline of the paper is as follows. Section 2 describes the various partonic processes 
contributing up to $\mathcal{O}(\alpha_s^2\alpha)$ and the strategy of the calculation. 
The numerical impact of the EW and QCD corrections are discussed in Section 3, 
followed by our conclusions in Section 4.

\section{Calculation method}
\label{Sec:Computation}
In our calculation we consider and treat independently all the possible light-flavor and chirality configurations for squark--antisquark pairs, as denoted in eq. \eqref{Eq:Processes}.
Since  we treat quarks of the first two generations as massless, the  squark chirality eigenstates are also the mass eigenstates, thus each squark--antisquark pair can separately be considered as a physical final state.
In total, the possible  two chirality/mass eigenstates  for each of the four  (anti)squark flavor ($u,d,c,s$) lead to eight distinguished types of (anti)squarks and thus to sixty-four possible $\tilde q_\alpha \tilde q'^\ast_\beta$ squark--antisquark pairs.

For each one of these sixty-four final states the LO cross section is of $\ord(\alpha_s^2)$. In our calculation, we take into account the complete set of NLO EW corrections, which are of $\ord(\alpha_s^2 \alpha)$, and all the contributions originating from tree-level diagrams only. The latter include the LO, but also $\ord(\alpha_s \alpha)$ contributions, which  arise from the interference of $\ord(\alpha_s)$ and $\ord(\alpha)$ amplitudes, and $\ord(\alpha^2)$ contributions from $\ord(\alpha)$ squared amplitudes. With the term ``EW corrections'' we will in general indicate the sum of $\ord(\alpha_s^2 \alpha)$, $\ord(\alpha_s \alpha)$ and $\ord(\alpha^2)$ contributions.

All these contributions are independently calculated for each one of the sixty-four final states. Thus, the masses of the eight different (anti)squarks   can be treated as completely non-degenerate. In our phenomenological analyses of Section \ref{sec:numerics}, we will include also NLO QCD corrections by the help of the code {\tt Prospino}~\cite{Beenakker:1996ed}, which is based on the calculation in~\cite{Beenakker:1996ch}. However, in order to obtain genuine NLO QCD corrections, the usage of {\tt Prospino} and thus the evaluation of NLO QCD corrections will be performed only for degenerate squark masses and only at the level of inclusive cross sections. \medskip

\begin{table}[t]
\begin{center}
\begin{tabular}{c     c    c    c    c   c}
\hline \hline 
Processes \phantom{a} &&  & Channels \\ 
 & $\ord(\alpha_s)$ && $\ord(\alpha)$ &&  $\ord(\alpha_s^{1/2}\alpha^{1/2})$ $\gamma$-ind.\\ \hline
\parbox{2.5cm}{\vspace{-1.5cm} \ulult}                      & \unitlength=.15bp%
\begin{feynartspicture}(300,300)(1,1)
\FADiagram{}
\FAProp(0.,15.)(10.,14.)(0.,){/Cycles}{0}
\FALabel(4.84577,13.4377)[t]{\tiny$g$}
\FAProp(0.,5.)(10.,6.)(0.,){/Cycles}{0}
\FALabel(5.15423,4.43769)[t]{\tiny$g$}
\FAProp(20.,15.)(10.,14.)(0.,){/ScalarDash}{-1}
\FALabel(14.8458,15.5623)[b]{\tiny$\tilde q_\alpha$}
\FAProp(20.,5.)(10.,6.)(0.,){/ScalarDash}{1}
\FALabel(15.1542,6.56231)[b]{\tiny$\tilde q_\alpha$}
\FAProp(10.,14.)(10.,6.)(0.,){/ScalarDash}{-1}
\FALabel(8.93,8.8)[r]{\tiny$\tilde q_\alpha$}
\FAVert(10.,14.){0}
\FAVert(10.,6.){0}
\end{feynartspicture}
%

  &   \phantom{a}    &  \unitlength=0.15bp%
\begin{feynartspicture}(300,300)(1,1)
\FADiagram{}
\FAProp(0.,15.)(6.,10.)(0.,){/Straight}{1}
\FALabel(2.48771,11.7893)[tr]{\tiny $q'$}
\FAProp(0.,5.)(6.,10.)(0.,){/Straight}{-1}
\FALabel(3.51229,6.78926)[tl]{\tiny $q'$}
\FAProp(20.,15.)(14.,10.)(0.,){/ScalarDash}{-1}
\FALabel(16.4877,13.2107)[br]{\tiny $\tilde q_\alpha$}
\FAProp(20.,5.)(14.,10.)(0.,){/ScalarDash}{1}
\FALabel(17.5123,8.21074)[bl]{\tiny $\tilde q_\alpha$}
\FAProp(6.,10.)(14.,10.)(0.,){/Sine}{0}
\FALabel(10.,8.93)[t]{\tiny $\gamma, Z$}
\FAVert(6.,10.){0}
\FAVert(14.,10.){0}
\end{feynartspicture}

  & \phantom{a}   &   \unitlength=.15bp%
%
%
\begin{feynartspicture}(300,300)(1,1)
\FADiagram{}
\FAProp(0.,15.)(10.,14.)(0.,){/Cycles}{0}
\FALabel(4.84577,13.4377)[t]{\tiny$g$}
\FAProp(0.,5.)(10.,6.)(0.,){/Sine}{0}
\FALabel(5.15423,4.43769)[t]{\tiny$\gamma$}
\FAProp(20.,15.)(10.,14.)(0.,){/ScalarDash}{-1}
\FALabel(14.8458,15.5623)[b]{\tiny$\tilde q_\alpha$}
\FAProp(20.,5.)(10.,6.)(0.,){/ScalarDash}{1}
\FALabel(15.1542,6.56231)[b]{\tiny$\tilde q_\alpha$}
\FAProp(10.,14.)(10.,6.)(0.,){/ScalarDash}{-1}
\FALabel(8.93,8.8)[r]{\tiny$\tilde q_\alpha$}
\FAVert(10.,14.){0}
\FAVert(10.,6.){0}
\end{feynartspicture}
 \\[-3.5ex]
   &   \unitlength=0.15bp%
\color{nero}
\begin{feynartspicture}(300,300)(1,1)
\FADiagram{}
\FAProp(0.,15.)(6.,10.)(0.,){/Straight}{1}
\FALabel(2.48771,11.7893)[tr]{\tiny $q'$}
\FAProp(0.,5.)(6.,10.)(0.,){/Straight}{-1}
\FALabel(3.51229,6.78926)[tl]{\tiny $q'$}
\FAProp(20.,15.)(14.,10.)(0.,){/ScalarDash}{-1}
\FALabel(16.4877,13.2107)[br]{\tiny $\tilde q_\alpha$}
\FAProp(20.,5.)(14.,10.)(0.,){/ScalarDash}{1}
\FALabel(17.5123,8.21074)[bl]{\tiny $\tilde q_\alpha$}
\FAProp(6.,10.)(14.,10.)(0.,){/Cycles}{0}
\FALabel(10.,8.93)[t]{\tiny $g$}
\FAVert(6.,10.){0}
\FAVert(14.,10.){0}
\end{feynartspicture}

\color{rosso}
\begin{feynartspicture}(300,300)(1,1)
\FADiagram{}
\FAProp(0.,15.)(10.,14.)(0.,){/Straight}{1}
\FALabel(4.84577,13.4377)[t]{\tiny $q$}
\FAProp(0.,5.)(10.,6.)(0.,){/Straight}{-1}
\FALabel(5.15423,4.43769)[t]{\tiny $q$}
\FAProp(20.,15.)(10.,14.)(0.,){/ScalarDash}{-1}
\FALabel(14.8458,15.5623)[b]{\tiny $\tilde q_\alpha$}
\FAProp(20.,5.)(10.,6.)(0.,){/ScalarDash}{1}
\FALabel(15.1542,6.56231)[b]{\tiny $\tilde q_\alpha$}
\FAProp(10.,14.)(10.,6.)(0.,){/Straight}{0}
\FALabel(9.18,10.)[r]{\tiny $\tilde g$}
\FAVert(10.,14.){0}
\FAVert(10.,6.){0}
\end{feynartspicture}

%
\color{nero}
 &        &  \unitlength=0.15bp%
\color{blu}
\begin{feynartspicture}(300,300)(1,1)
\FADiagram{}
\FAProp(0.,15.)(10.,14.)(0.,){/Straight}{1}
\FALabel(4.84577,13.4377)[t]{\tiny $p_q$}
\FAProp(0.,5.)(10.,6.)(0.,){/Straight}{-1}
\FALabel(5.15423,4.43769)[t]{\tiny $p_q$}
\FAProp(20.,15.)(10.,14.)(0.,){/ScalarDash}{-1}
\FALabel(14.8458,15.5623)[b]{\tiny $\tilde q_\alpha$}
\FAProp(20.,5.)(10.,6.)(0.,){/ScalarDash}{1}
\FALabel(15.1542,6.56231)[b]{\tiny $\tilde q_\alpha$}
\FAProp(10.,14.)(10.,6.)(0.,){/Straight}{0}
\FALabel(9.18,8.6)[r]{\tiny $\tilde \chi_m^\pm$}
\FAVert(10.,14.){0}
\FAVert(10.,6.){0}
\end{feynartspicture}
\color{nero}
\color{rosso}
\begin{feynartspicture}(300,300)(1,1)
\FADiagram{}
\FAProp(0.,15.)(10.,14.)(0.,){/Straight}{1}
\FALabel(4.84577,13.4377)[t]{\tiny $q$}
\FAProp(0.,5.)(10.,6.)(0.,){/Straight}{-1}
\FALabel(5.15423,4.43769)[t]{\tiny $q$}
\FAProp(20.,15.)(10.,14.)(0.,){/ScalarDash}{-1}
\FALabel(14.8458,15.5623)[b]{\tiny $\tilde q_\alpha$}
\FAProp(20.,5.)(10.,6.)(0.,){/ScalarDash}{1}
\FALabel(15.1542,6.56231)[b]{\tiny $\tilde q_\alpha$}
\FAProp(10.,14.)(10.,6.)(0.,){/Straight}{0}
\FALabel(9.18,8.6)[r]{\tiny $\tilde \chi_m^0$}
\FAVert(10.,14.){0}
\FAVert(10.,6.){0}
\end{feynartspicture}
\color{nero}

\color{nero}

  &    &  \\[-3.5ex]
                        &     				  &        &    &    & \\
 \parbox{2.5cm}{\vspace{-1.5cm} \ulclt}  &  \unitlength=0.15bp%
\begin{feynartspicture}(300,300)(1,1)
\FADiagram{}
\FAProp(0.,15.)(10.,14.)(0.,){/Straight}{1}
\FALabel(4.84577,13.4377)[t]{\tiny $q$}
\FAProp(0.,5.)(10.,6.)(0.,){/Straight}{-1}
\FALabel(5.15423,4.43769)[t]{\tiny $q$}
\FAProp(20.,15.)(10.,14.)(0.,){/ScalarDash}{-1}
\FALabel(14.8458,15.5623)[b]{\tiny $\tilde q_\alpha$}
\FAProp(20.,5.)(10.,6.)(0.,){/ScalarDash}{1}
\FALabel(15.1542,6.56231)[b]{\tiny $\tilde q_\beta$}
\FAProp(10.,14.)(10.,6.)(0.,){/Straight}{0}
\FALabel(9.18,10.)[r]{\tiny $\tilde g$}
\FAVert(10.,14.){0}
\FAVert(10.,6.){0}
\end{feynartspicture}

    &   &  \unitlength=0.15bp%
\begin{feynartspicture}(300,300)(1,1)
\FADiagram{}
\FAProp(0.,15.)(10.,14.)(0.,){/Straight}{1}
\FALabel(4.84577,13.4377)[t]{\tiny $p_q$}
\FAProp(0.,5.)(10.,6.)(0.,){/Straight}{-1}
\FALabel(5.15423,4.43769)[t]{\tiny $p_{q'}$}
\FAProp(20.,15.)(10.,14.)(0.,){/ScalarDash}{-1}
\FALabel(14.8458,15.5623)[b]{\tiny $\tilde q_\alpha$}
\FAProp(20.,5.)(10.,6.)(0.,){/ScalarDash}{1}
\FALabel(15.1542,6.56231)[b]{\tiny $\tilde q'_\beta$}
\FAProp(10.,14.)(10.,6.)(0.,){/Straight}{0}
\FALabel(9.18,8.6)[r]{\tiny $\tilde \chi^\pm_m$}
\FAVert(10.,14.){0}
\FAVert(10.,6.){0}
\end{feynartspicture}
\begin{feynartspicture}(300,300)(1,1)
\FADiagram{}
\FAProp(0.,15.)(10.,14.)(0.,){/Straight}{1}
\FALabel(4.84577,13.4377)[t]{\tiny $q$}
\FAProp(0.,5.)(10.,6.)(0.,){/Straight}{-1}
\FALabel(5.15423,4.43769)[t]{\tiny $q'$}
\FAProp(20.,15.)(10.,14.)(0.,){/ScalarDash}{-1}
\FALabel(14.8458,15.5623)[b]{\tiny $\tilde q_\alpha$}
\FAProp(20.,5.)(10.,6.)(0.,){/ScalarDash}{1}
\FALabel(15.1542,6.56231)[b]{\tiny $\tilde q'_\beta$}
\FAProp(10.,14.)(10.,6.)(0.,){/Straight}{0}
\FALabel(9.18,8.6)[r]{\tiny $\tilde \chi_m^0$}
\FAVert(10.,14.){0}
\FAVert(10.,6.){0}
\end{feynartspicture}
   \\
%
 \parbox{2.5cm}{\vspace{-1.5cm} \uldlt} &  \unitlength=0.15bp%
\begin{feynartspicture}(300,300)(1,1)
\FADiagram{}
\FAProp(0.,15.)(10.,14.)(0.,){/Straight}{1}
\FALabel(4.84577,13.4377)[t]{\tiny $q$}
\FAProp(0.,5.)(10.,6.)(0.,){/Straight}{-1}
\FALabel(5.15423,4.43769)[t]{\tiny $q'$}
\FAProp(20.,15.)(10.,14.)(0.,){/ScalarDash}{-1}
\FALabel(14.8458,15.5623)[b]{\tiny $\tilde q_\alpha$}
\FAProp(20.,5.)(10.,6.)(0.,){/ScalarDash}{1}
\FALabel(15.1542,6.56231)[b]{\tiny $\tilde q'_\beta$}
\FAProp(10.,14.)(10.,6.)(0.,){/Straight}{0}
\FALabel(9.18,10.)[r]{\tiny $\tilde g$}
\FAVert(10.,14.){0}
\FAVert(10.,6.){0}
\end{feynartspicture}
%
\color{nero}   &  &   \unitlength=0.15bp%
\begin{feynartspicture}(300,300)(1,1)
\FADiagram{}
\FAProp(0.,15.)(6.,10.)(0.,){/Straight}{1}
\FALabel(2.48771,11.7893)[tr]{\tiny $q$}
\FAProp(0.,5.)(6.,10.)(0.,){/Straight}{-1}
\FALabel(3.51229,6.78926)[tl]{\tiny $q'$}
\FAProp(20.,15.)(14.,10.)(0.,){/ScalarDash}{-1}
\FALabel(16.4877,13.2107)[br]{\tiny $\tilde q_\alpha$}
\FAProp(20.,5.)(14.,10.)(0.,){/ScalarDash}{1}
\FALabel(17.5123,8.21074)[bl]{\tiny $\tilde q'_\beta$}
\FAProp(6.,10.)(14.,10.)(0.,){/Sine}{0}
\FALabel(10.,8.93)[t]{\tiny $W$}
\FAVert(6.,10.){0}
\FAVert(14.,10.){0}
\end{feynartspicture}
\begin{feynartspicture}(300,300)(1,1)
\FADiagram{}
\FAProp(0.,15.)(6.,10.)(0.,){/Straight}{1}
\FALabel(2.48771,11.7893)[tr]{\tiny $r_q$}
\FAProp(0.,5.)(6.,10.)(0.,){/Straight}{-1}
\FALabel(3.51229,6.78926)[tl]{\tiny $r_{q'}$}
\FAProp(20.,15.)(14.,10.)(0.,){/ScalarDash}{-1}
\FALabel(16.4877,13.2107)[br]{\tiny $\tilde q_\alpha$}
\FAProp(20.,5.)(14.,10.)(0.,){/ScalarDash}{1}
\FALabel(17.5123,8.21074)[bl]{\tiny $\tilde q'_\beta$}
\FAProp(6.,10.)(14.,10.)(0.,){/Sine}{0}
\FALabel(10.,8.93)[t]{\tiny $W$}
\FAVert(6.,10.){0}
\FAVert(14.,10.){0}
\end{feynartspicture}

   \\[-3.5ex]
                                                            &  &   &  \unitlength=0.15bp%

\begin{feynartspicture}(300,300)(1,1)
\FADiagram{}
\FAProp(0.,15.)(10.,14.)(0.,){/Straight}{1}
\FALabel(4.84577,13.4377)[t]{\tiny $q$}
\FAProp(0.,5.)(10.,6.)(0.,){/Straight}{-1}
\FALabel(5.15423,4.43769)[t]{\tiny $q'$}
\FAProp(20.,15.)(10.,14.)(0.,){/ScalarDash}{-1}
\FALabel(14.8458,15.5623)[b]{\tiny $\tilde q_\alpha$}
\FAProp(20.,5.)(10.,6.)(0.,){/ScalarDash}{1}
\FALabel(15.1542,6.56231)[b]{\tiny $\tilde q'_\beta$}
\FAProp(10.,14.)(10.,6.)(0.,){/Straight}{0}
\FALabel(9.18,8.6)[r]{\tiny $\tilde \chi_m^0$}
\FAVert(10.,14.){0}
\FAVert(10.,6.){0}
\end{feynartspicture}   \\ 
%
 \parbox{2.5cm}{\vspace{-1.5cm} \ulurtulslt} & \unitlength=0.15bp%
\begin{feynartspicture}(300,300)(1,1)
\FADiagram{}
\FAProp(0.,15.)(10.,14.)(0.,){/Straight}{1}
\FALabel(4.84577,13.4377)[t]{\tiny $q$}
\FAProp(0.,5.)(10.,6.)(0.,){/Straight}{-1}
\FALabel(5.15423,4.43769)[t]{\tiny $q$}
\FAProp(20.,15.)(10.,14.)(0.,){/ScalarDash}{-1}
\FALabel(14.8458,15.5623)[b]{\tiny $\tilde q_\alpha$}
\FAProp(20.,5.)(10.,6.)(0.,){/ScalarDash}{1}
\FALabel(15.1542,6.56231)[b]{\tiny $\tilde q_\beta$}
\FAProp(10.,14.)(10.,6.)(0.,){/Straight}{0}
\FALabel(9.18,10.)[r]{\tiny $\tilde g$}
\FAVert(10.,14.){0}
\FAVert(10.,6.){0}
\end{feynartspicture}

 & & \unitlength=0.15bp%

\begin{feynartspicture}(300,300)(1,1)
\FADiagram{}
\FAProp(0.,15.)(10.,14.)(0.,){/Straight}{1}
\FALabel(4.84577,13.4377)[t]{\tiny $q$}
\FAProp(0.,5.)(10.,6.)(0.,){/Straight}{-1}
\FALabel(5.15423,4.43769)[t]{\tiny $q$}
\FAProp(20.,15.)(10.,14.)(0.,){/ScalarDash}{-1}
\FALabel(14.8458,15.5623)[b]{\tiny $\tilde q_\alpha$}
\FAProp(20.,5.)(10.,6.)(0.,){/ScalarDash}{1}
\FALabel(15.1542,6.56231)[b]{\tiny $\tilde q_\beta$}
\FAProp(10.,14.)(10.,6.)(0.,){/Straight}{0}
\FALabel(9.18,8.6)[r]{\tiny $\tilde \chi_m^0$}
\FAVert(10.,14.){0}
\FAVert(10.,6.){0}
\end{feynartspicture} \\
%
\hline 
\hline 
\end{tabular}
\end{center}
\caption[.]{Representative tree-level diagrams for the different partonic subprocesses entering the various squark--antisquark production processes.
The diagrams in red enter only if $q= q'$, while the diagram in blue contributes only if  $q$ and $q'$ belong to the same SU$(2)$
doublet. The quarks $p_q$ and $r_q$ are defined in eqs.~(\ref{Eq:AncDef}).}
\label{Tab:PP}
\end{table}

In the following we describe the organization of the calculation of the EW corrections and we explain in detail how the different perturbative orders enter for the sixty-four final states. To this end, it is useful to identify classes of processes that share the same type of partonic initial states for the different perturbative order under consideration. 
Among the sixty-four final states  we may distinguish  five  different classes of processes:
%
%
%
%
%
%
\begin{subequations}
\begin{itemize}
\item $\mathbf{\tilde u_L \tilde u_L^\ast }${\bf-type} processes  producing squarks with same flavor and chirality,
\begin{align}
PP &\to 
\tilde u_\alpha \, \tilde u^\ast_\alpha, \; \;
\tilde d_\alpha \, \tilde d^\ast_\alpha, \;  \; 
\tilde c_\alpha \, \tilde c^\ast_\alpha, \;   \;
\tilde s_\alpha \, \tilde s^\ast_\alpha . \;   \; 
&  \phantom{\alpha \neq \beta \, .} 
\label{Eq:Type1} 
\end{align}
\item $\mathbf{\tilde u_L \tilde u_R^\ast }${\bf-type} processes  producing squarks with same flavor and different chirality,
\begin{align}
PP &\to 
\tilde u_\alpha \, \tilde u^\ast_\beta, \; \;
\tilde d_\alpha \, \tilde d^\ast_\beta, \;  \; 
\tilde c_\alpha \, \tilde c^\ast_\beta, \;   \;
\tilde s_\alpha \, \tilde s^\ast_\beta, \;   \; 
 &  \alpha \neq \beta \,  .
\label{Eq:Type1} 
\end{align}
\item $\mathbf{\tilde u_L \tilde c_L^\ast }${\bf-type} processes  producing two up-type or two down-type squarks with different flavor,
\begin{align}
PP &\to 
\tilde u_\alpha \, \tilde c^\ast_\beta, \; \;
\tilde c_\alpha \, \tilde u^\ast_\beta, \;  \; 
\tilde d_\alpha \, \tilde s^\ast_\beta, \;   \;
\tilde s_\alpha \, \tilde d^\ast_\beta. \;    
&  \phantom{\alpha \neq \beta \, .} 
\label{Eq:Type2} 
\end{align}
\item $\mathbf{\tilde u_L \tilde s_L^\ast }${\bf-type} processes  producing one up-type and one down-type squark  of different families,
\begin{align}
PP &\to 
\tilde u_\alpha \, \tilde s^\ast_\beta, \; \;
\tilde s_\alpha \, \tilde u^\ast_\beta, \;  \; 
\tilde d_\alpha \, \tilde c^\ast_\beta, \;   \;
\tilde c_\alpha \, \tilde d^\ast_\beta. \;    
&  \phantom{\alpha \neq \beta \, .} 
\label{Eq:Type2} 
\end{align}
\item $\mathbf{\tilde u_L \tilde d_L^\ast }${\bf-type} processes  producing one up-type squark and one down-type squark of the same family,
\begin{align}
PP &\to 
\tilde u_\alpha \, \tilde d^\ast_\beta, \; \;
\tilde d_\alpha \, \tilde u^\ast_\beta, \;  \; 
\tilde c_\alpha \, \tilde s^\ast_\beta, \;   \;
\tilde s_\alpha \, \tilde c^\ast_\beta. \;    
&  \phantom{\alpha \neq \beta \, .} 
\end{align}
\end{itemize}
\label{Eq:Type2} 
\end{subequations}
This classification is based only on the technical aspects of the computation. In particular it  does not  consider 
 the dependence of the partonic contributions on the chirality and the flavor of the produced squarks, e.g., 
$\tilde d_R \tilde d_R^\ast$   production    is considered to be a \ulult{} process. Therefore  members of the same class 
entail  numerically different contributions. For example, all necessary matrix elements for $\tilde d_R \tilde d_R^\ast$   production can be obtained directly from $\tilde u_L \tilde u_L^\ast$ matrix elements by flipping the chirality and SU$(2)$ states. In general, a global flip of chiralities, families and/or SU$(2)$ states project a specific squark--antisquark pair, and its corresponding matrix elements, into another one of the same class. In order to correctly perform such a flip, the Feynman rules for all interaction vertices are written in the most general form allowing for arbitrary chirality and SU(2) states. Correspondingly, for a specific chirality or flavor configuration within a given class individual diagrams might vanish. The presented classification allows for a dramatic reduction of the number of matrix elements that have to be practically computed. 
\medskip


In order to specify the structure of the EW corrections, for any partonic subprocess  $X$ we will denote  the contribution of $\Order{a}{b}$ to the total hadronic cross section as $\psec{a}{b}{X}$. This notation can be easily generalized to any observable and differential distribution for the squark--antisquark system.
\medskip

We start considering the  contributions to the cross section from tree-level diagrams only. As already noted,
these contributions can be of $\mathcal{O}(\alpha^2_s)$, 
$\mathcal{O}(\alpha_s\alpha)$ and $\mathcal{O}(\alpha^2)$. Since quark--antiquark initial states can  produce a
squark--antisquark pair with the same flavor configuration via a QCD
or a EW gauge boson or gaugino in the t-channel, the
amplitudes of  $\ord(\alpha_s)$ and $\ord(\alpha)$ for $q \bar q'\to
\tilde q_\alpha \tilde q'^\ast_\beta$ are always present. The corresponding Feynman diagrams are
shown for each class in the first two columns of Table \ref{Tab:PP}. Consequentially,  $\mathcal{O}(\alpha^2_s)$ 
and $\mathcal{O}(\alpha^2)$ cross sections are always nonzero for each squark--antiquark chiral and flavor configuration. On the contrary, $\mathcal{O}(\alpha_s\alpha)$ cross sections can be zero, since the interference of $\ord(\alpha_s)$ and $\ord(\alpha)$ amplitudes can vanish in particular classes due to the color flow or chirality structure. 

For this reason, it is convenient to separate  the contribution
$\psec{2}{0}{q \bar q'\to \tilde q_\alpha \tilde q'^\ast_\beta}$ and $\psec{0}{2}{q \bar q'\to \tilde q_\alpha \tilde q'^\ast_\beta} $ from
all the others in the sum over all parton channels,
\begin{align}
\psec{2}{0}{PP\to \tilde q_\alpha  \tilde q'^\ast_\beta}     & =   \psec{2}{0}{q \bar q'\to \tilde q_\alpha \tilde q'^\ast_\beta}  +    \csec{2}{0}{PP \to \tilde q_\alpha  \tilde q'^\ast_\beta} \, , \nonumber \\
\psec{1}{1}{PP\to \tilde q_\alpha  \tilde q'^\ast_\beta}     & = \phantom{  \psec{2}{0}{q \bar q'\to \tilde q_\alpha \tilde q'^\ast_\beta}  + \;\,}\csec{1}{1}{PP \to \tilde q_\alpha  \tilde q'^\ast_\beta}  \, ,   \nonumber \\
\psec{0}{2}{PP\to \tilde q_\alpha  \tilde q'^\ast_\beta}     & =   \psec{0}{2}{q \bar q'\to \tilde q_\alpha \tilde q'^\ast_\beta}  +    \csec{0}{2}{PP \to \tilde q_\alpha  \tilde q'^\ast_\beta}   \, .
\label{Eq:Tree}
\end{align}
In eq. \eqref{Eq:Tree} all the process-dependent  contributions are  included in the barred quantities $\csec{a}{b}{PP \to \tilde q_\alpha  \tilde q'^\ast_\beta}$. The specific structure of $\csec{a}{b}{PP \to \tilde q_\alpha  \tilde q'^\ast_\beta}$ for each one of the five classes is listed in the following,
\begin{subequations}
\begin{align}
& \text{\ulult: }    &&  
\left \{ \begin{array}{l} 
	\csec{2}{0}{PP \to \tilde q_\alpha  \tilde q^\ast_\alpha}  =  \psec{2}{0}{gg \to \tilde q_\alpha  \tilde q^\ast_\alpha}  
	 										+ \displaystyle \sum_{q'' \neq q} \psec{2}{0}{q'' \bar q''\to \tilde q_\alpha  \tilde q^\ast_\alpha}      \\
	\csec{1}{1}{PP \to \tilde q_\alpha  \tilde q^\ast_\alpha}  =  \psec{1}{1}{q \bar q\to \tilde q_\alpha  \tilde q^\ast_\alpha} + \psec{1}{1}{g\gamma\to \tilde q_\alpha  \tilde q^\ast_\alpha}    \phantom{ \displaystyle  \sum_{q'' \neq q} }\\
	\csec{0}{2}{PP \to \tilde q_\alpha  \tilde q^\ast_\alpha}  =   \displaystyle  \sum_{q'' \neq q} \psec{0}{2}{q'' \bar q''\to \tilde q_\alpha  \tilde q^\ast_\alpha}  \\
\end{array}  \right .  \, , \phantom{123456789} \\[1.5ex]					     
& \text{\ulclt: }   && 
\left \{ \begin{array}{l} 
	   \csec{2}{0}{PP \to \tilde q_\alpha \tilde q'^\ast_\beta}   =  0  \\
	     \csec{1}{1}{PP \to \tilde q_\alpha \tilde q'^\ast_\beta}  =  0  \\
	    \csec{0}{2}{PP \to \tilde q_\alpha \tilde q'^\ast_\beta}  =   \psec{0}{2}{p_q \bar p_{q'}\to \tilde q_\alpha  \tilde q'^\ast_\beta} \\
\end{array}  \right .   \, , \\[1.5ex]	
& \text{\uldlt: }   &&
\left \{ \begin{array}{l}
	\csec{2}{0}{PP \to \tilde q_\alpha \tilde q'^\ast_\beta}  =  0      \\
	\csec{1}{1}{PP \to \tilde q_\alpha \tilde q'^\ast_\beta}  = \psec{1}{1}{q \bar q'\to \tilde q_\alpha  \tilde q'^\ast_\beta} \\	
	\csec{0}{2}{PP \to \tilde q_\alpha \tilde q'^\ast_\beta}  =   \psec{0}{2}{r_q \bar r_{q'}\to \tilde q_\alpha  \tilde q'^\ast_\beta} 		\\
\end{array}  \right .   \, , \\[1.5ex]	
& \text{$\tilde u_L  \tilde u^\ast_R$ \mbox{ and } \ulslt: }   &&
\left \{ \begin{array}{l}
	\csec{2}{0}{PP \to \tilde q_\alpha \tilde q'^\ast_\beta}  =  0      \\
	\csec{1}{1}{PP \to \tilde q_\alpha \tilde q'^\ast_\beta}  = 0 \\	
	\csec{0}{2}{PP \to \tilde q_\alpha \tilde q'^\ast_\beta}  = 0		\\
\end{array}  \right .   \, .
%
\end{align}\label{bar-tree}
\end{subequations}
In the previous equations the  quark   $p_q$ denotes the SU$(2)$ partner of the quark $q$,
\begin{subequations}
\begin{align}
p_u & = d\, ,  & p_d & = u\, ,  & p_c & = s\, ,  & p_s & = c   \, ,
\end{align}
while the quark  $r_q$  is defined as follows,
\begin{align}
r_u & = c\, ,  & r_d & = s\, ,  & r_c & = u\, ,  & r_s & = d   \, ,
\end{align}
\label{Eq:AncDef}
\end{subequations}
i.e. ~$r_q$ and $q$ are same-charge quarks belonging  to different generations.  All these contributions have been successfully checked at numerical level against both {\tt Madgraph}~\cite{Alwall:2014hca} and 
 {\tt Prospino}~\cite{Beenakker:1996ed}.
\medskip

The   NLO EW corrections, i.e.~the $\mathcal{O}(\alpha_s^2\alpha)$ contribution, constitute the original computation of this paper.  Besides virtual corrections,  this order receives contributions also from the real 
radiation of gluons, photons and (anti-)quarks. The $\mathcal{O}(\alpha_s^2\alpha)$ corrections to the cross section can be written as follows,
\begin{align}
\psec{2}{1}{PP\to \tilde q_\alpha  \tilde q'^\ast_\beta}     & =   \psec{2}{1}{q \bar q'\to \tilde q_\alpha \tilde q'^\ast_\beta}  +  
\psec{2}{1}{q \bar q'\to \tilde q_\alpha \tilde q'^\ast_\beta g} + \psec{2}{1}{q \bar q'\to \tilde q_\alpha \tilde q'^\ast_\beta \gamma} \nonumber \\
& \qquad + \psec{2}{1}{q g \to \tilde q_\alpha \tilde q'^\ast_\beta  q'} + \psec{2}{1}{\bar q' g \to \tilde q_\alpha \tilde q'^\ast_\beta  \bar q}  +
  \csec{2}{1}{PP \to \tilde q_\alpha  \tilde q'^\ast_\beta} \, .
  \label{Eq:Cor21}
\end{align}
Again, with the barred quantity $\csec{2}{1}{PP \to \tilde q_\alpha
  \tilde q'^\ast_\beta}$ we denote the set of initial states that are class-dependent.
The term  $\csec{2}{1}{PP \to \tilde q_\alpha  \tilde q'^\ast_\beta} $
is nonzero only  for \ulult{} processes, i.e., 
\begin{subequations}
\begin{align}
&\text{\ulult: }   &    \csec{2}{1}{PP \to \tilde q_\alpha \tilde q'^\ast_\alpha} &=       \psec{2}{1}{gg \to \tilde q_\alpha  \tilde q^\ast_\alpha}  + \psec{2}{1}{gg \to \tilde q_\alpha  \tilde q^\ast_\alpha\gamma} \nonumber \\
                                    && &\quad +                        \sum_{q'' \neq q}  \bigg (  \psec{2}{1}{q'' \bar q''\to \tilde q_\alpha  \tilde q^\ast_\alpha}  + 
 						   \psec{2}{1}{q'' \bar q''\to \tilde q_\alpha  \tilde q^\ast_\alpha g} \nonumber  \\
                                  && &\quad +   \psec{2}{1}{q'' \bar q''\to \tilde q_\alpha  \tilde q^\ast_\alpha \gamma} +  \psec{2}{1}{q'' g \to \tilde q_\alpha  \tilde q^\ast_\alpha q''}  + 
                                  \psec{2}{1}{\bar q'' g \to \tilde
                                    q_\alpha  \tilde q^\ast_\alpha \bar q''} \bigg )  \, , \\[1.5ex] 
&\text{other processes: } & \csec{2}{1}{PP \to \tilde q_\alpha \tilde q'^\ast_\beta} &= 0     \, .
\end{align}   \label{bar-NLOEW}  
\end{subequations}
At the same order of perturbation theory, i.e. at $\ord(\alpha_s^2\alpha)$ in principle also NLO QCD corrections to the $\ord(\alpha_s\alpha)$ photon-induced squark--antisquark production enter. We do not consider these contribution here.

The NLO EW corrections~(\ref{Eq:Cor21}) depend on the full set of the MSSM parameters and have been evaluated by using  {\tt FeynArts}~\cite{Hahn:2000kx,Hahn:2006qw},  {\tt FormCalc}~\cite{Hahn:2006qw,Hahn:2001rv} and,
for the evaluation of the one-loop integrals, {\tt LoopTools}~\cite{Hahn:2006qw}. The cancellations of the ultraviolet divergences involves both $\mathcal{O}(\alpha_s)$ and $\mathcal{O}(\alpha)$ one-loop renormalization and has been extensively   described in  ~\cite{Hollik:2008yi,Germer:2010vn}.
Infrared singularities (IR)  are regularized   by introducing a  small mass for the  photon and the 
gluon, while quark masses are kept as regulators for the collinear singularities. 
In the processes considered, the IR singularities of gluonic origin  are Abelian-like and can be safely treated  by using  mass regularization.
Infrared and collinear singularities are handled by  the double cut-off phase-space-slicing method~\cite{Yennie:1961ad,Weinberg:1965nx,Baier:1973ms}, along the lines of~\cite{Germer:2010vn}. The
initial-state collinear singularities of gluonic (photonic) origin are factorized and absorbed
in the parton distribution functions by using the $\overline{\mbox{MS}}$ (DIS) scheme. We carefully checked  that, on the level of inclusive cross sections and of individual distributions, all our numerical results do not depend on the two phase-space slicing parameters and on the fictitious gluon and light-flavor quark masses.

 The amplitudes for partonic processes  with  (anti)quark--gluon in the initial state 
may exhibit an internal gluino, neutralinos or  charginos that can go on-shell. This happens when one of these particles is heavier than one of the (anti)squarks produced. However, the corresponding Breit-Wigner distributions appear only at $\ord(\alpha_s^3)$ or $\ord(\alpha_s \alpha^2)$, but not at $\ord(\alpha_s^2 \alpha)$.  Thus, similarly to the squark-squark production case~\cite{Hollik:2008yi,Germer:2010vn}, in  NLO EW corrections these singularities do not correspond to a physical resonance. In order to avoid numerical instabilities,  the poles have been regularized by including the width for the resonant particle in the corresponding propagators. Practically, the width plays the role of a regulator parameter, numerical results do not depend on its value.

Technical details of our calculation have already been extensively discussed  in \cite{Hollik:2008yi}, where  NLO EW corrections were calculated for the $\tilde u_L \tilde u_L^\ast$-type processes only. 
\medskip

The NLO QCD corrections, of $\mathcal{O}(\alpha_s^3)$, depend only on the mass of the  squarks and of the gluino
and   have been computed in~\cite{Beenakker:1996ch}.  For the phenomenological studies  of section \ref{sec:numerics}, we evaluated them  for total cross sections in the degenerate squark-mass case with the code  {\tt Prospino}~\cite{Beenakker:1996ed}.
For completeness we show the general structure of the NLO QCD contributions, using the same notation adopted in eqs. \eqref{Eq:Cor21} and \eqref{bar-NLOEW} for NLO EW corrections:
\begin{align}
\psec{3}{0}{PP\to \tilde q_\alpha  \tilde q'^\ast_\beta}     & =   \psec{3}{0}{q \bar q'\to \tilde q_\alpha \tilde q'^\ast_\beta}  +  
\psec{3}{0}{q \bar q'\to \tilde q_\alpha \tilde q'^\ast_\beta g} + \psec{3}{0}{q g \to \tilde q_\alpha \tilde q'^\ast_\beta  q'} \nonumber \\
& \qquad + \psec{3}{0}{\bar q' g \to \tilde q_\alpha \tilde q'^\ast_\beta  \bar q}  +
  \csec{3}{0}{PP \to \tilde q_\alpha  \tilde q'^\ast_\beta} \, .
\end{align}
Also in this case, the barred quantity $\psec{3}{0}{\bar q' g \to \tilde q_\alpha \tilde q'^\ast_\beta  \bar q}$  depends on the flavor and the chirality of the final states:
\begin{subequations}
\begin{align}
&\text{\ulult: }   &    \csec{3}{0}{PP \to \tilde q_\alpha \tilde q'^\ast_\alpha} &=    \psec{3}{0}{gg \to \tilde q_\alpha  \tilde q^\ast_\alpha}  +  \psec{3}{0}{gg \to \tilde q_\alpha  \tilde q^\ast_\alpha g}  
				+  \sum_{q'' \neq q}  \bigg (  \psec{3}{0}{q'' \bar q''\to \tilde q_\alpha  \tilde q^\ast_\alpha}  \nonumber  \\
                                  && &\quad + \psec{3}{0}{q'' \bar q''\to \tilde q_\alpha  \tilde q^\ast_\alpha g}  +   \psec{3}{0}{q''g \to \tilde q_\alpha  \tilde q^\ast_\alpha q''} 
                                        +  \psec{3}{0}{\bar q'' g \to \tilde q_\alpha  \tilde q^\ast_\alpha \bar q'' }  \bigg )  \, ,   \\[1.5ex] 
& \text{other processes: }  &                    \csec{3}{0}{PP \to \tilde q_\alpha \tilde q'^\ast_\beta} &= 0   \, .         
\end{align} 
\end{subequations}

\section{Numerical results}
\label{sec:numerics}
The results presented in this paper  are obtained for a 
hadronic center-of-mass energy of $\sqrt{S} = 13$ TeV.
The numerical evaluation of the hadronic cross sections
has been performed by using the   {\tt NNPDF2.3QED\_mc}
PDF set~\cite{Ball:2013hta}, which includes the photon PDF and LO QED effects in the evolution of all the  PDF  members. 
Moreover, this set of positive-definite
PDFs  provides an  excellent description of all hard scattering
data~\cite{Deans:2013mha} avoiding  the negative cross sections 
for high-mass particle production described  in~\cite{Ball:2014uwa}.

The relevant Standard Model input parameters are taken from~\cite{Agashe:2014kda} and, consistently with the specific PDF set used, we set $\alpha_s(m_{{\rm Z}})=0.119 $ in the numerical evaluations.
Fermions of the first two generations are considered as massless.
All the MSSM masses and couplings  are determined 
 by eleven independent TeV-scale parameters,
 \begin{align}
m_{A_0} \, ,     && 
\tan {\beta} \, ,  &&  
X_t  \, ,             &&  
\mu  \, ,            &&  
M_2  \, ,           && 
m_{\tilde g}  \, ,               &&
M_{\tilde q, \rm L} \, ,     && 
M_{\tilde q, \rm R} \, ,     && 
M_{\tilde \ell }  \, , &&
M_{\tilde q _{ 3}} \, ,      &&   
M_{\tilde \ell_3}\,  . 
\label{Eq:Inputs}
 \end{align}
where, for the soft-breaking sfermion-mass parameters the relations
\begin{subequations}
\begin{IEEEeqnarray}{rCcCl C  rCcCl  C    rCl}
M_{\tilde t,  \rm L/R} &=&  M_{\tilde b ,  \rm L/R}  &=& M_{\tilde q_3}\, ,   &\phantom{abcdef}&
M_{\tilde f,  \rm L/R} &=&   M_{\tilde q,   \rm L/R}\, , &&   &\phantom{abcde}&  
 (f& =& u,d,c,s)\, ,  
\nonumber \\
 M_{\tilde \tau, \rm L/R} &=&   M_{\tilde \nu_\tau , \rm L} &=&  M_{\tilde \ell_3}\, ,    &&
 M_{\tilde f,  \rm L/R} &=&  M_{\tilde \nu_f, \rm L } &=&  M_{\tilde \ell}\,  ,   &&
(f& =& e,\mu) \, ,
\label{MSSMpar}
\end{IEEEeqnarray}
are assumed, and the bino mass $M_1$ and the sfermion trilinear couplings are set to 
\begin{equation}
M_1  =  \frac{5}{3}  \frac{   \sin^2 \theta_{\rm W}  }{   \cos^2 \theta_{\rm W} }  \, M_2 \, ,  \qquad \qquad 
A_t  =    X_t - \frac{\mu}{\tan\beta}   \, , \qquad A_f  = 0 \, , \qquad
(f = e,\mu,\tau,u,d,c,s,b)  \, .
\label{MSSMpar2}
\end{equation}
\end{subequations}
In eq.~(\ref{MSSMpar2}), $M_1$  is obtained  from gaugino-mass unification at the GUT scale.
The physical spectrum and all the Feynman rules are derived from this parameter set  by using 
tree-level relations. The only exception is the physical mass of 
the Higgs bosons, which is computed with the help of  {\tt FeynHiggs 2.10}~\cite{Heinemeyer:1998np,Heinemeyer:1998yj,Frank:2006yh,Hahn:2009zz,Hahn:2013ria}. 
\medskip

\begin{figure}[t]
\centering
\includegraphics[width=10.cm]{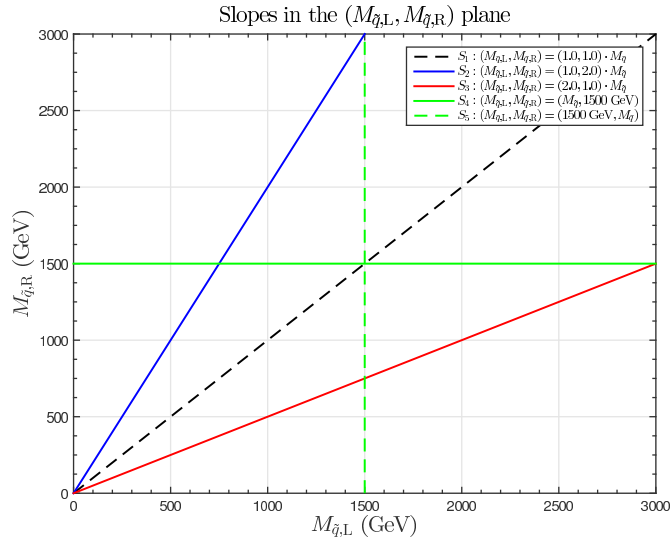}
\caption{Slopes in the  $(M_{\tilde q,  \rm L}, M_{\tilde q,  \rm R})$ plane.}
\label{Fig:slopes}
\end{figure}

In order to study both degenerate and non-degenerate scenarios, in the numerical discussion we will consider particular {\it slopes}   of the $(M_{\tilde q,  \rm L}, M_{\tilde q,  \rm R})$ plane, parametrized in 
terms of a mass $M_{\tilde q}$ as follows (see also Fig.~\ref{Fig:slopes}):
\begin{IEEEeqnarray}{rCc  rCl}
&\mbox{slope } &S_1:& \phantom{abcde}     &(M_{\tilde q,  \rm L}, M_{\tilde q,  \rm R})=& (1,1) \cdot  M_{\tilde q}  \, ,\nonumber \\ 
&\mbox{slope } &S_2:& \phantom{abcde}     &(M_{\tilde q,  \rm L}, M_{\tilde q,  \rm R})=& (1,2) \cdot  M_{\tilde q} \, , \nonumber \\ 
&\mbox{slope } &S_3:& \phantom{abcde}     &(M_{\tilde q,  \rm L}, M_{\tilde q,  \rm R})=& (2,1) \cdot  M_{\tilde q} \, , \nonumber \\ 
&\mbox{slope } &S_4:& \phantom{abcde}     &(M_{\tilde q,  \rm L}, M_{\tilde q,  \rm R})=& (  M_{\tilde q}, 1500\mbox{ GeV}) \, , \nonumber \\ 
&\mbox{slope } &S_5:& \phantom{abcde}     &(M_{\tilde q,  \rm L}, M_{\tilde q,  \rm R})=& (1500\mbox{ GeV},   M_{\tilde q} ) \, .  
\label{eq:slopes}
\end{IEEEeqnarray}

The slope $S_1$ represents the degenerate case, whereas the other slopes correspond to different possible non-degenerate spectra. The motivation for considering non-degenerate scenarios is twofold.
First, EW corrections depend on the chiralities of the produced particles. Thus, a mass gap between the different-chirality squarks is expected to modify the impact of EW corrections on the inclusive cross section for squark--antisquark production. Second, squarks of different chiralities may be  experimentally distinguishable by their different decay products~\cite{Hollik:2013xwa}. This is already relevant for the degenerate case\footnote{Experimental analyses typically assume simplified models, where all the squarks decay directly into the lightest neutralinos. Allowing for different decays, very different signatures would emerge from left- and right-handed squarks. The corresponding bounds from direct searches would be consequentially modified.} and  plays an even more important role in case of large mass gaps, where the production of heavy states is suppressed and the experimental signature is determined by the decays of the lighter states.
\medskip

In order to identify the dependence of the cross sections on  the chirality of the produced (anti-)squarks, we will consider 
four different classes of final states:
\begin{itemize}
\item ``LL" : production of a left-handed squark and a left-handed antisquark,
\item ``RR" : production of a right-handed squark and a right-handed antisquark,
\item ``LR+RL" : production of a squark and an antisquark with different chiralities,
\item ``incl." : inclusive production of a   squark--antisquark pair,  
\end{itemize} 
where the inclusive production is by definition given by the sum of all the final states in the ``LL", ``RR" and ``LR+RL" classes. 
Depending on the class and the point of the slope considered, the masses of a given squark--antisquark pair can be considerably different.
As a general approach, we set the  factorization scale $\mu_{\rm F}$ and the renormalization scale $\mu_{\rm R}$    equal to the average of the mass of the
produced particles, $\mu_{\rm F} = \mu_{\rm R} = 
( m_{\tilde q_\alpha}  + m_{\tilde q'_\beta} ) / 2$, independently for each final state.
It is worth to note that each one of the classes of processes in eq. \eqref{Eq:Type2} can in general contribute to LL, RR and LR+RL production. The only exceptions are {\ulult} processes, which do not contribute to LR+RL production, and {\ulurt} processes which do not contribute to LL and RR production.
\medskip

In the numerical results we will refer to any contribution either specifying explicitly its perturbative order (and partonic initial-state) or using the following standard notation:
\begin{itemize}
\item ``LO"  refers to  $\mathcal{O}(\alpha_s^2)$ contributions,
\item ``NLO QCD"  refers to the sum of  $\mathcal{O}(\alpha_s^2)$  and  $\mathcal{O}(\alpha_s^3)$ contributions, 
\item ``EW corr." refers to the corrections of EW origin, i.e. to  the sum of  $\mathcal{O}(\alpha_s\alpha)$, $\mathcal{O}(\alpha^2)$ and $\mathcal{O}(\alpha_s^2 \alpha)$ contributions,
\item ``EW corr. (no $g \gamma$)" refers to the EW corrections {\it without} the contribution of the $g\gamma$ channel. 
\end{itemize} 

Before presenting our numerical results, we want to comment on the choice and the motivations of presenting EW corrections with and without $\ord(\alpha_s \alpha)$ contributions from the $g \gamma$ initial state. 

\subsection{PDF uncertainties in the gluon--photon channel}
\begin{figure}[t]
\begin{subfigure}[b]{0.5\textwidth}
\includegraphics[width=7.2cm, height=6.3cm]{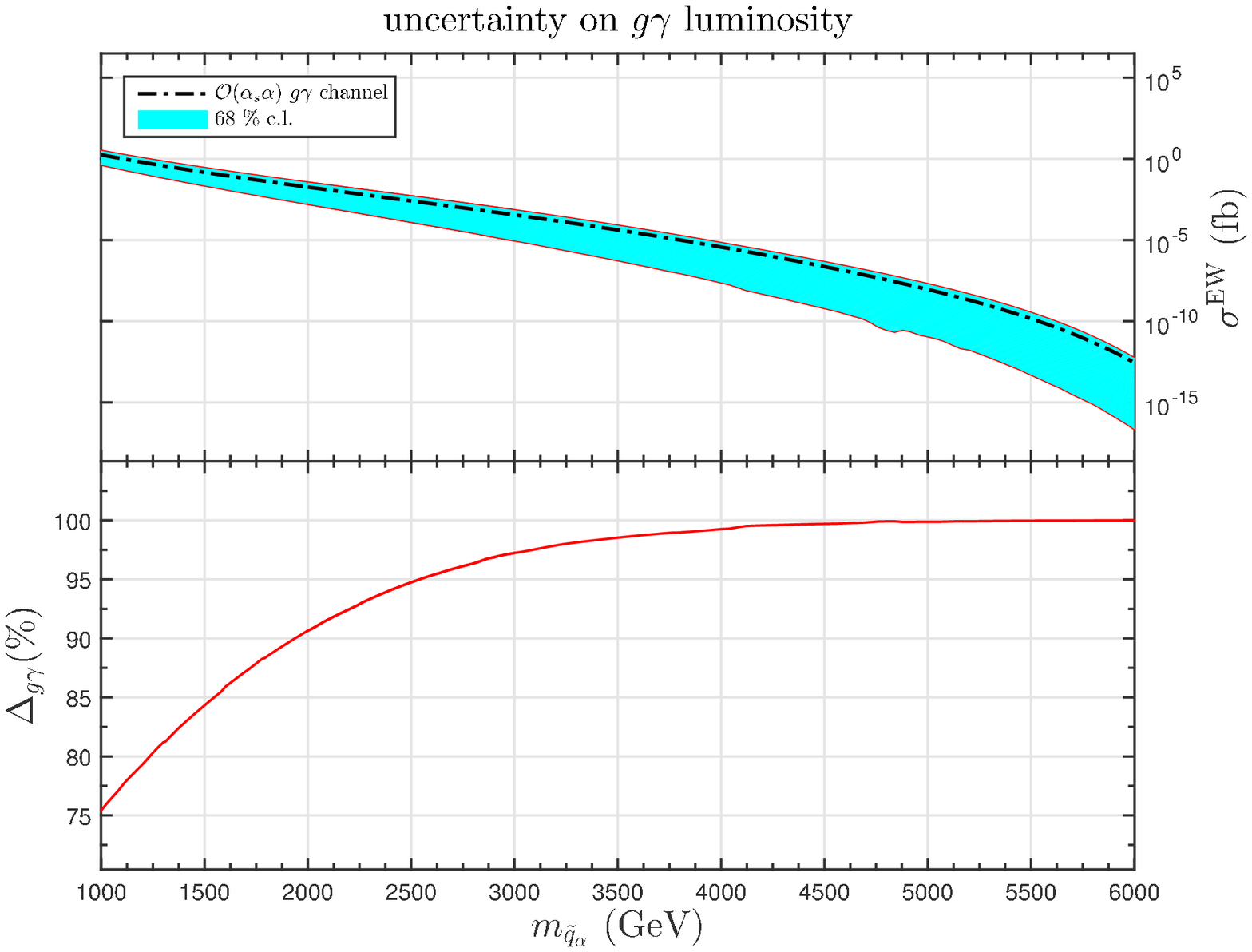}
\caption{}
\end{subfigure}
\begin{subfigure}[b]{0.5\textwidth}
\includegraphics[width=7.2cm, height=6.3cm]{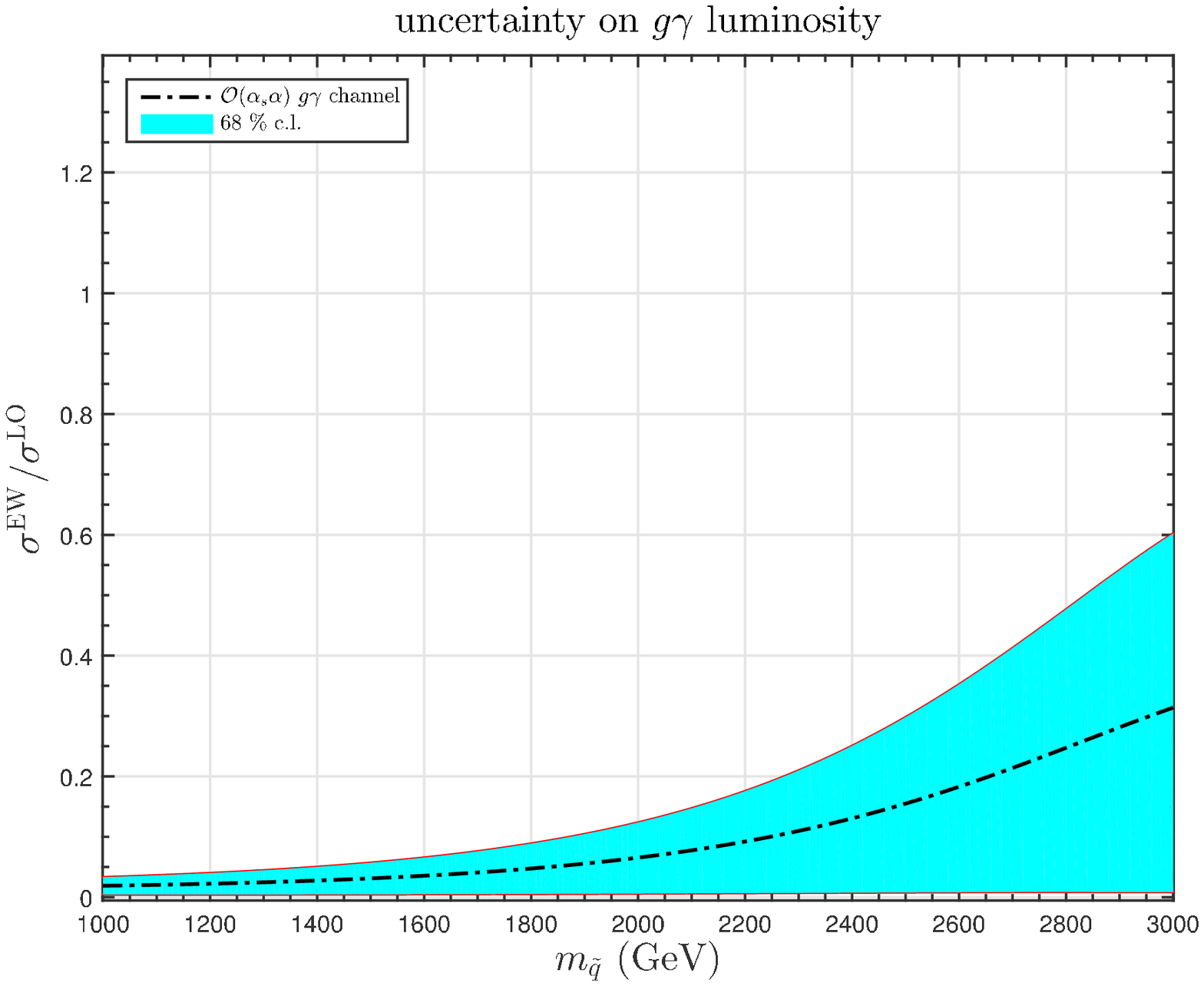}
\caption{}
\end{subfigure}
\phantom{} \\
\begin{subfigure}[b]{0.5\textwidth}
\includegraphics[width=7.2cm, height=6.3cm]{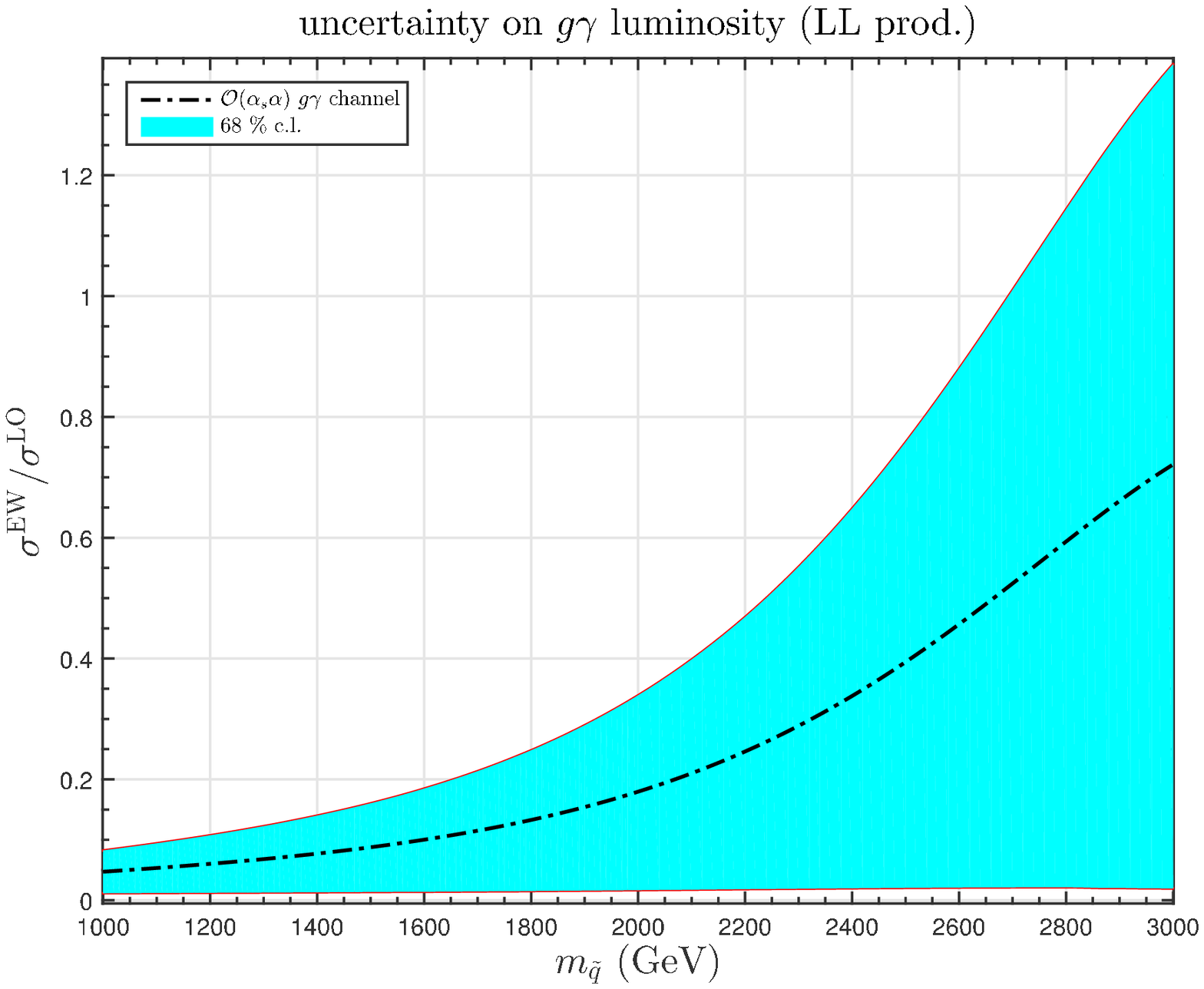}
\caption{}
\end{subfigure}
\begin{subfigure}[b]{0.5\textwidth}
\includegraphics[width=7.2cm, height=6.3cm]{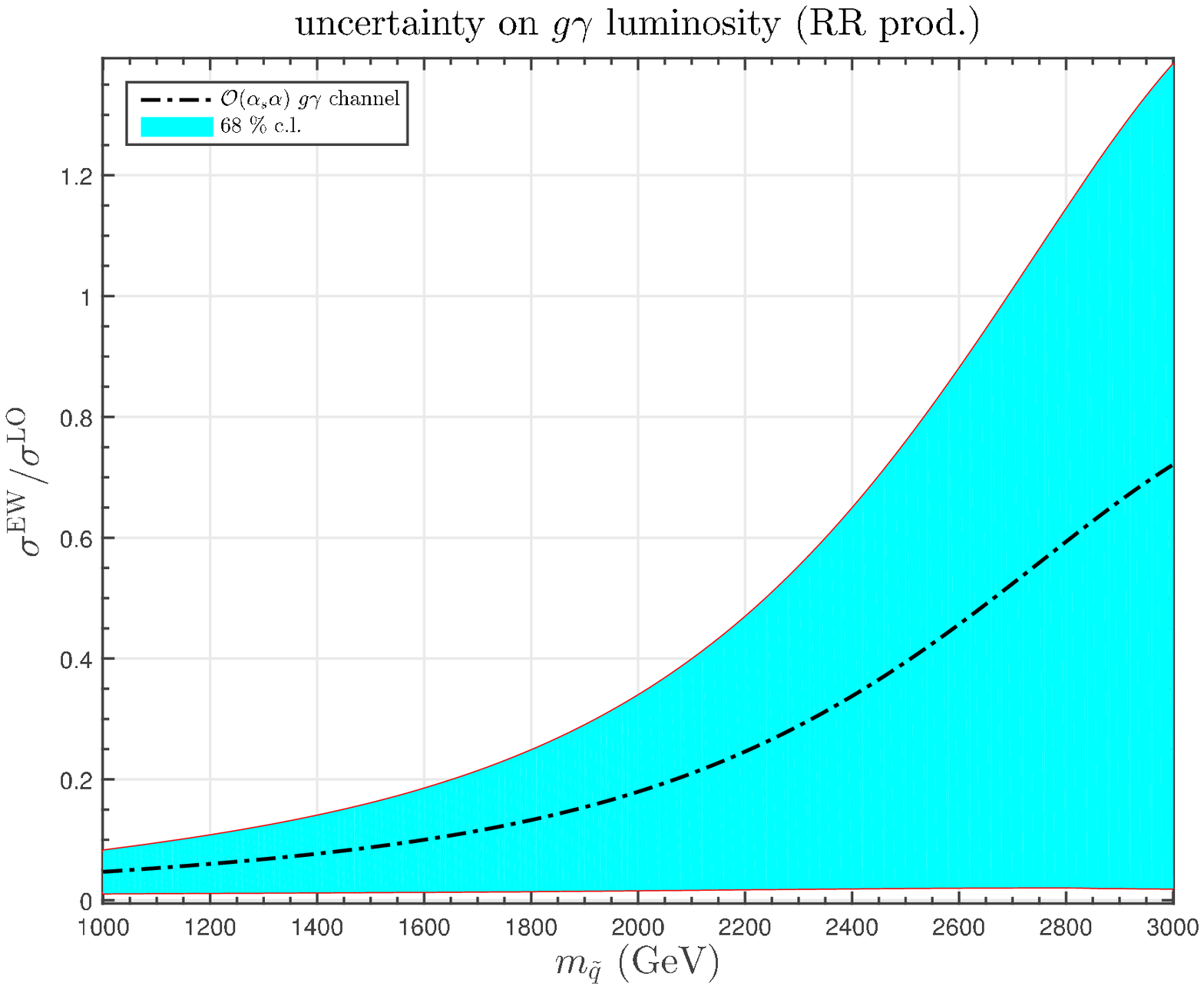}
\caption{}
\end{subfigure}
\caption[.]{
Absolute and relative contributions of the photon--gluon channel to the total hadronic cross section. Central values (black dash-dotted lines) and PDF uncertainties (blue bands) are plotted as function 
of a common squark mass $m_{\tilde q}$  ($m_{\tilde g} = 1500$ GeV). 
In panel (a) $\Delta_{g \gamma \to \tilde q_\alpha \tilde q_\alpha^\ast}$ denotes
the ratio between the PDF uncertainties 
of the cross section of
the $g \gamma$ channel and its absolute central value.
}
\label{Fig:GYunc}
\end{figure}

Direct production of  massive particles, i.e. with masses above $1$ TeV,   probes
PDFs at large  Bjorken-$x$, where they are  poorly constrained by experimental data and exhibit large uncertainties. 
In the case of  the direct production of  $1$ TeV ($2$ TeV)  squark--antisquark pairs, e.g., the  intrinsic uncertainties on the gluon and the quark PDFs
lead to an uncertainty on the  total NLO QCD cross section of the order of $4 \%$ ($17 \%$)~\cite{Ball:2014uwa}.
EW corrections originate in general from processes with the same initial states of those from QCD corrections. Thus, the uncertainty on the total cross section induced by intrinsic PDF uncertainties from EW corrections is expected to be a fraction of the analogue contribution from the NLO QCD cross section.

An important exception in our calculation is the $\ord(\alpha_s \alpha)$ contribution from gluon--photon initial state, which appears in same-flavor same-chirality squark--antisquark production (\ulult{}  processes). This initial state contributes to the EW corrections, but it does not contribute to LO cross sections and NLO QCD corrections.
At $\ord(\alpha_s \alpha)$, the only MSSM parameters entering the cross sections in the $g \gamma$ channel are the masses of squarks and the gluino.
Moreover, the photon PDF in {\tt NNPDF2.3QED} is affected by large uncertainties ~\cite{Ball:2013hta}, which have to be taken into account to correctly identify the impact of higher-order corrections  \cite{Boughezal:2013cwa,Carrazza:2014cpa}. For this reason, before discussing the effect of EW corrections, we consider the PDF uncertainties arising from  $g\gamma$-induced production and we describe their impact on our calculation and phenomenological studies.
Here, we do not want to perform a complete analysis on the effects of PDF uncertainties from EW corrections\footnote{This kind of study is typically not  done even for SM processes. In case of discovery, it can be easily performed if it is necessary.}, instead we want to show how the evaluation of photon-induced production without PDF uncertainty can give qualitatively misleading results in the context of TeV-scale supersymmetric-particle  production.

Fig.~\ref{Fig:GYunc}(a) depicts the contribution of the $g\gamma$-channel to squark--antisquark production as a function of
the  common mass of the produced squarks, $m_{\tilde q}$ (degenerate case). The blue band  around the central value (black solid line) corresponds to the PDF $68 \%$ c.l. error computed along the lines of~\cite{Ball:2014uwa}. In particular, since the photon PDF is positive by construction, replicas distribute in a very non-gaussian way. The $68 \%$ c.l. error around the central value corresponds to the symmetric error that includes 68 of the 100 results given by the different PDF replica from {\tt NNPDF2.3QED}. In the lower panel, we plot as function of $m_{\tilde q}$ the value of $\Delta_{g\gamma}$, which is defined as the ratio of the $68 \%$ c.l.~error and the central value for the $g\gamma$-channel.
As expected, the PDF uncertainty is very large. It ranges from $75 \%$  at $m_{\tilde q} = 1$~TeV to almost $100\%$ at $m_{\tilde q} \ge 4$~TeV.
Confronted with such a huge uncertainty it has to be understood that for heavy squarks the contribution from the $g\gamma$-channel is compatible both, with zero and the double of its central value. In a new version of {\tt NNPDF} or in another future   set of PDFs with QED evolution the size of the error could be very different. Thus, it is wise to keep the $g\gamma$-channel always as a separate contribution to the overall EW corrections.
 
The importance of keeping this term separate is even more evident when its size is compared to that of the LO cross section. 
An illustrative example may be found in Fig.~\ref{Fig:GYunc}(b), which shows the impact of the $g \gamma$ channel relative to the leading order contribution, computed assuming  $m_{\tilde g} = 1500$~GeV. 
In the following we will focus on the productions of squark with masses below $2500$ GeV, where  the $g\gamma$ channel is supposed to be smaller than $30 \%$ of the LO cross-section.  This mass range  will be probed by  Run-II of the LHC~\cite{ATL-PHYS-PUB-2014-010}. 

Actually, the relative contribution of the $g\gamma$ channel to the overall squark--antisquark production cross section is diluted by the ``LR+RL'' processes, which do not involve the $g\gamma$ initial state.  In the case of only LL or RR production, which do include the $g\gamma$ initial-state contribution from \ulult{}  processes, the impact of the photon PDF uncertainty is even more important,  cfr. Figs.~\ref{Fig:GYunc}(c) and~\ref{Fig:GYunc}(d). With $m_{\tilde q} = 2.5$ TeV, the $g\gamma$ channel can induce corrections from zero up to $80 \%$ of the LO results. 
Clearly, this means that the precise contribution from the $g \gamma$ channel is largely unknown, but it is potentially large. Conversely, without taking into account PDF uncertainties, one would be tempted to claim large  effects of order 40\% from photon-induced EW corrections. 
In the following discussion, we will not show  the uncertainty band of the  $g\gamma$ channel. However, as general rule, one has to bare in mind  an $\ord(100\%)$ uncertainty is always associated to its contribution.  
%

\subsection{Inclusive cross sections}

\begin{table}[t]
%
\centering
\begin{tabular}{ c | c   ||  c |  c }
\hline
\hline
 $m_{A_0}$  & $700$    GeV & $\tan \beta$ & $20$ \\
  $X_t$ & $2 M_{\tilde q_3}$       &   $\mu$ & $350$ GeV  \\
 $M_2$          & $350$     GeV   &  $m_{\tilde g}$ & $1500$   GeV  \\
 $M_{\tilde q, \rm L}$ & $1500$ GeV &     $M_{\tilde q, \rm R}$ & $1500$    GeV \\
  $M_{\tilde q_{3}}$ & $500$ GeV   &  $M_{\tilde \ell_3}$ & $1000$    GeV \\
  $M_{\tilde \ell}$ & $500$    GeV   & & \\
\hline
\hline
\end{tabular}
%
%
\caption{Default parameters within our eleven-parameter phenomenological MSSM.}
\label{Tab:Bench}
\end{table}

For our numerical evaluation we consider  the   benchmark scenario
defined  in Table~\ref{Tab:Bench}, which is a slight modification of
the  ``light-stop'' scenario defined
in~\cite{Carena:2013qia}. Starting from this scenario we perform
various one-dimensional scans, by varying the parameters in the
list~(\ref{Eq:Inputs}). Specifically, we considered scans over $M_2$,
$m_{\tilde g}$, $\mu$, $\tan \beta$ and  five different slopes in the
$(M_{\tilde q, {\rm L}},M_{\tilde q, {\rm R}})$ plane, which have been
defined in~(\ref{eq:slopes}). Since we always consider $M_{\tilde q, {\rm L/R}}>1~\text{TeV}$, the soft breaking parameter  $M_{\tilde q, {\rm L/R}}$ can be safely considered as equal to the physical masses  $m_{\tilde q, {\rm L/R}}$.\footnote{The physical masses are given by
$ m_{\tilde q_{L/R}}  \simeq M_{\tilde q, {\rm L/R }} \,  \left ( 1 + \mathcal{O}\left (  \frac{m_{\rm Z}^2 }{    M^2_{\tilde q, {\rm L/R}}  } \right ) \right )$. 
Thus, when
e.g.  $ M_{\tilde q, {\rm L/R}}  = 1 / 1.5 /  2$ TeV, the physical mass and the soft mass parameters are equal at the $0.8 / 0.4 / 0.2\%$ level.}
The regions considered in the scans are within the limits from the
Higgs sector, i.e. they exhibit  a Higgs-state close to the observed one, 
with couplings compatible with the observed rates in the Higgs search channels. 
The compatibility between the scenarios  and the experimental results 
has been checked by using the  codes  {\tt HiggsBounds}~\cite{Bechtle:2008jh,Bechtle:2011sb,Bechtle:2013gu,Bechtle:2013wla}
 and {\tt HiggsSignals}~\cite{Bechtle:2013xfa}, by following the procedure described in~\cite{Germer:2014jpa}.

\medskip
The results of the scans are shown in Figures~\ref{Fig:LSS1_M2}-\ref{Fig:LSS5_MSQ12}. Since the scans over $\mu$ and $\tan \beta$ show that the total cross section is very insensitive to their values, we did not include them here. 
Each figure collects  six plots related to a particular  scan.  Plot~(a)  shows  the LO and NLO  cross section predictions, inclusive over the sixty-four squark--antisquark pairs.   In the lower panel of plot~(a) we show the size of the corresponding relative corrections to the LO result.   NLO QCD corrections are included only for degenerate squarks.
The plot~(b)  shows the contributions of the individual channels and perturbative orders to the relative EW corrections and their sum. Thus, the blue line in plot~(b) and the lower panel of plot~(a) refer to the same quantity.
 Plot~(c) shows the LO contributions of LL, RR, LR+RL processes and their sum; in the lower panel we display the relative EW corrections for each one of these classes.  Panels~(d), (e) and (f) 
respectively show the same kind of plot depicted in panels~(b), but
for  the individual cases of LL, RR and LR+RL squark--antisquark production.

\subsubsection*{Degenerate squark masses}

\begin{figure}[t]
\begin{subfigure}[b]{0.5\textwidth}
\includegraphics[width=7.2cm,height=6.3cm]{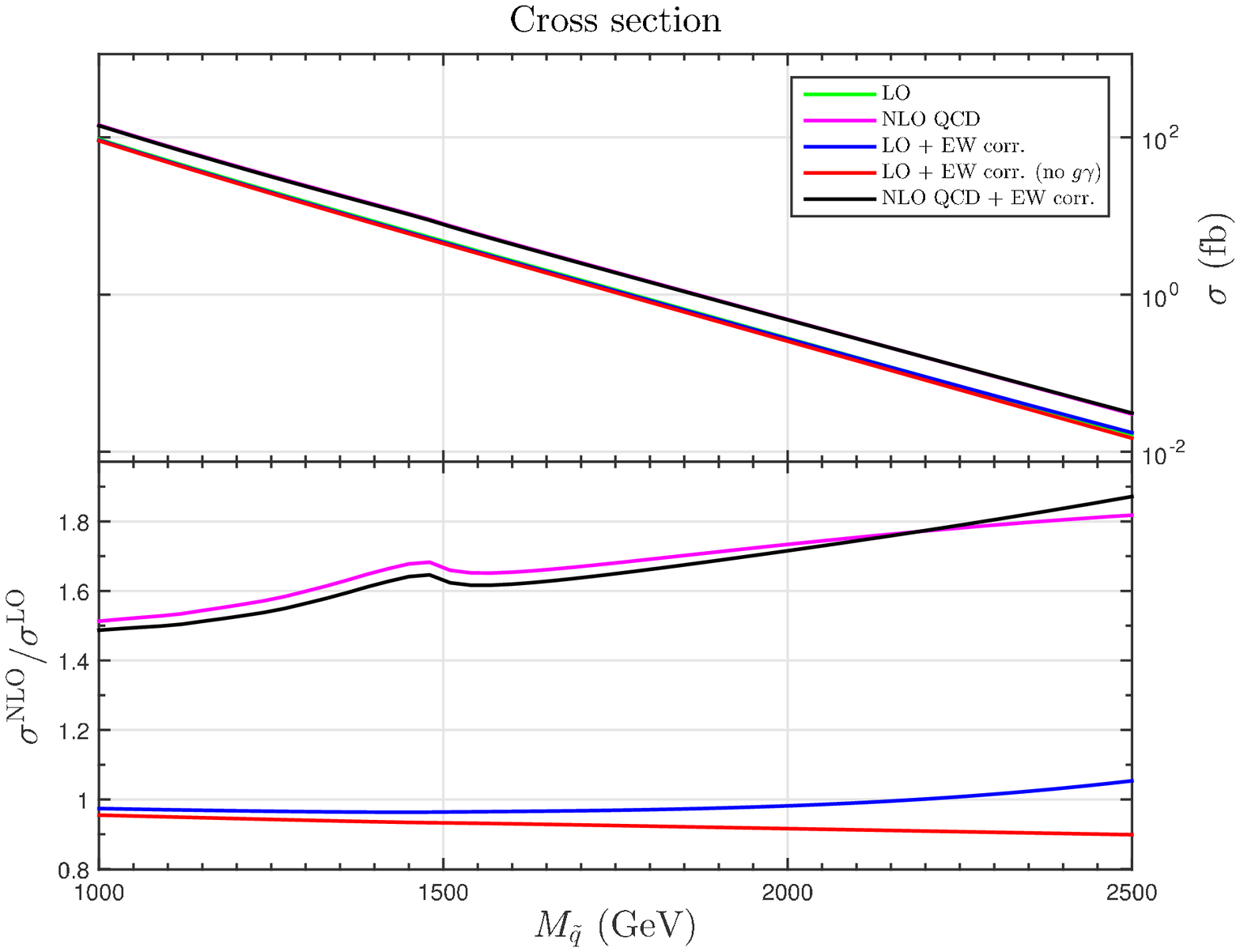}
\caption{}
\end{subfigure}
\begin{subfigure}[b]{0.5\textwidth}
\includegraphics[width=7.2cm,height=6.3cm]{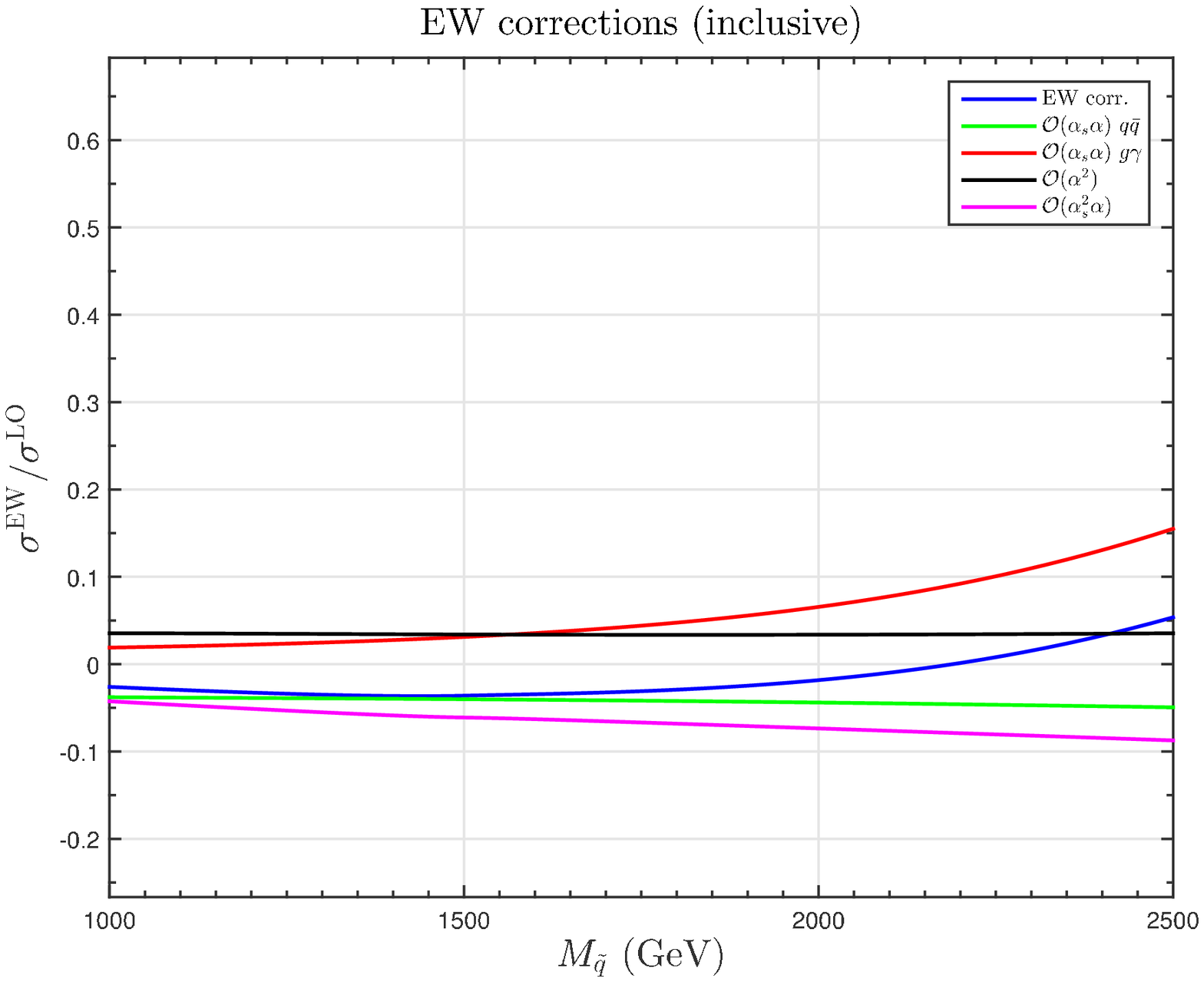}
\caption{}
\end{subfigure}
\phantom{pic}  \\
\begin{subfigure}[b]{0.5\textwidth}
\includegraphics[width=7.2cm,height=6.3cm]{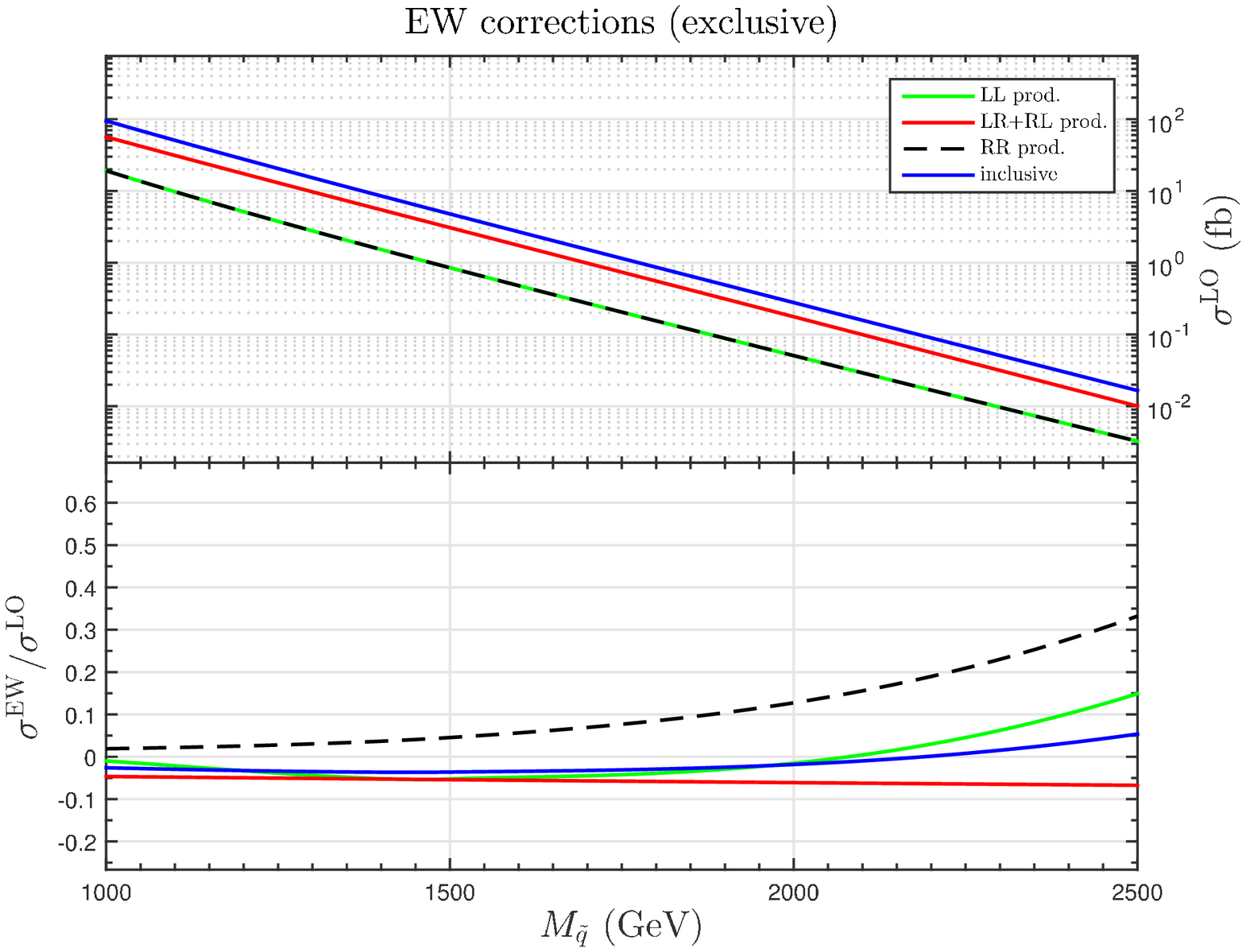}
\caption{}
\end{subfigure}
\begin{subfigure}[b]{0.5\textwidth}
\includegraphics[width=7.2cm,height=6.3cm]{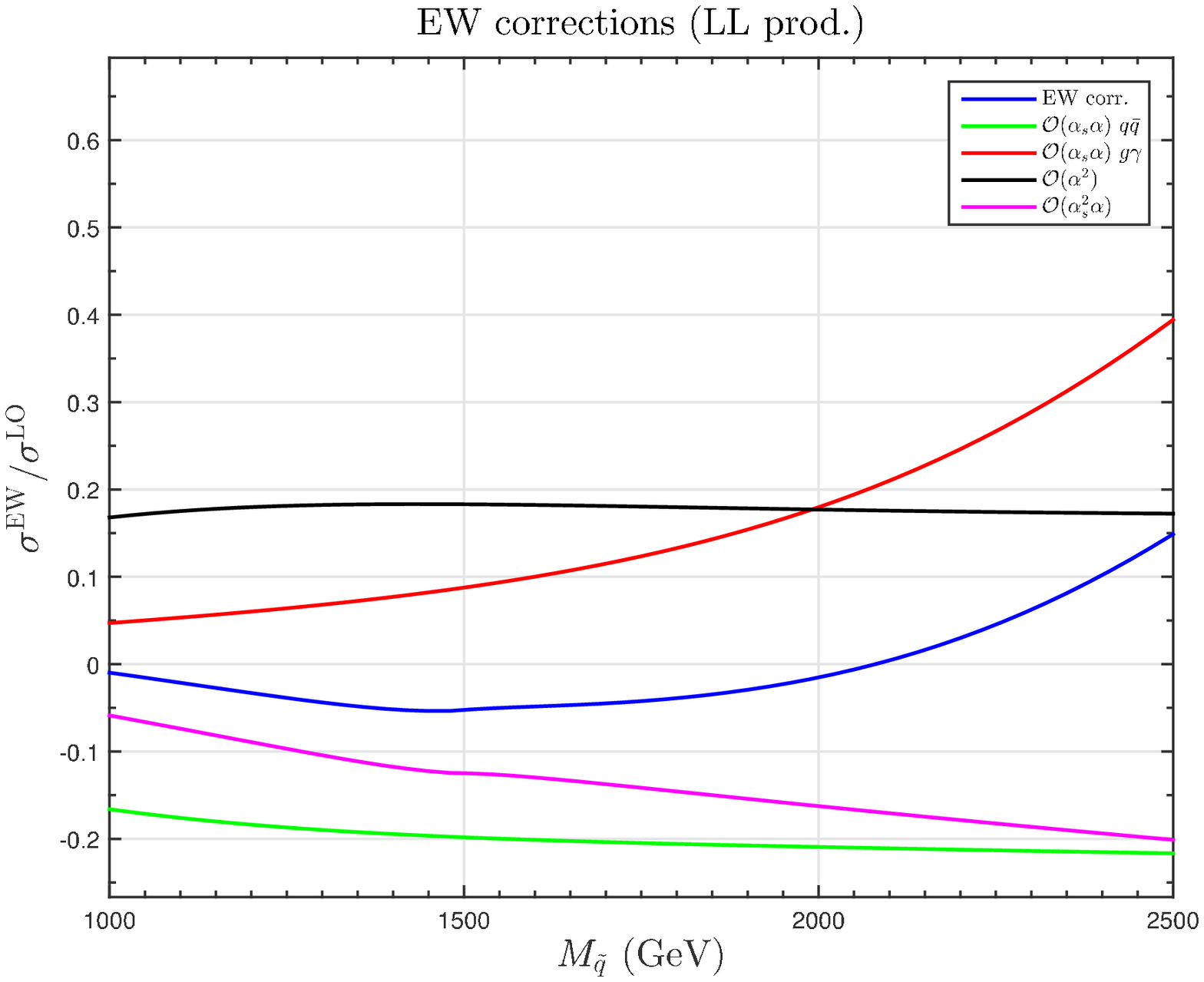}
\caption{}
\end{subfigure}
\phantom{pic}  \\
\begin{subfigure}[b]{0.5\textwidth}
\includegraphics[width=7.2cm,height=6.3cm]{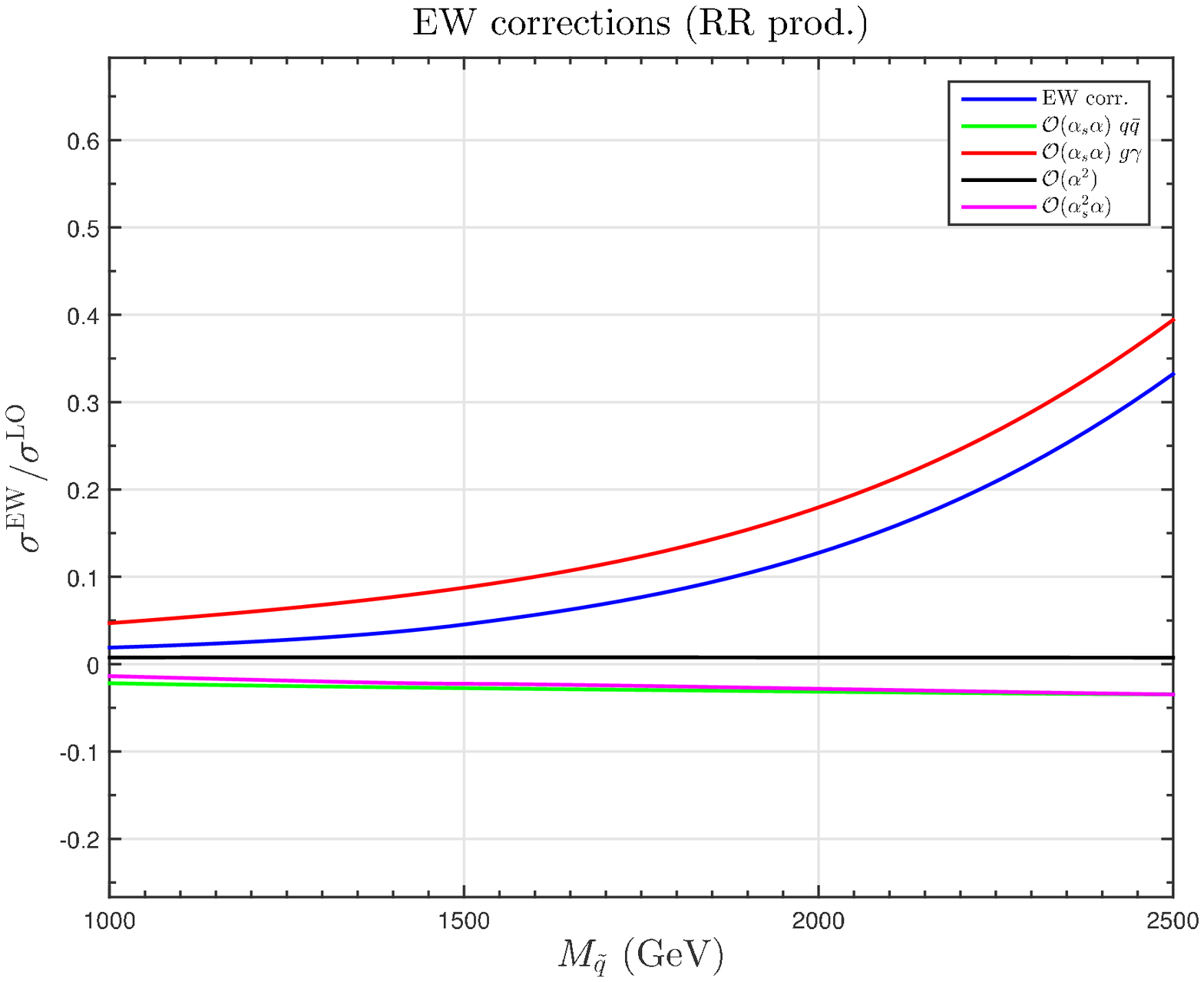}
\caption{}
\end{subfigure}
\begin{subfigure}[b]{0.5\textwidth}
\includegraphics[width=7.2cm,height=6.3cm]{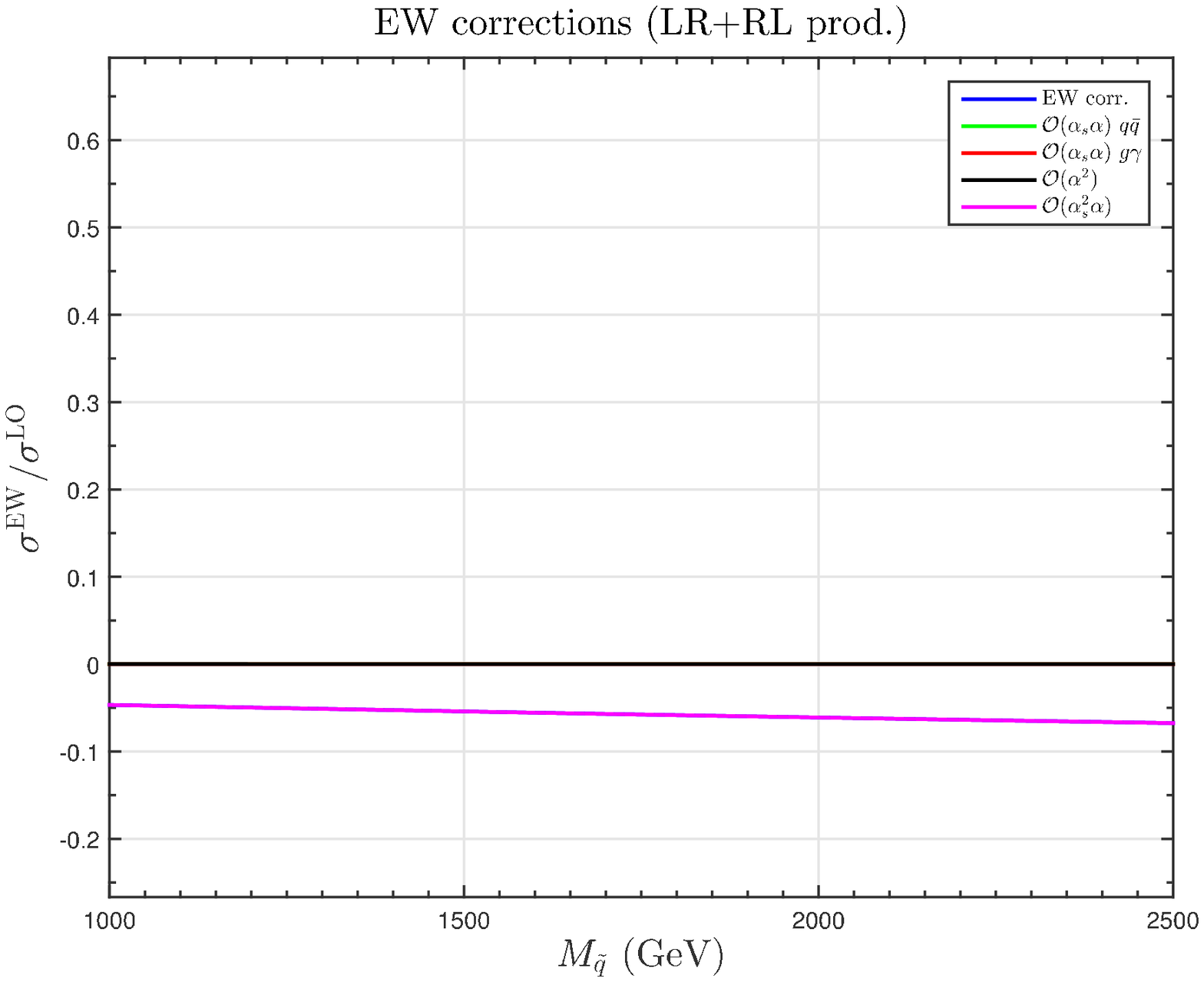}
\caption{}
\end{subfigure}
\caption[.]{Scan over $M_{\tilde q}$, which corresponds to Slope $S_1$. The value of the other parameters are collected  in Table~\ref{Tab:Bench}. 
}
\label{Fig:LSS1_MSQ12}
 \end{figure}

\textbf{Slope $\mathbf{S_1}$}: 
Fig.~\ref{Fig:LSS1_MSQ12} corresponds to the case of degenerate squarks, showing 
the dependence of squark--antisquark production on the common mass
$M_{\tilde q}$ of all light flavor squarks. The inclusive cross
section varies over two orders of magnitude in the considered mass
range. The QCD corrections vary from $50 \%$ to $70\%$ as $M_{\tilde q}$ varies from $1000$ to $2500$ GeV.  The relative size of the total EW corrections increases with $M_{\tilde q}$, and Fig.~\ref{Fig:LSS1_MSQ12}(b) clearly shows that the total amount of the EW corrections is the result of substantial cancellations among the different  channels. 
This cancellation may be jeopardized by the photon PDF uncertainty, since the $g\gamma$-channel (with its large uncertainties) has a large impact on the total size of the EW corrections, as can be seen from the comparison of red and blue lines in the lower panel of Fig.~\ref{Fig:LSS1_MSQ12}(a).
Looking at the different chirality combinations separately, in the case of LL production, mutual cancellations among the various channels keep the total EW corrections small in the entire $M_{\tilde q}$ range. On the contrary, in RR production the total EW corrections are positive and increase as $M_{\tilde q}$ increases. However, without the (possibly) large positive $g\gamma$-channel the EW corrections are negative in the entire considered mass range, and they reach $-20\%(-5\%)$ for $M_{\tilde q}=2500$~GeV for LL(RR) production.
The total EW contributions to LR+RL production, which gives the largest part of the ``incl.''~cross section (see Fig.~\ref{Fig:LSS1_MSQ12}(c)), depend mildly on $M_{\tilde q}$; they are negative and of the order of $-5 \%$ in the entire range considered.  As the $g\gamma$-channel does not contribute to LR+RL production these predictions are not affected by large photon PDF uncertainties. It is worth to notice that in LR+RL production the entire $\ord(\alpha_s \alpha)$ contribution is in general zero, due to the different chiralities in the final state. Moreover, the size of the $\ord(\alpha^2)$ contribution is negligible. Thus, in LR+RL production, the total EW corrections can be identified with the  $\ord(\alpha_s^2 \alpha)$ NLO EW corrections.

\begin{figure}[t]
\begin{subfigure}[b]{0.5\textwidth}
\includegraphics[width=7.2cm,height=6.3cm]{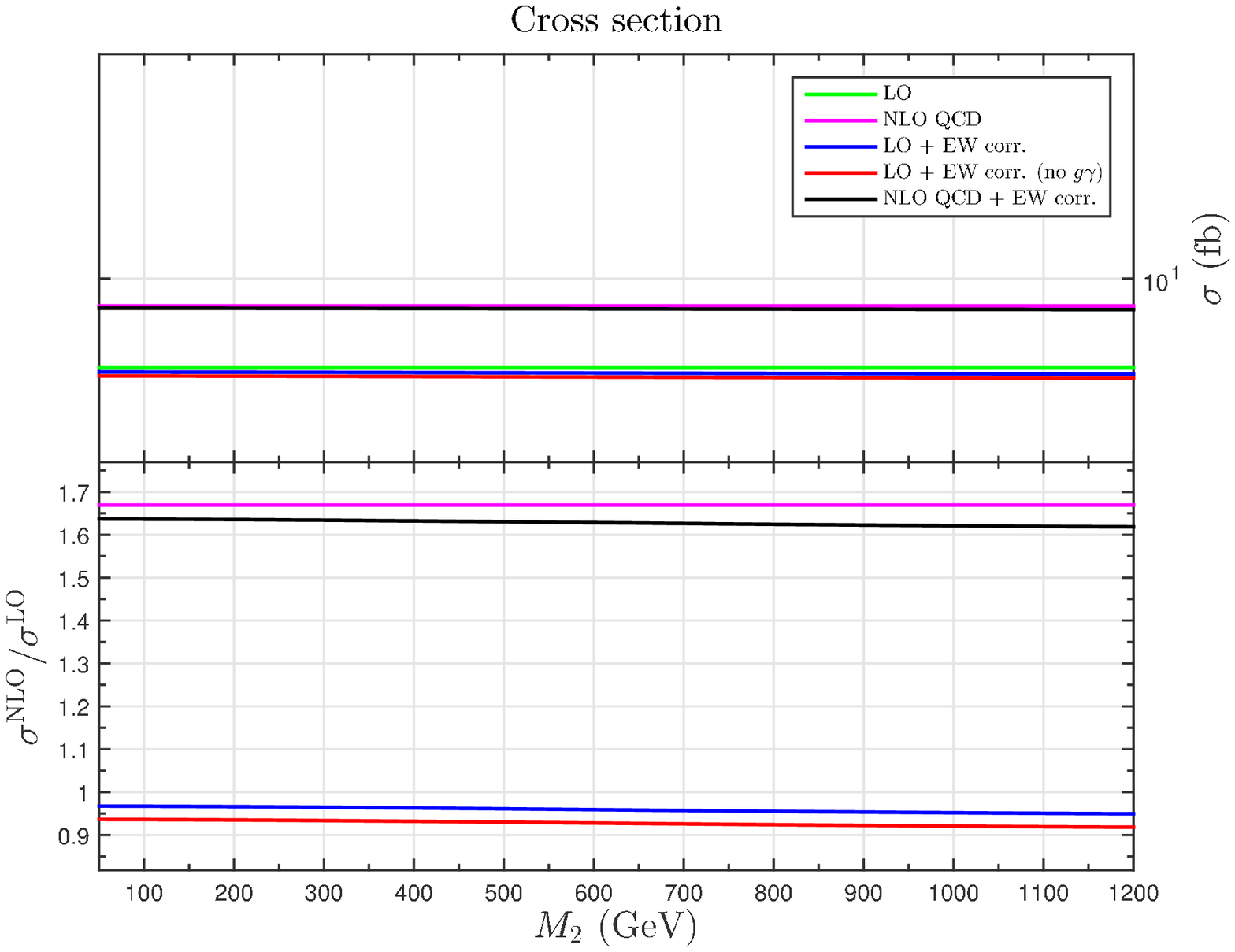}
\caption{}
\end{subfigure}
\begin{subfigure}[b]{0.5\textwidth}
\includegraphics[width=7.2cm,height=6.3cm]{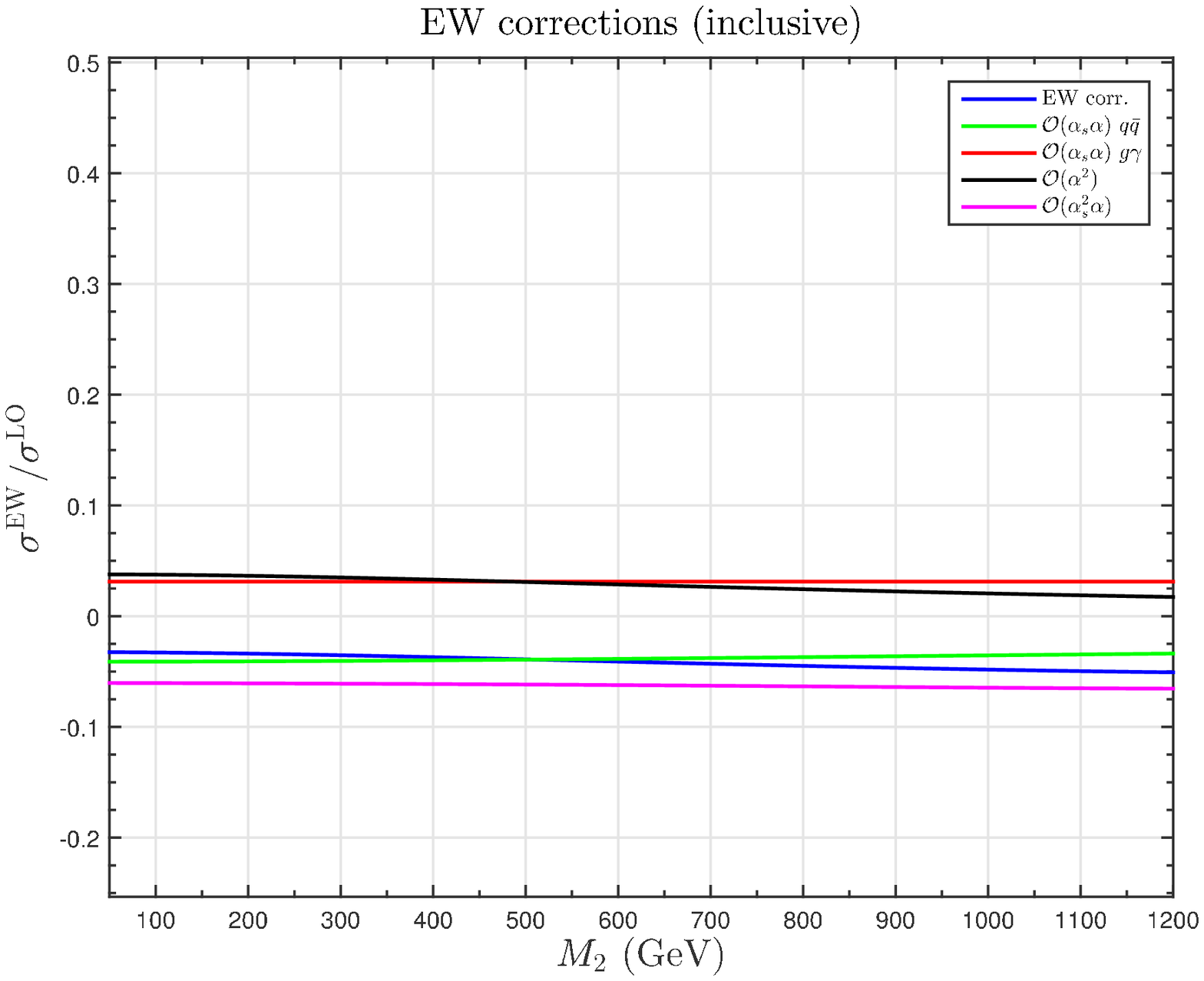}
\caption{}
\end{subfigure}
\phantom{pic}  \\
\begin{subfigure}[b]{0.5\textwidth}
\includegraphics[width=7.2cm,height=6.3cm]{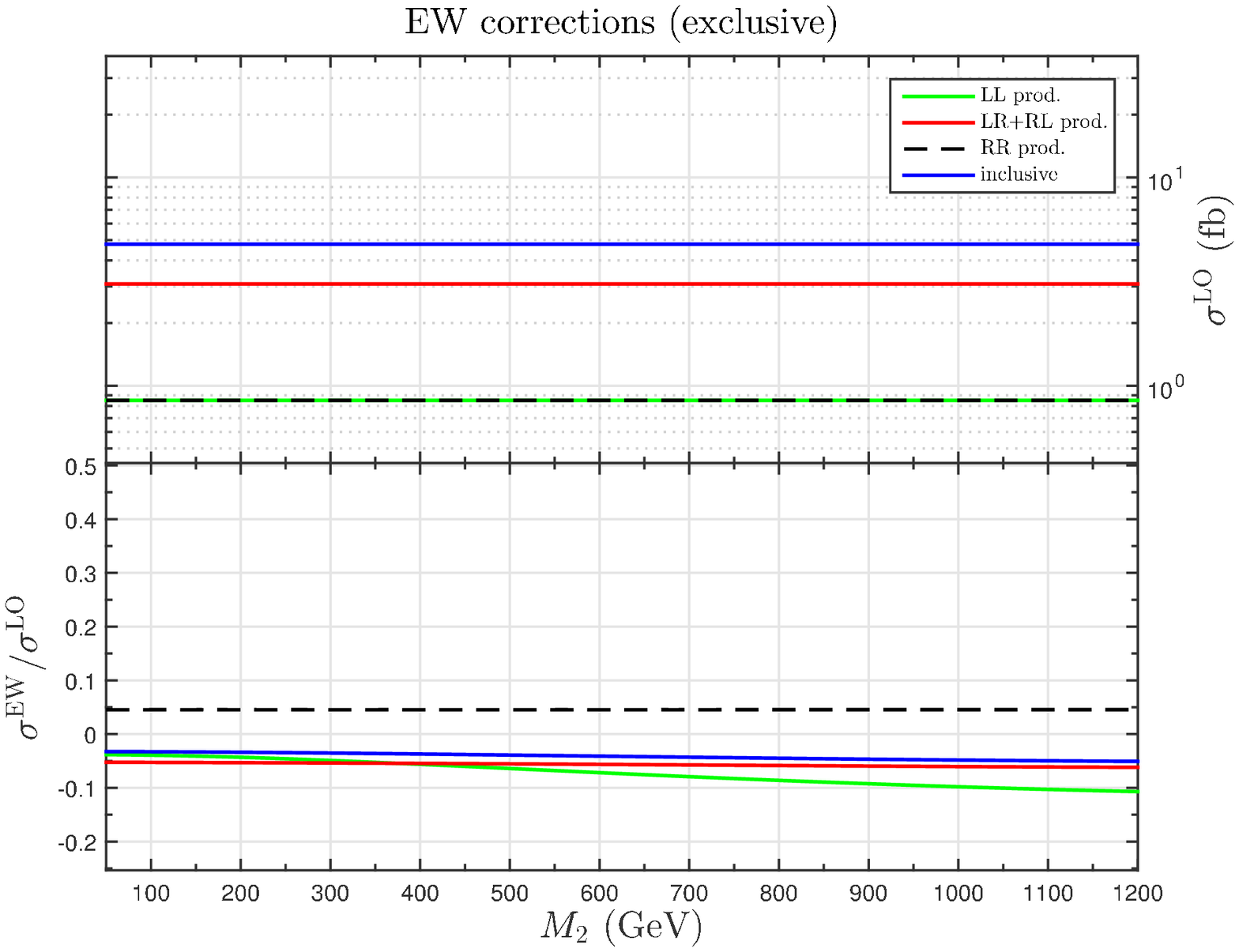}
\caption{}
\end{subfigure}
\begin{subfigure}[b]{0.5\textwidth}
\includegraphics[width=7.2cm,height=6.3cm]{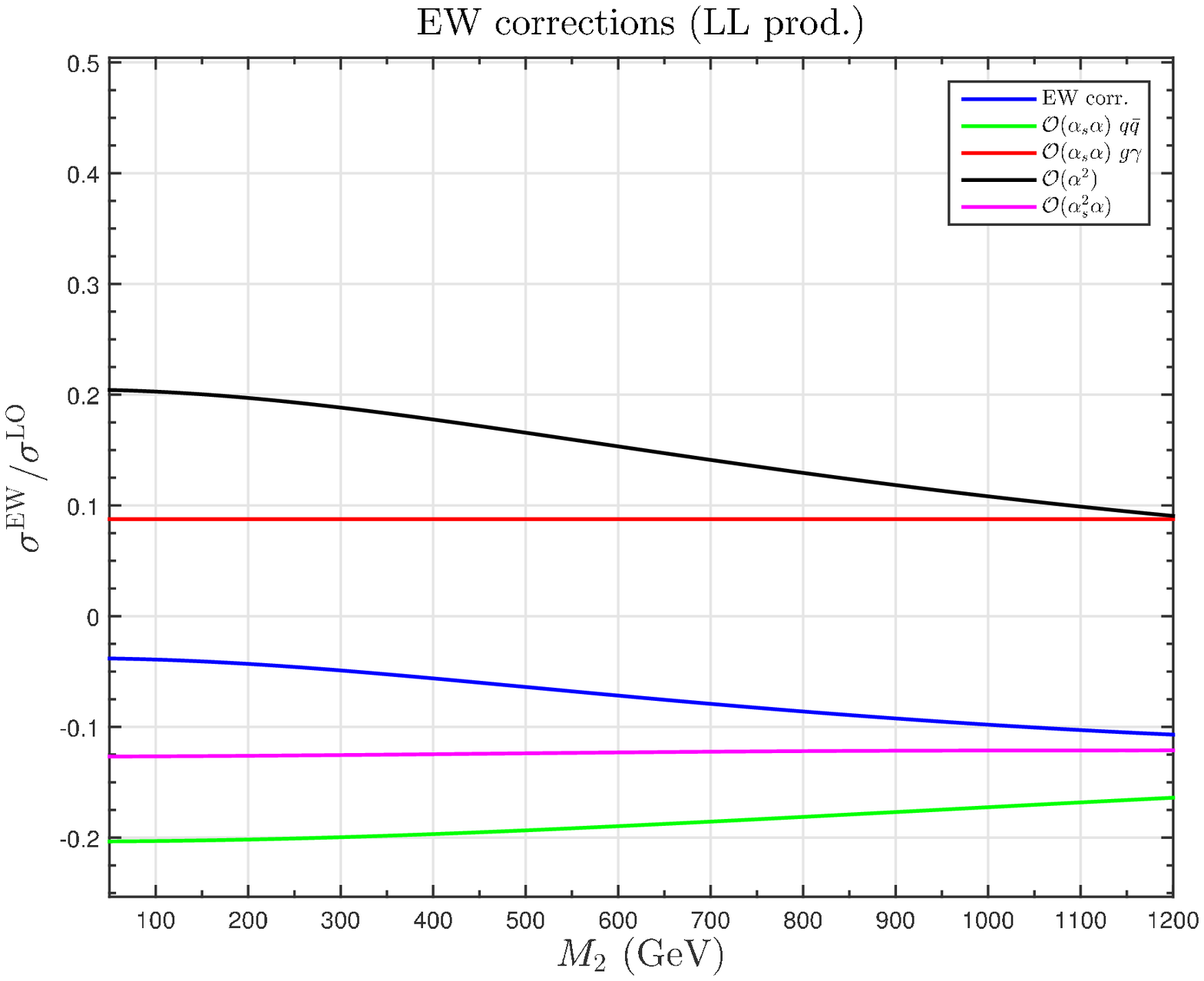}
\caption{}
\end{subfigure}
\phantom{pic}  \\
\begin{subfigure}[b]{0.5\textwidth}
\includegraphics[width=7.2cm,height=6.3cm]{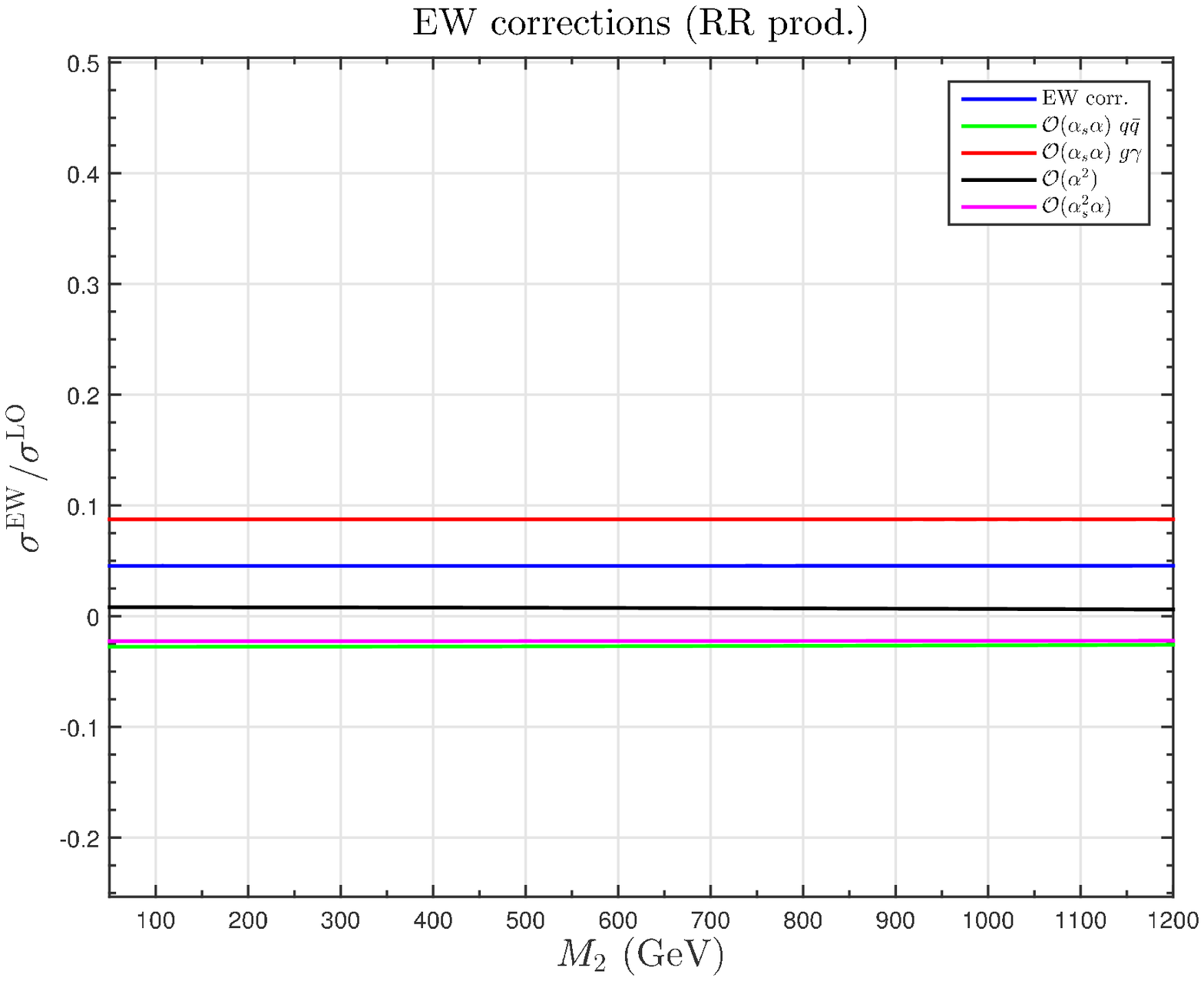}
\caption{}
\end{subfigure}
\begin{subfigure}[b]{0.5\textwidth}
\includegraphics[width=7.2cm,height=6.3cm]{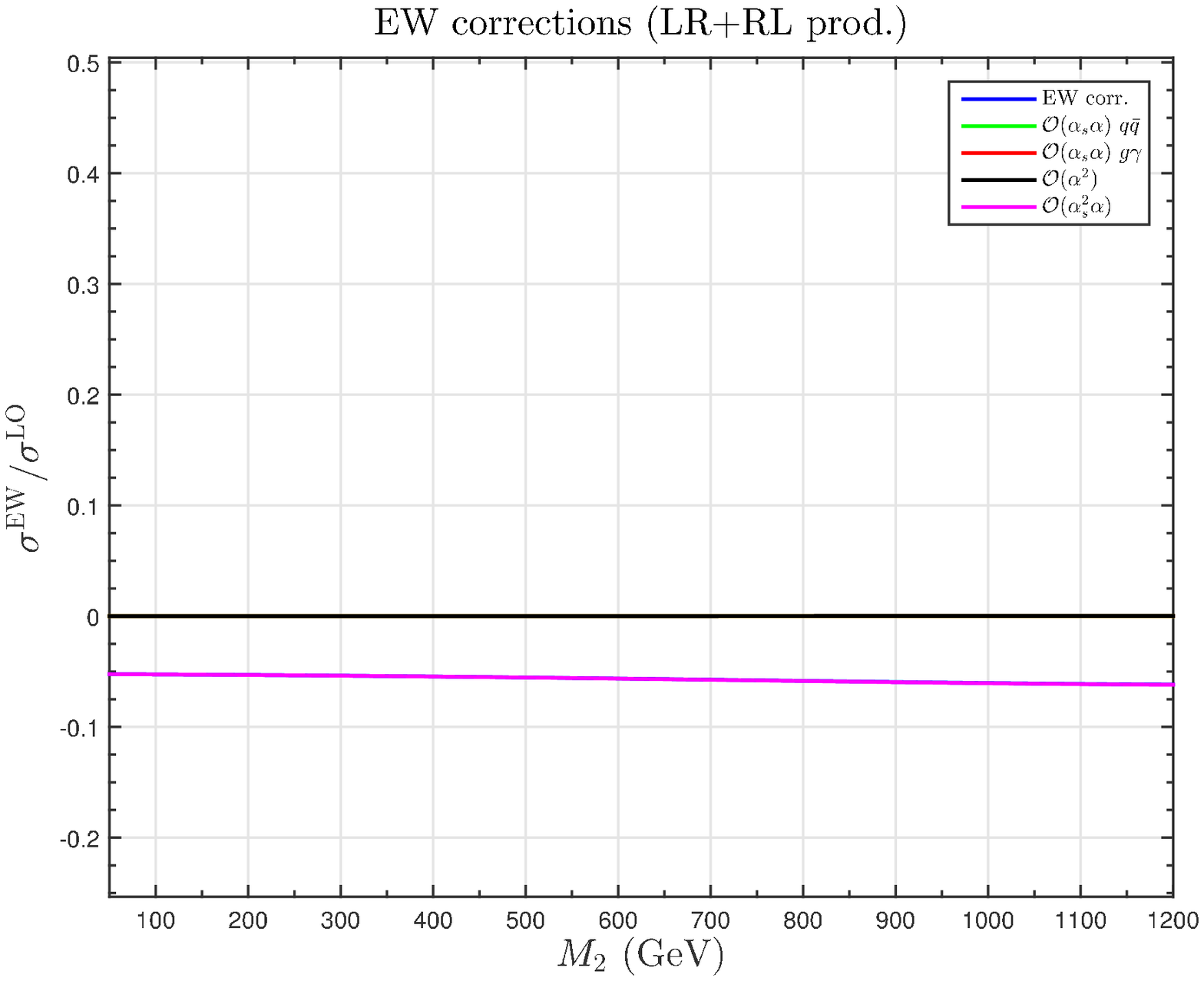}
\caption{}
\end{subfigure}
\caption[.]{Scan over $M_2$. The value of the other parameters are collected  in Table~\ref{Tab:Bench}.  }
\label{Fig:LSS1_M2}
 \end{figure}

\textbf{Scan in $\mathbf{M_2}$}: Results for the scan in $M_2$ are shown in Fig.~\ref{Fig:LSS1_M2}. The QCD  corrections to the inclusive cross section are independent of  $M_2$ and of the order of $65 \%$, while the total EW corrections are negative and small, of the order of $3\%$. The NLO EW corrections, $\ord(\alpha_s^2 \alpha)$, are negative and constant; the mild dependence on $M_2$ is induced by the EW tree-level induced contributions. 
 As can be inferred from Figs.~\ref{Fig:LSS1_M2}(c)--\ref{Fig:LSS1_M2}(f), the EW corrections  do depend on the chirality of the produced squarks. In the case of LL production, they strongly depend 
on $M_2$ ranging from -3\% to -10\% as $M_2$ varies from 50 to 1200 GeV.   This dependence is mainly induced by the $\ord(\alpha^2)$ contribution, which is suppressed for large values of $M_2$. Both in LL and RR production the relative contribution of the $g\gamma$-channel is enhanced to 10\%.  At variance with LL production, the EW corrections in RR production are independent of $M_2$, positive 
and of the order of $5\%$. 
As can be noticed in  Fig.~\ref{Fig:LSS1_M2}(f), for LR+RL production the only non-vanishing contribution to the total EW corrections is the $\ord(\alpha_s^2 \alpha)$ NLO EW, which  is of the order of -5\%.      

\begin{figure}[t]
\begin{subfigure}[b]{0.5\textwidth}
\includegraphics[width=7.2cm,height=6.3cm]{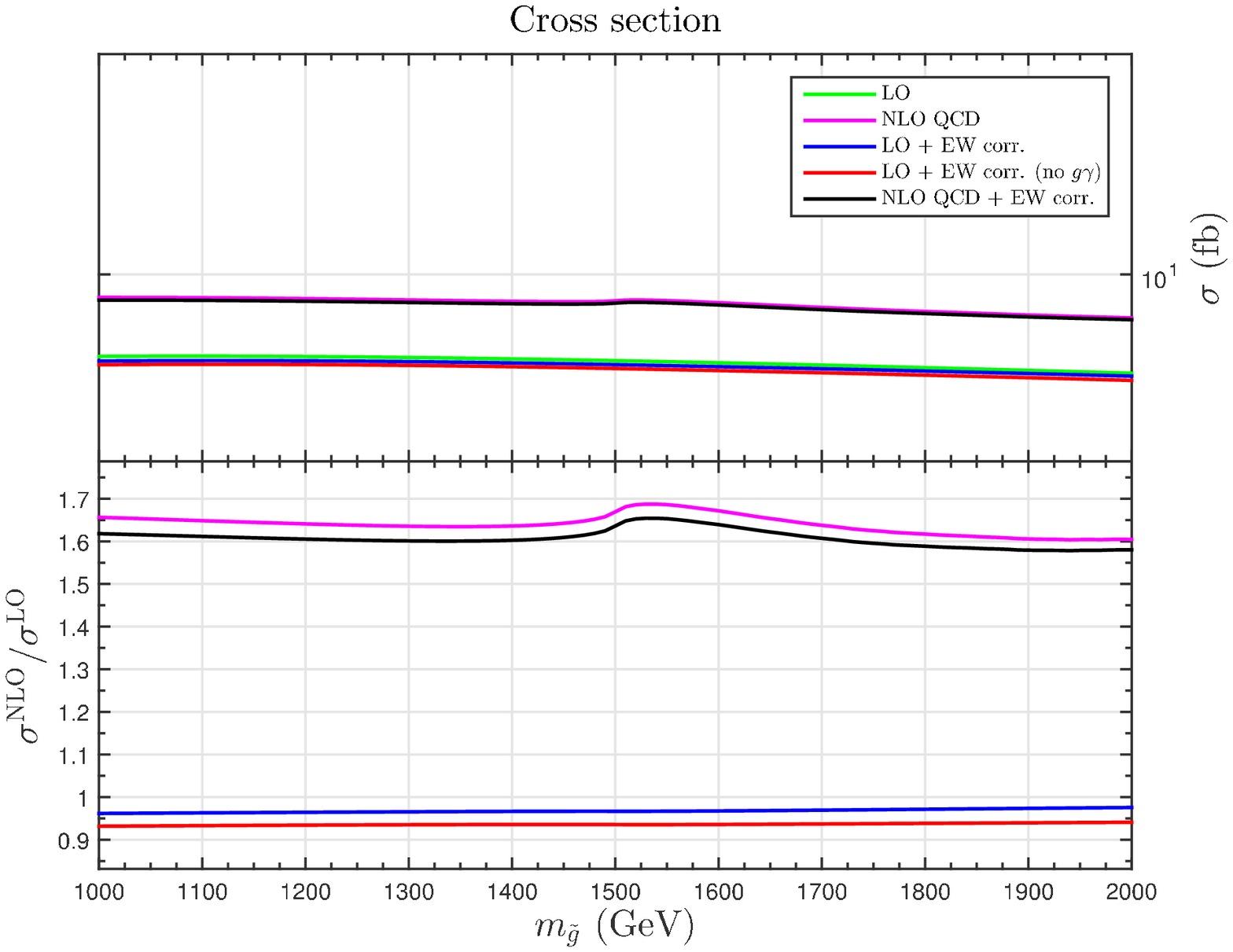}
\caption{}
\end{subfigure}
\begin{subfigure}[b]{0.5\textwidth}
\includegraphics[width=7.2cm,height=6.3cm]{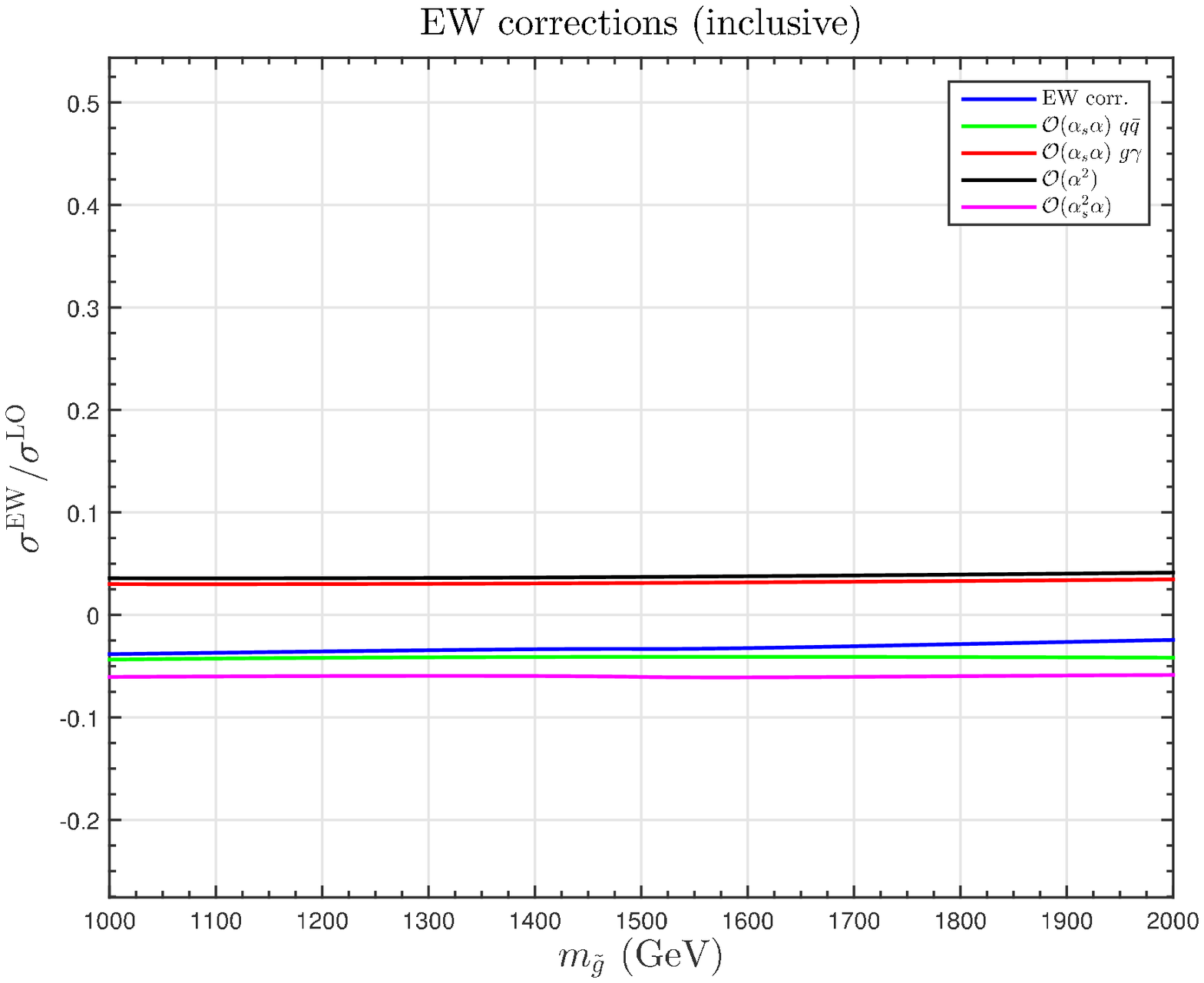}
\caption{}
\end{subfigure}
\phantom{pic}  \\
\begin{subfigure}[b]{0.5\textwidth}
\includegraphics[width=7.2cm,height=6.3cm]{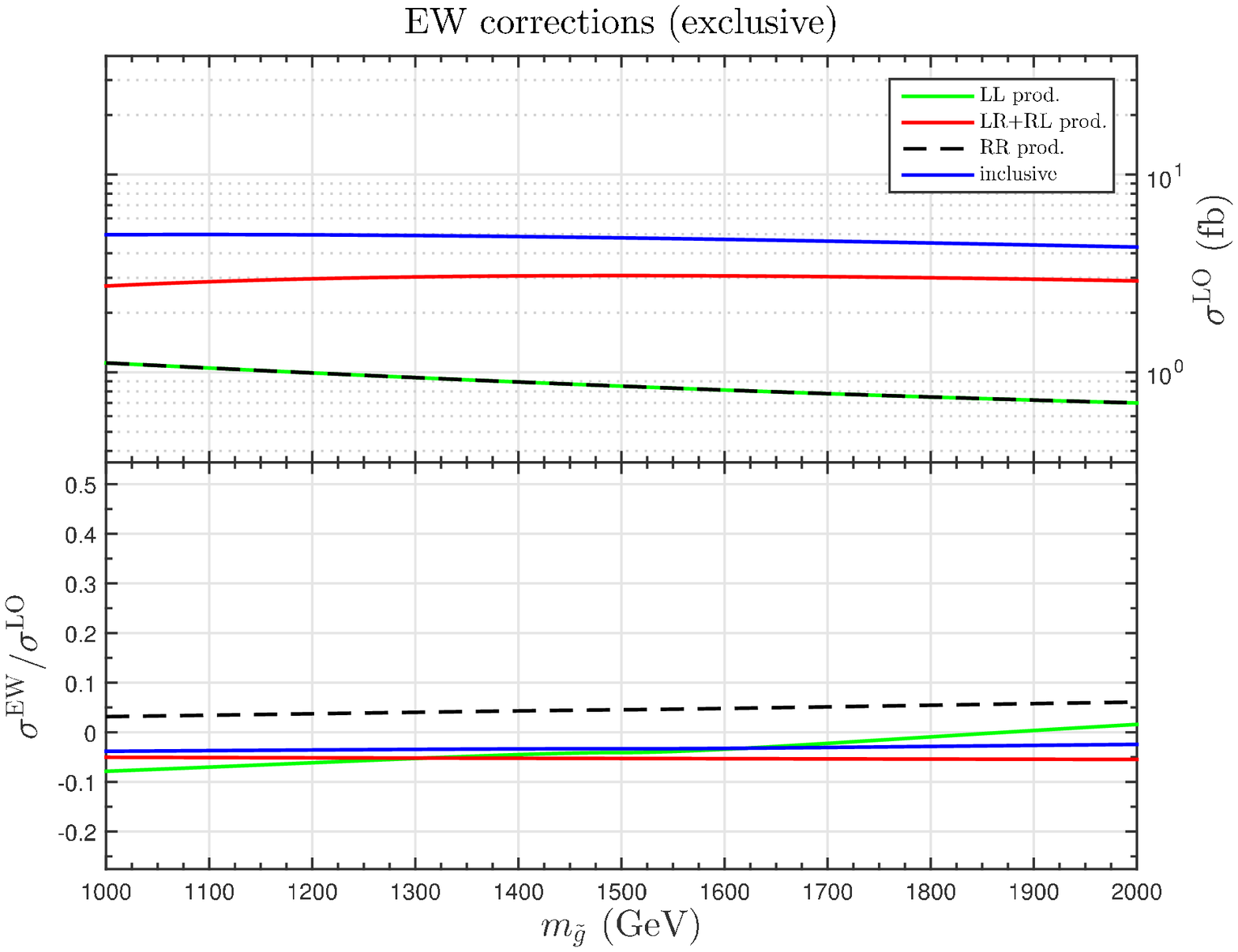}
\caption{}
\end{subfigure}
\begin{subfigure}[b]{0.5\textwidth}
\includegraphics[width=7.2cm,height=6.3cm]{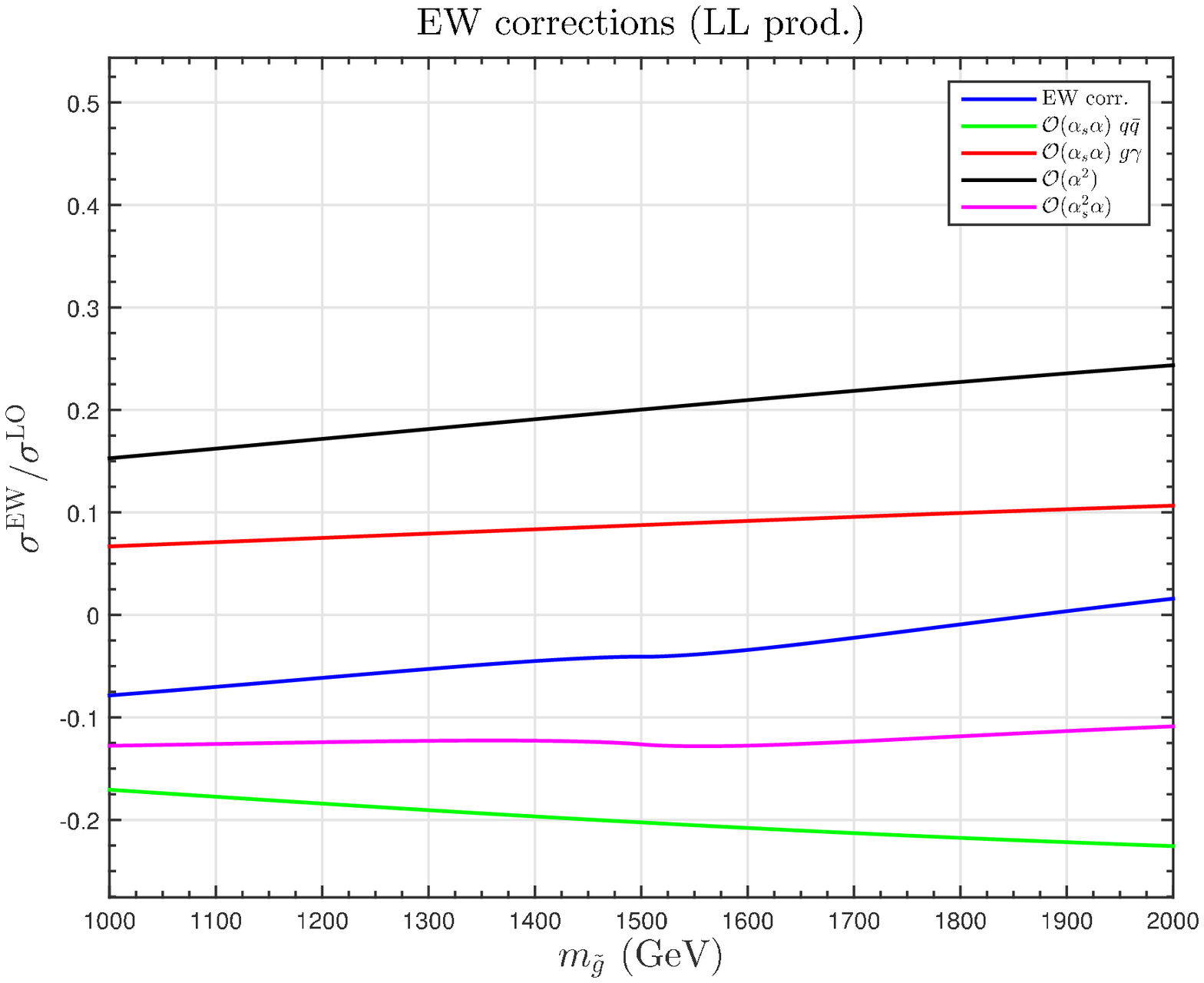}
\caption{}
\end{subfigure}
\phantom{pic}  \\
\begin{subfigure}[b]{0.5\textwidth}
\includegraphics[width=7.2cm,height=6.3cm]{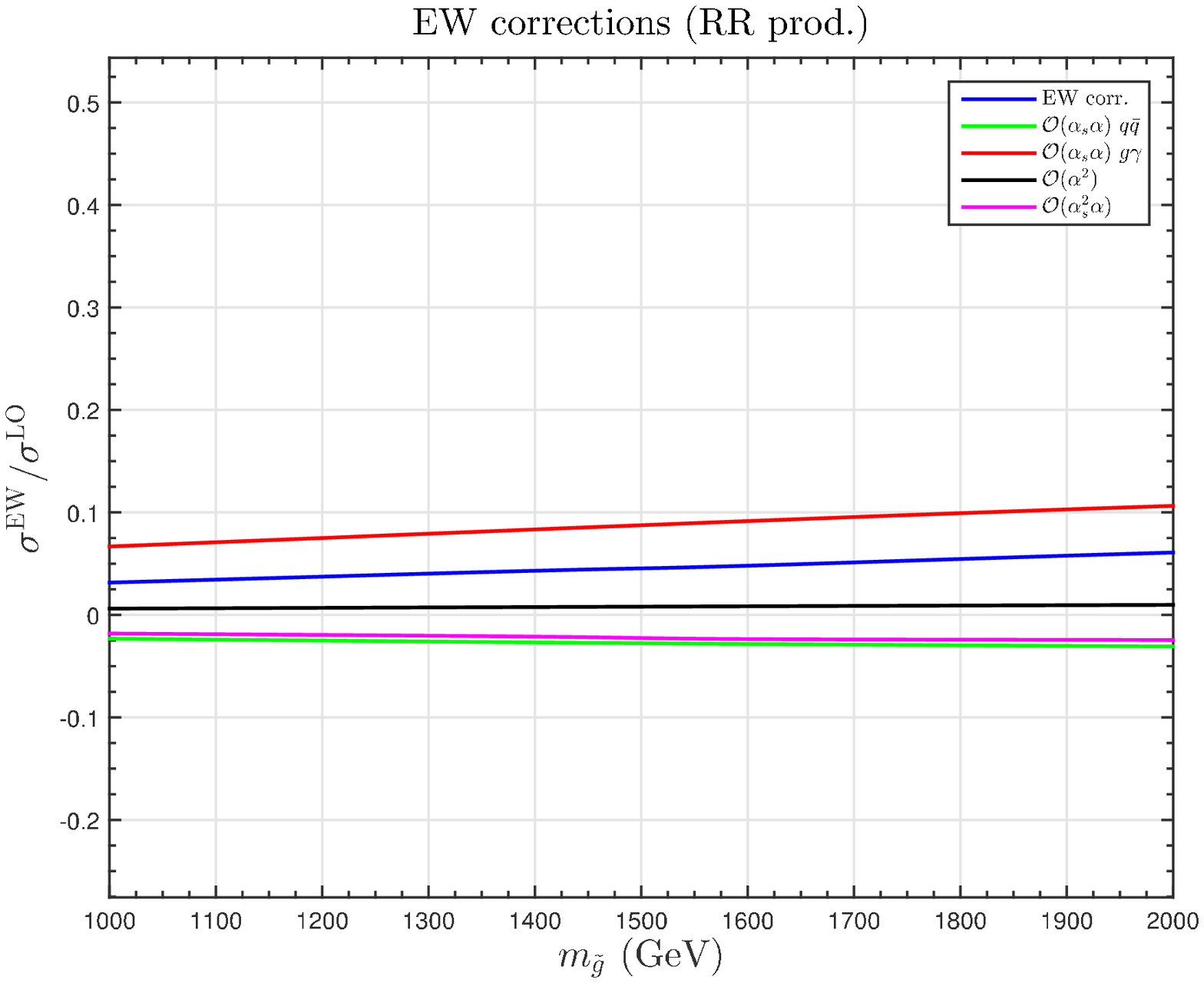}
\caption{}
\end{subfigure}
\begin{subfigure}[b]{0.5\textwidth}
\includegraphics[width=7.2cm,height=6.3cm]{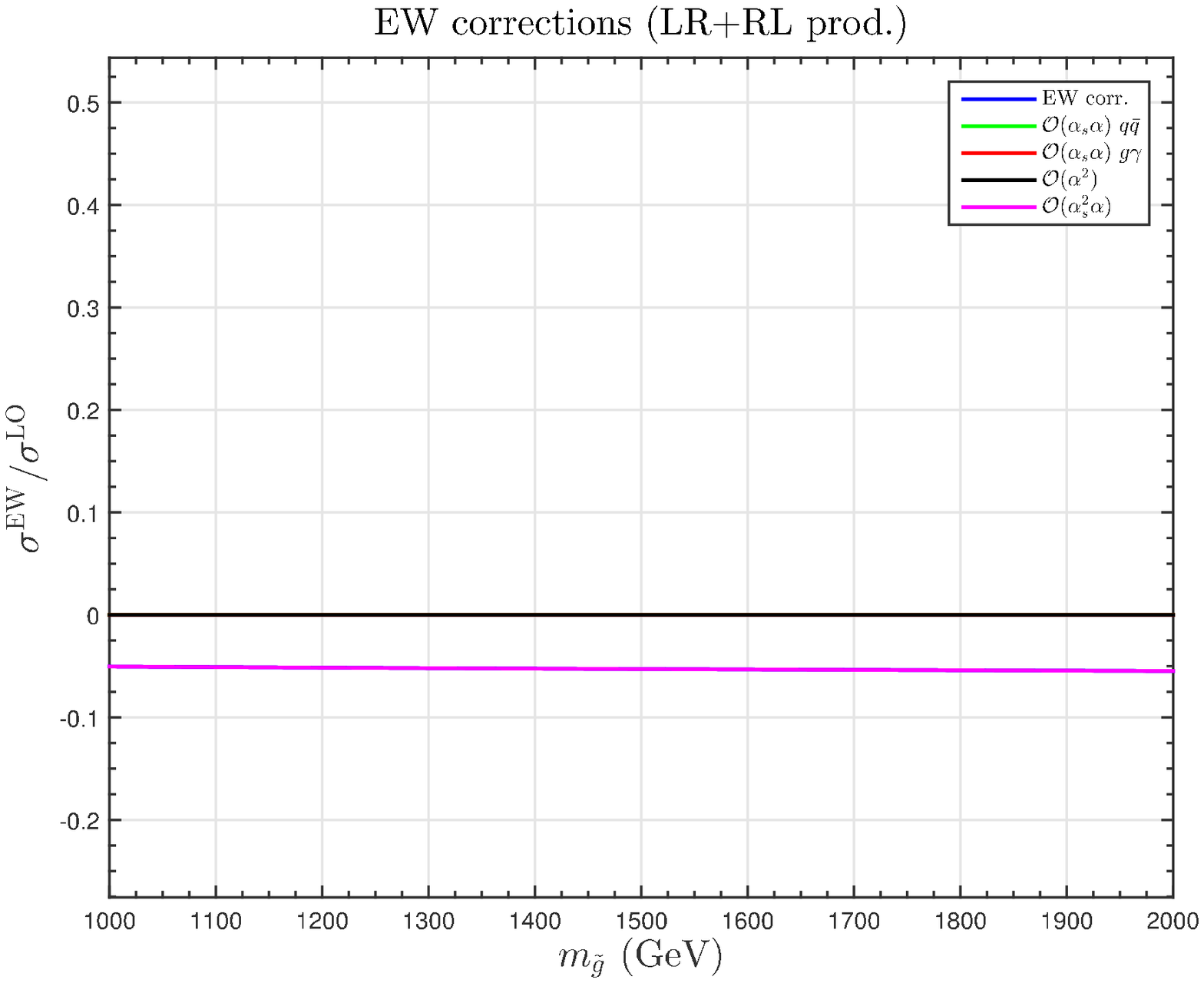}
\caption{}
\end{subfigure}
\caption[.]{Scan over  $m_{\tilde g}$. The value of the other parameters are collected  in Table~\ref{Tab:Bench}.}
\label{Fig:LSS1_MGL}
 \end{figure}

\textbf{Scan in $\mathbf{m_{\tilde g}}$}: The dependence of the QCD corrections  on $m_{\tilde g}$ is mild and  of the order of $60 \%$ in the entire range considered, as shown in Fig.~\ref{Fig:LSS1_MGL}. The EW corrections are small, negative and of the order 
of  $-4 \%$ in the low-$m_{\tilde g}$ region. They increase to $-2.5\%$ for  $m_{\tilde g} \simeq 2000$ GeV.  As can be inferred from Figs.~\ref{Fig:LSS1_MGL}(b), 
the increment is mainly due to the  positive yield from the $\mathcal{O}(\alpha^2)$ contribution and the $g\gamma$-channel.
Again, the  $g\gamma$-channel and the corresponding PDF uncertainty are of the same order as the EW corrections themselves.
As shown in Figs.~\ref{Fig:LSS1_M2}(c) and~\ref{Fig:LSS1_M2}(d), the EW corrections are more important in the case of LL production. In the large $m_{\tilde g}$ region, the $\mathcal{O}(\alpha^2)$ and the $g\gamma$-channel contributions become dominant and render the EW corrections positive and small, of the order of $1\%$. 
As shown in Fig.~\ref{Fig:LSS1_MGL}(e) and~\ref{Fig:LSS1_MGL}(f) the
total EW corrections to RR and LR+RL production  are respectively of
the order of $-2\%$ and $-5\%$ in the entire region considered. Again,
in LL and RR production  the photon PDF could substantially alter the
size of the EW corrections, while for LR+RL production the photon PDF
does not contribute.

\subsubsection*{Non-degenerate squark masses}
In the following scenarios the left- and right-handed soft squark masses $m_{\tilde q _{R}}$ and $m_{\tilde q _{L}}$ are treated as non-degenerate parameters. Here, NLO QCD corrections cannot be computed by {\tt Prospino} and are not included in the analysis. \\

\begin{figure}[t]
\begin{subfigure}[b]{0.5\textwidth}
\includegraphics[width=7.2cm,height=6.3cm]{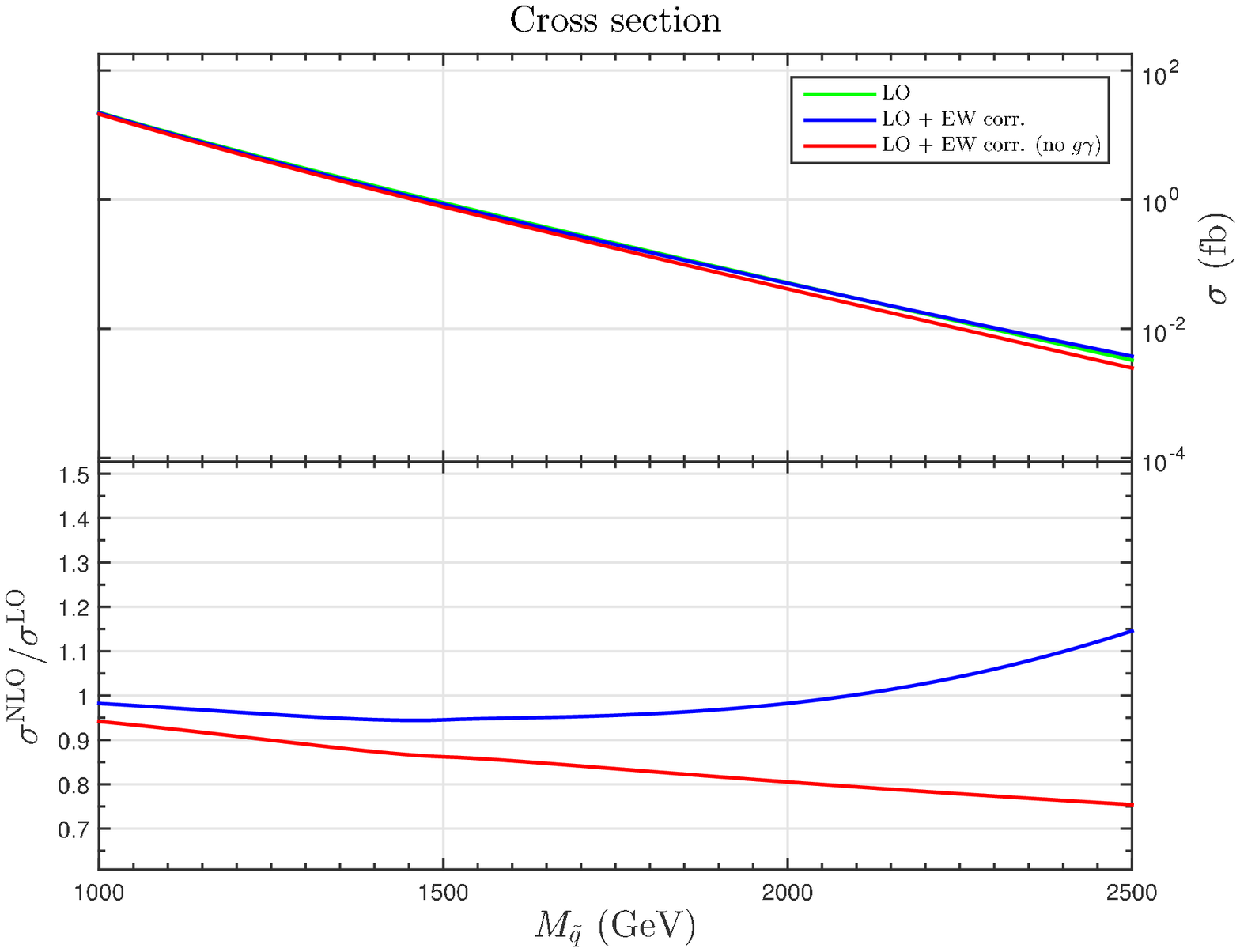}
\caption{}
\end{subfigure}
\begin{subfigure}[b]{0.5\textwidth}
\includegraphics[width=7.2cm,height=6.3cm]{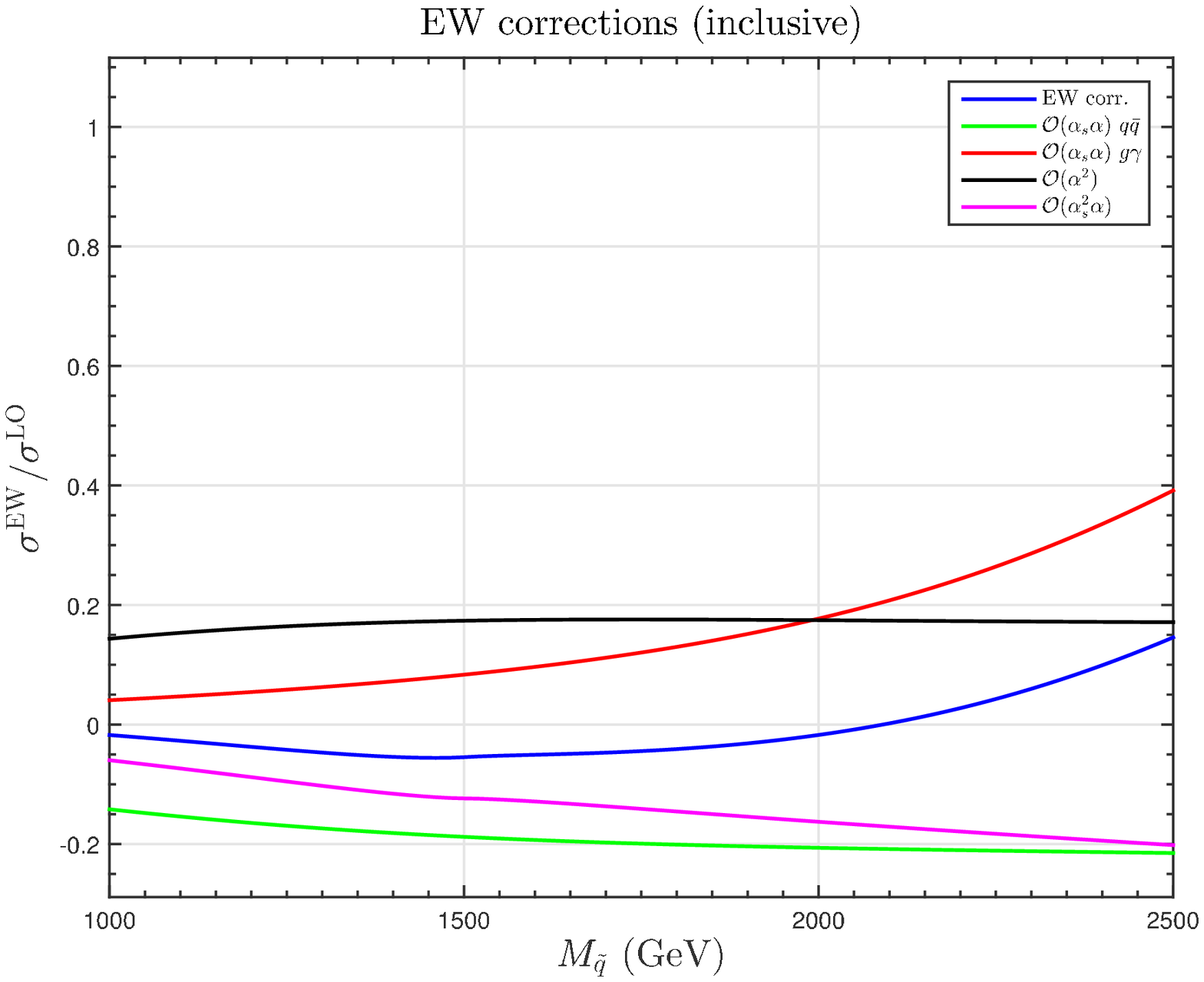}
\caption{}
\end{subfigure}
\phantom{pic}  \\
\begin{subfigure}[b]{0.5\textwidth}
\includegraphics[width=7.2cm,height=6.3cm]{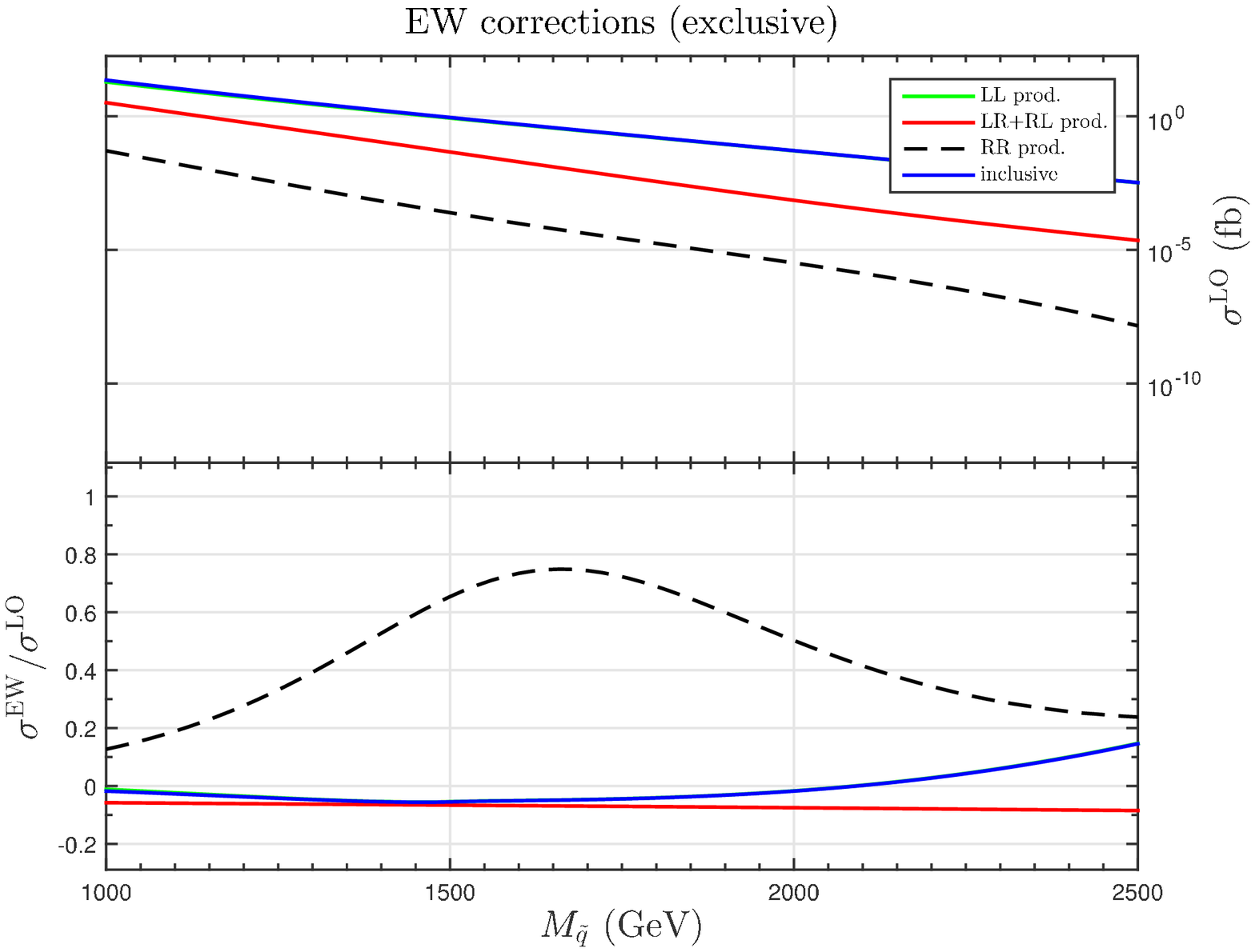}
\caption{}
\end{subfigure}
\begin{subfigure}[b]{0.5\textwidth}
\includegraphics[width=7.2cm,height=6.3cm]{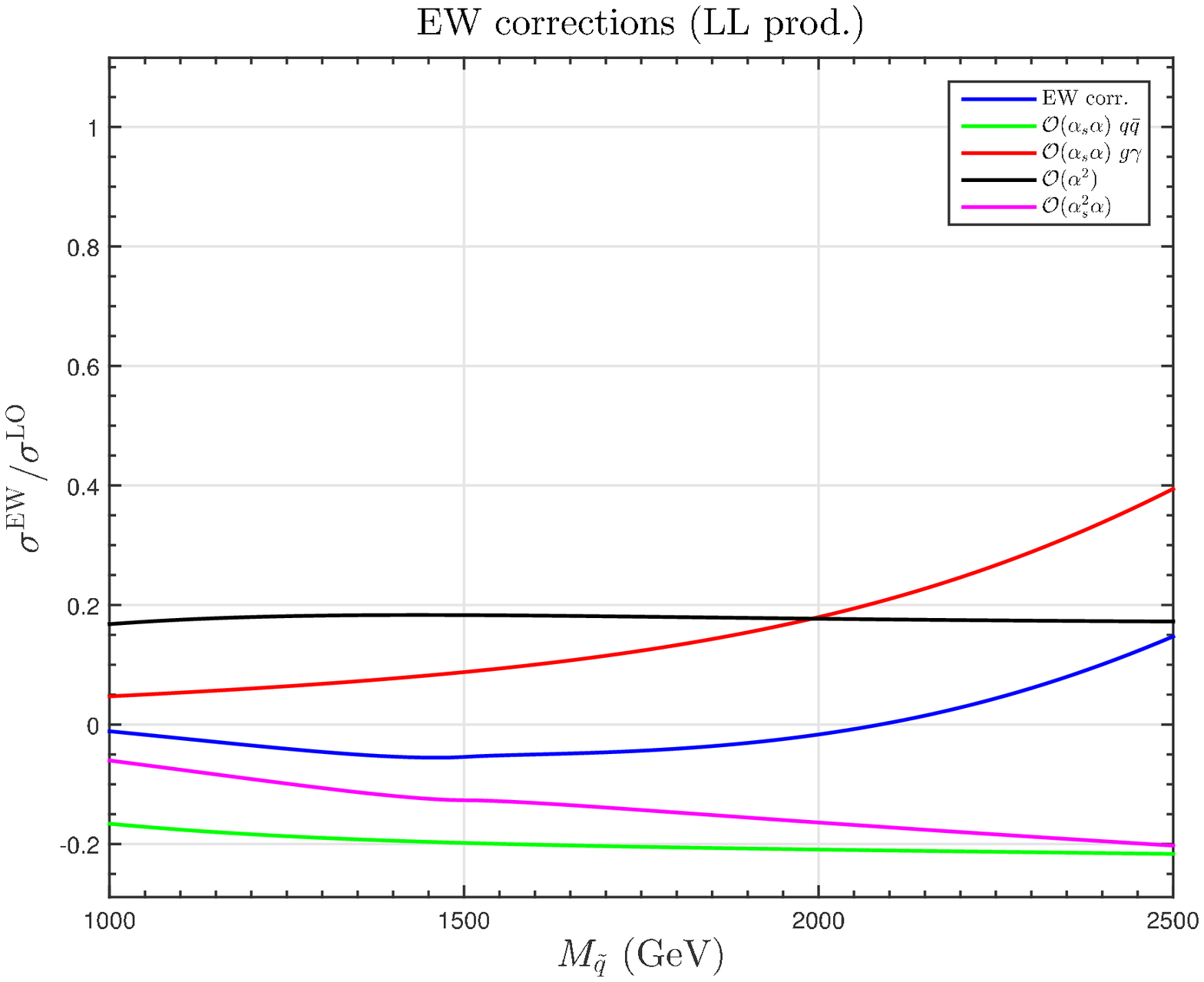}
\caption{}
\end{subfigure}
\phantom{pic}  \\
\begin{subfigure}[b]{0.5\textwidth}
\includegraphics[width=7.2cm,height=6.3cm]{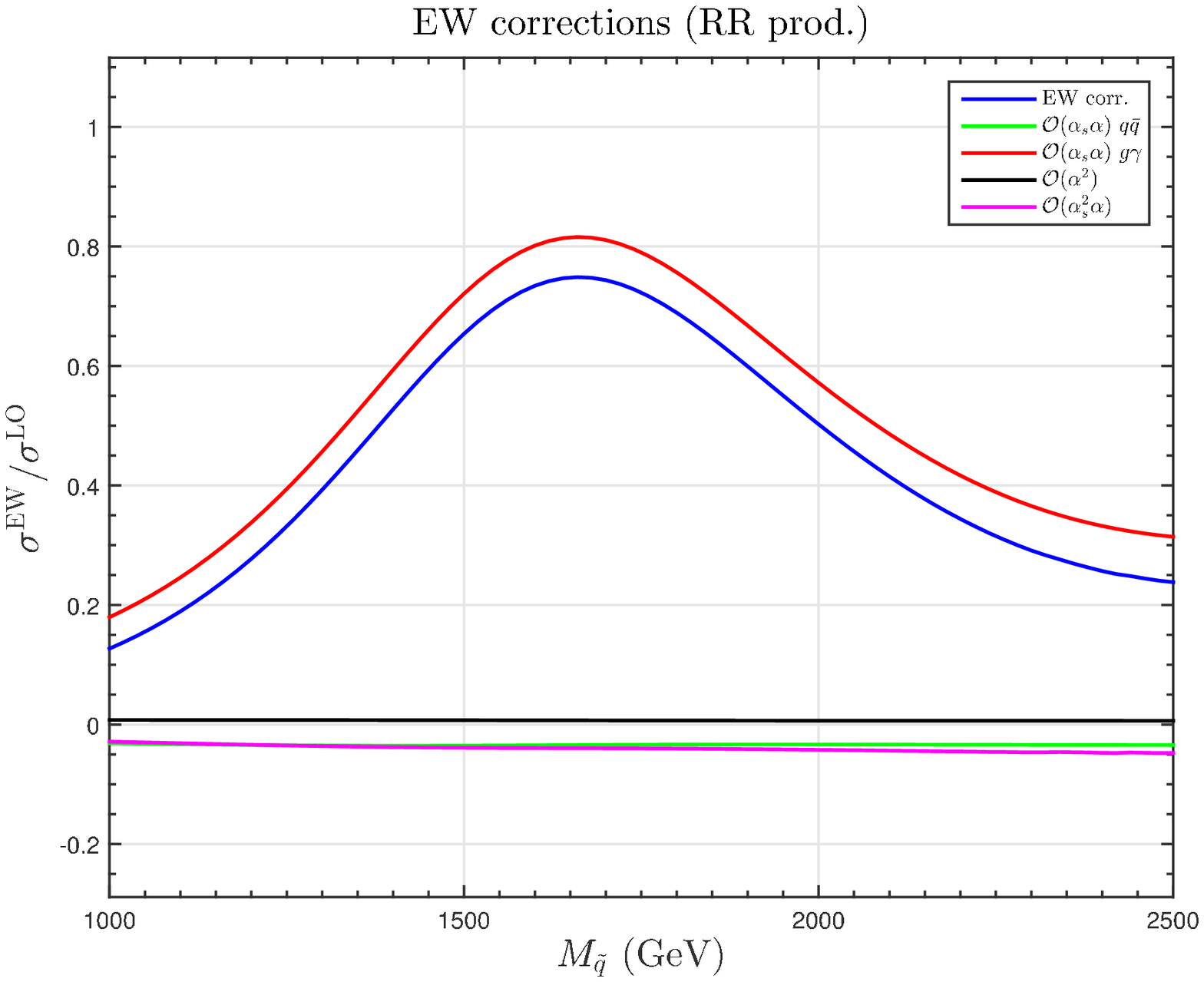}
\caption{}
\end{subfigure}
\begin{subfigure}[b]{0.5\textwidth}
\includegraphics[width=7.2cm,height=6.3cm]{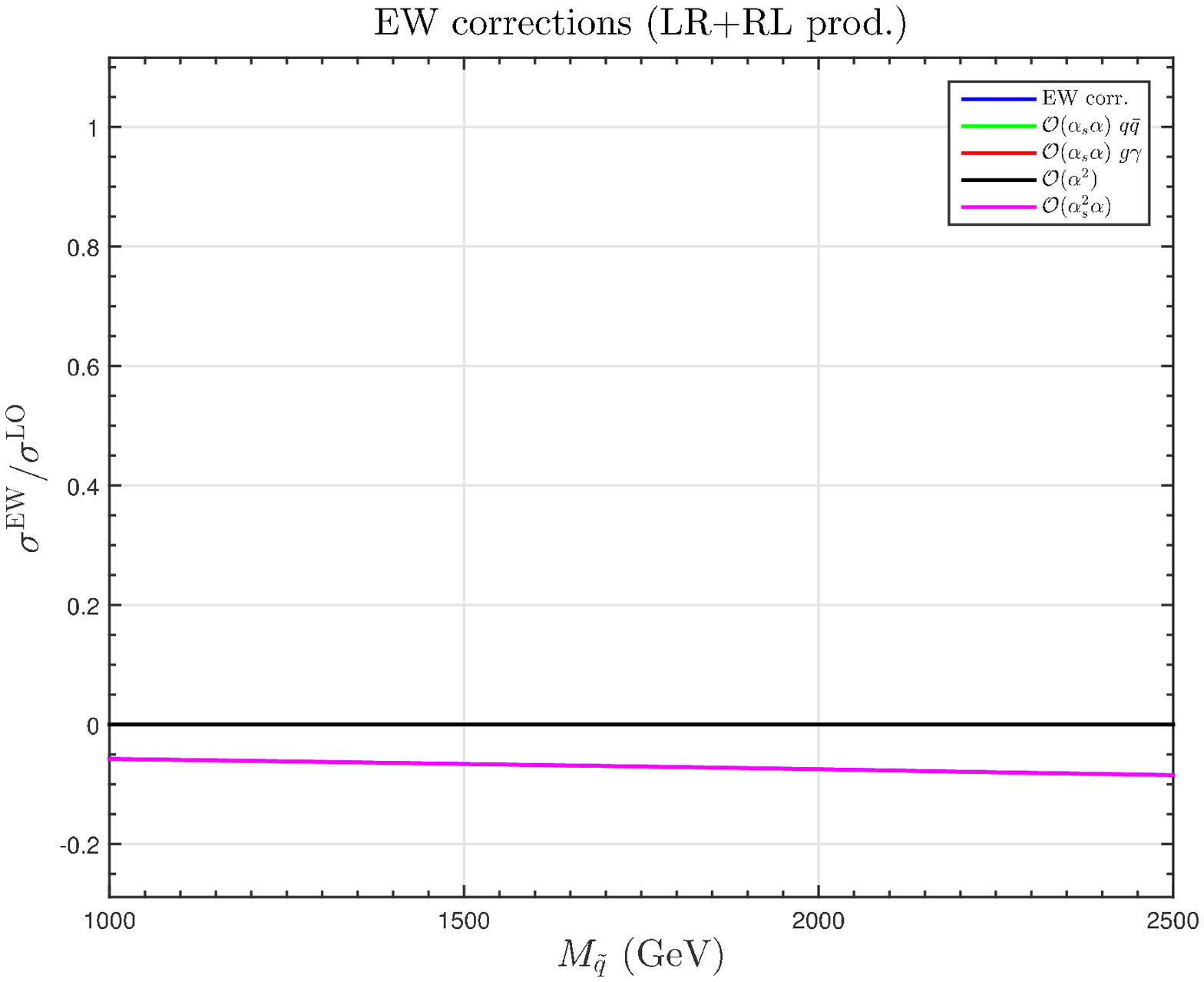}
\caption{}
\end{subfigure}
\caption[.]{Slope $S_2$: scan over $M_{\tilde q}$.   The value of the parameters not involved in the scan are collected in Table~\ref{Tab:Bench}. }
\label{Fig:LSS2_MSQ12}
 \end{figure}

\textbf{Slope $\mathbf{S_2}$}: 
For this scan, displayed in Fig.~\ref{Fig:LSS2_MSQ12}, 
we set $m_{\tilde q _{R}}=2m_{\tilde q _{L}}$; hence,  RR production
is negligible and the cross section for  LR+RL  is much smaller than for  LL production.
Therefore EW corrections to the inclusive squark--antisquark production are qualitatively very similar to the EW corrections to LL production in the slope $S_1$.
The overall size of the EW corrections is the result of mutual cancellations among the different channels and their dependence on $M_{\tilde q}$ is mostly determined by the $g\gamma$-channel. Thus, the photon PDF uncertainty may substantially alter these corrections. RR production is strongly suppressed by the high mass of the produced squarks, i.e.
the scan covers values of $m_{\tilde q _{R}}$ in the range $[2~{\rm TeV}, 5~{\rm TeV}]$. In this regime, the photon PDF uncertainty renders the EW corrections to RR production unreliable. The clear sign of their unphysical behavior is the ``bump'' at $  M_{\tilde{q}} \equiv m_{\tilde q _{R}}/2 =1.7~{\rm TeV}$, which is due to the photon PDF and not from matrix elements.

\begin{figure}[t]
\begin{subfigure}[b]{0.5\textwidth}
\includegraphics[width=7.2cm,height=6.3cm]{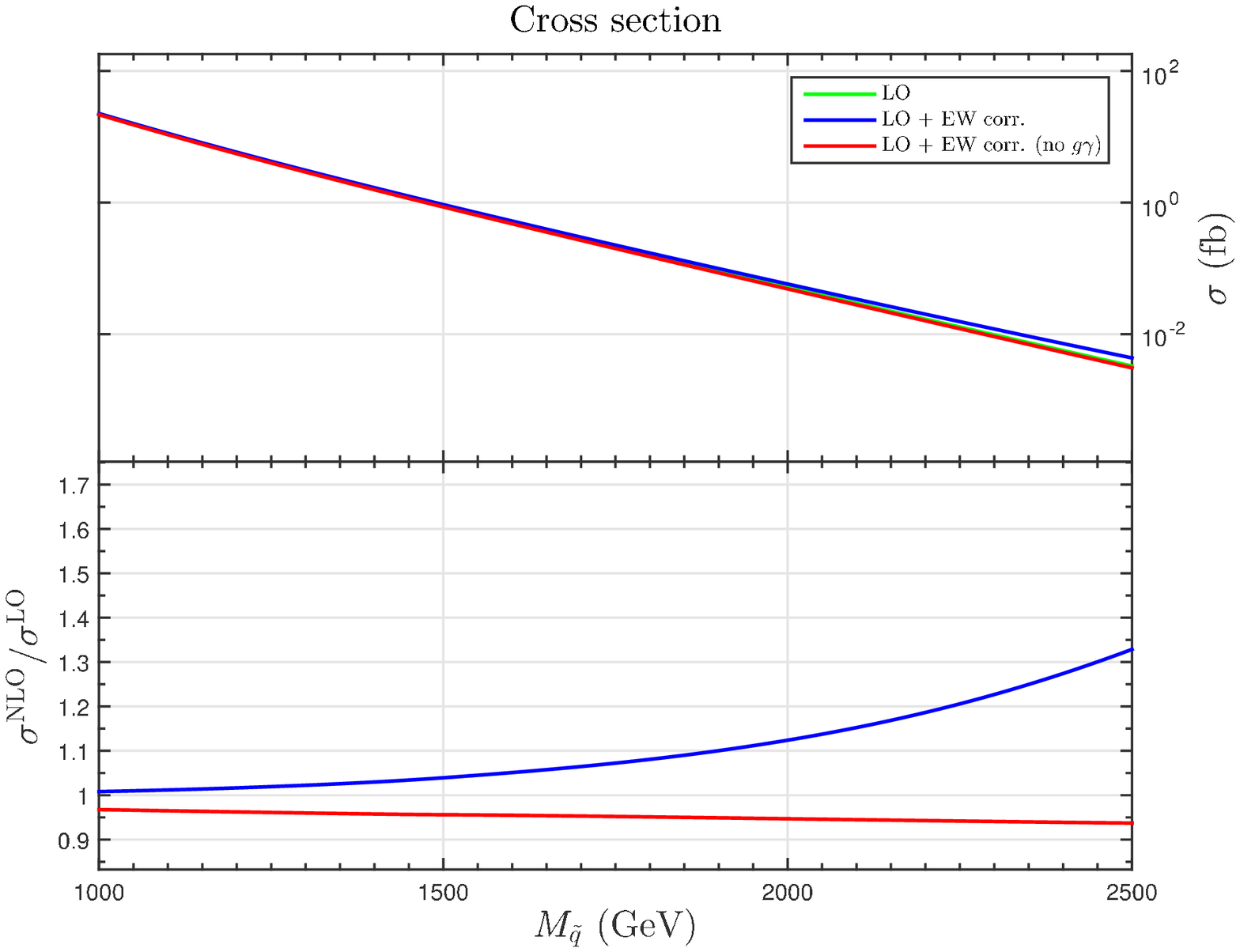}
\caption{}
\end{subfigure}
\begin{subfigure}[b]{0.5\textwidth}
\includegraphics[width=7.2cm,height=6.3cm]{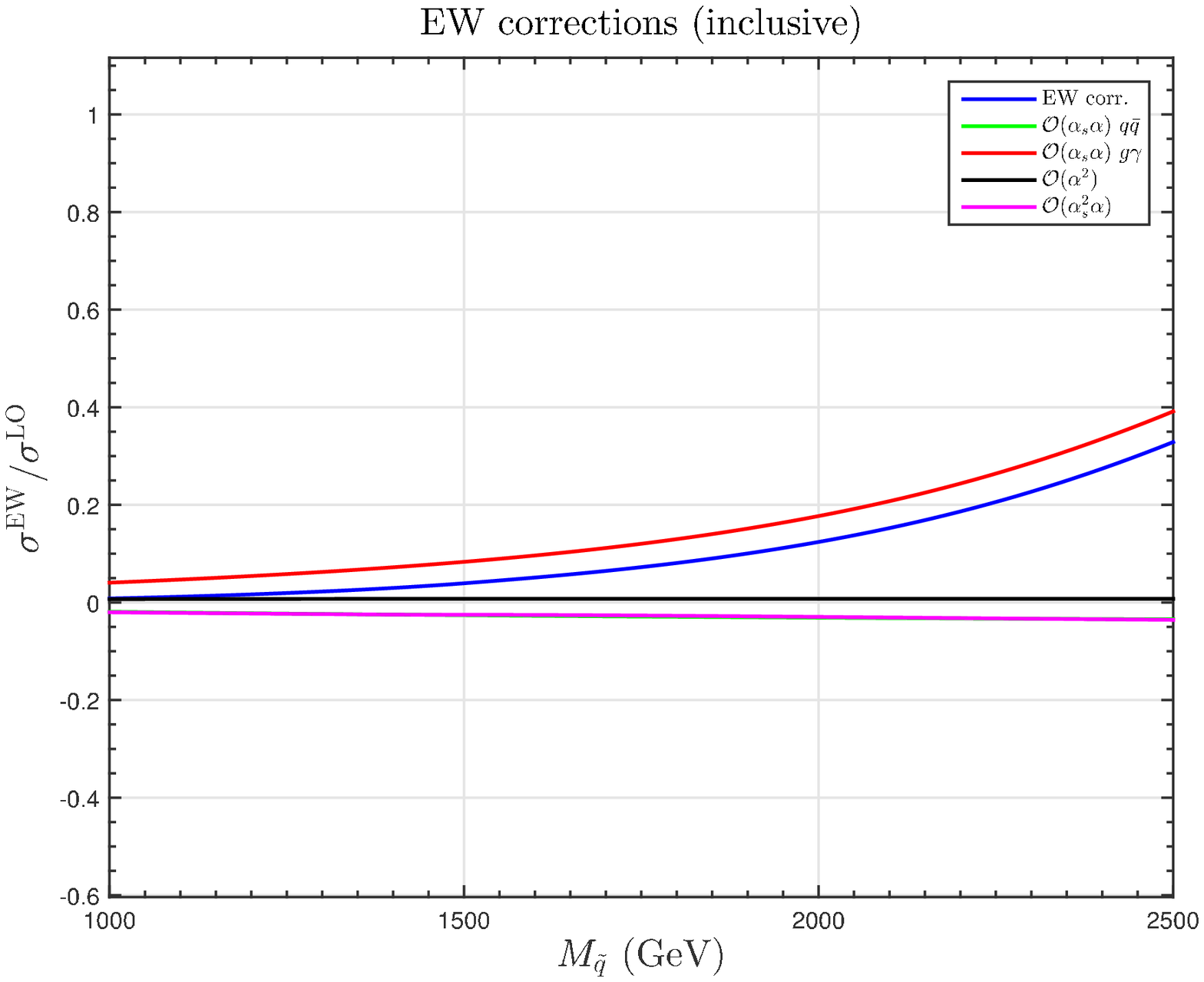}
\caption{}
\end{subfigure}
\phantom{pic}  \\
\begin{subfigure}[b]{0.5\textwidth}
\includegraphics[width=7.2cm,height=6.3cm]{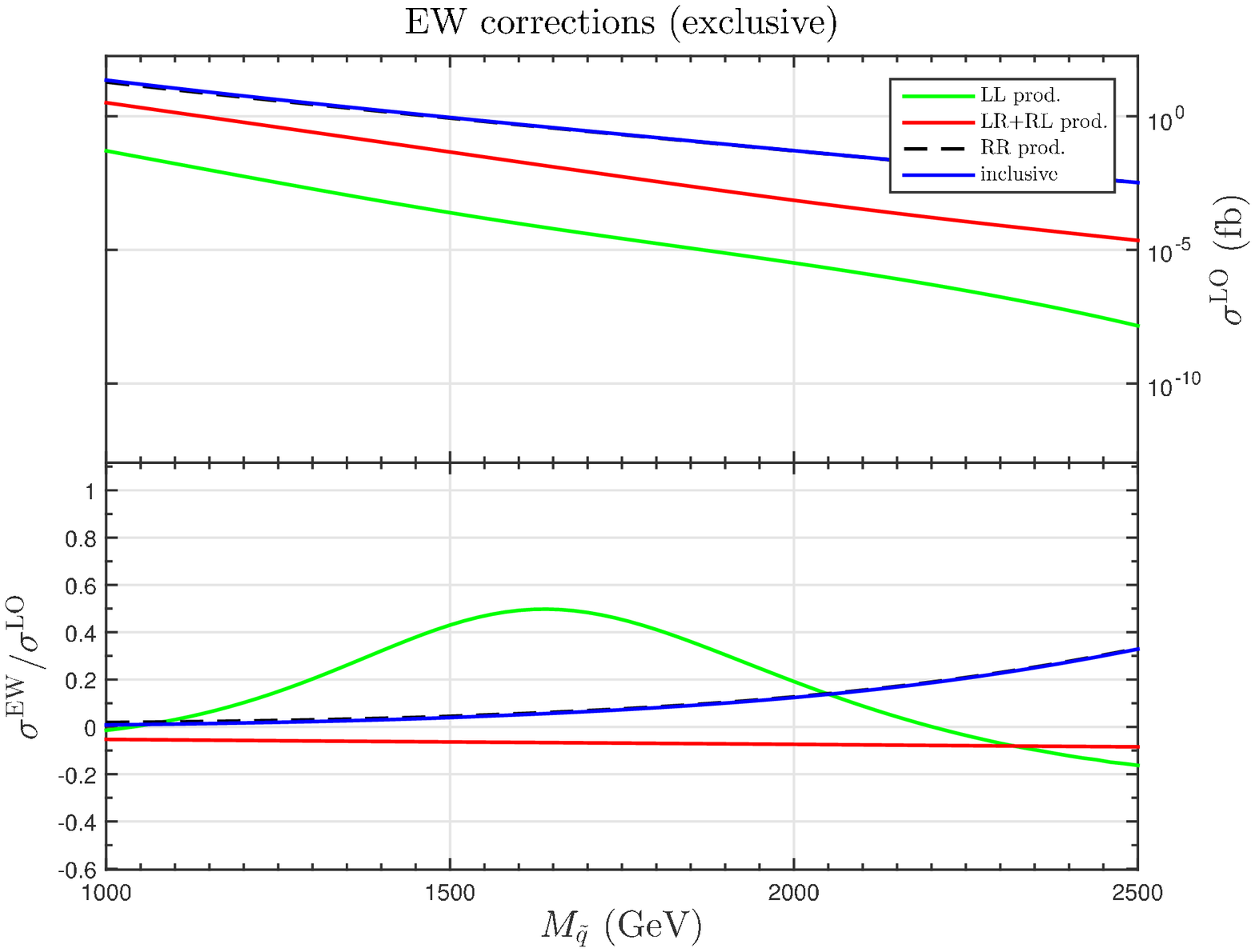}
\caption{}
\end{subfigure}
\begin{subfigure}[b]{0.5\textwidth}
\includegraphics[width=7.2cm,height=6.3cm]{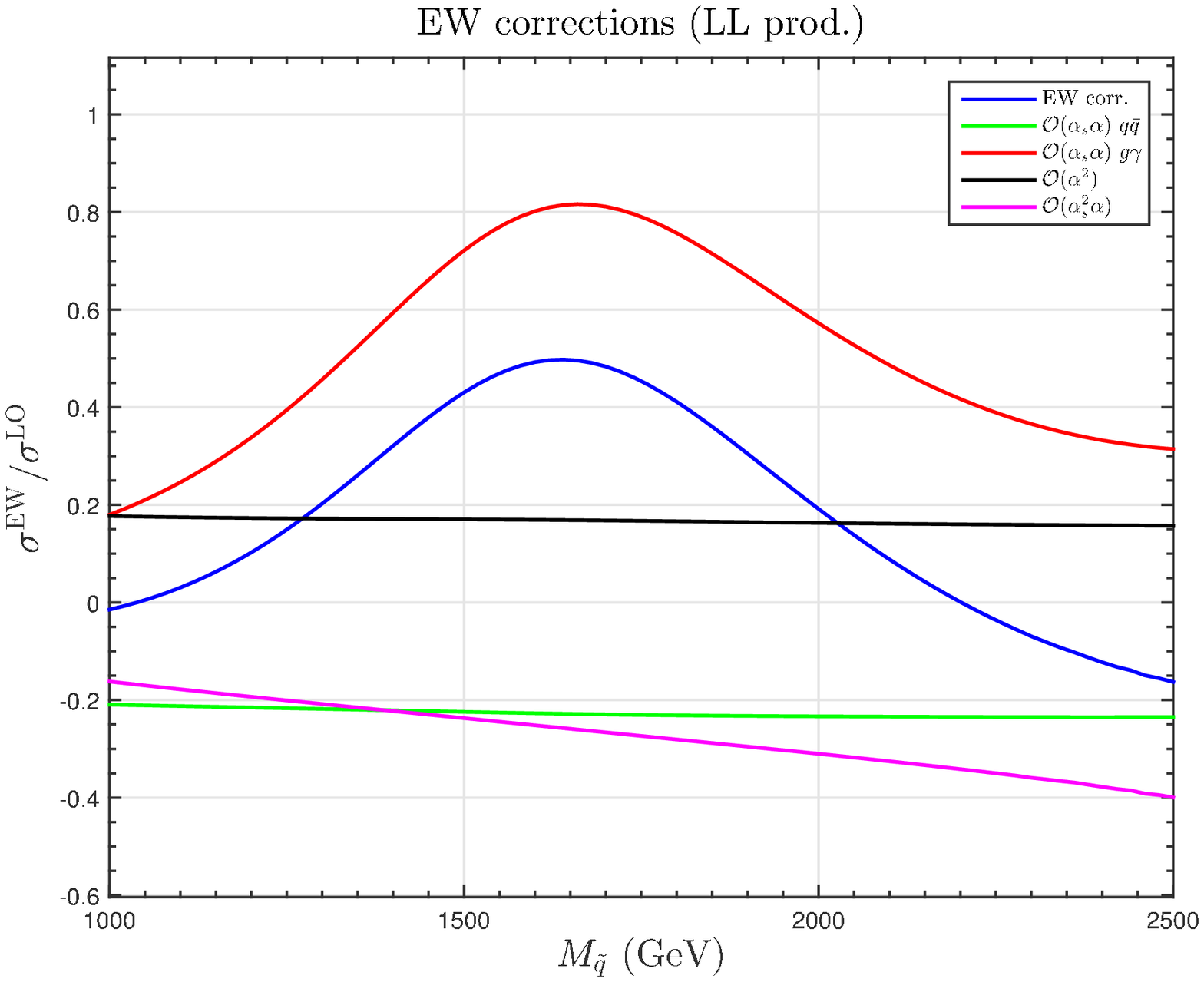}
\caption{}
\end{subfigure}
\phantom{pic}  \\
\begin{subfigure}[b]{0.5\textwidth}
\includegraphics[width=7.2cm,height=6.3cm]{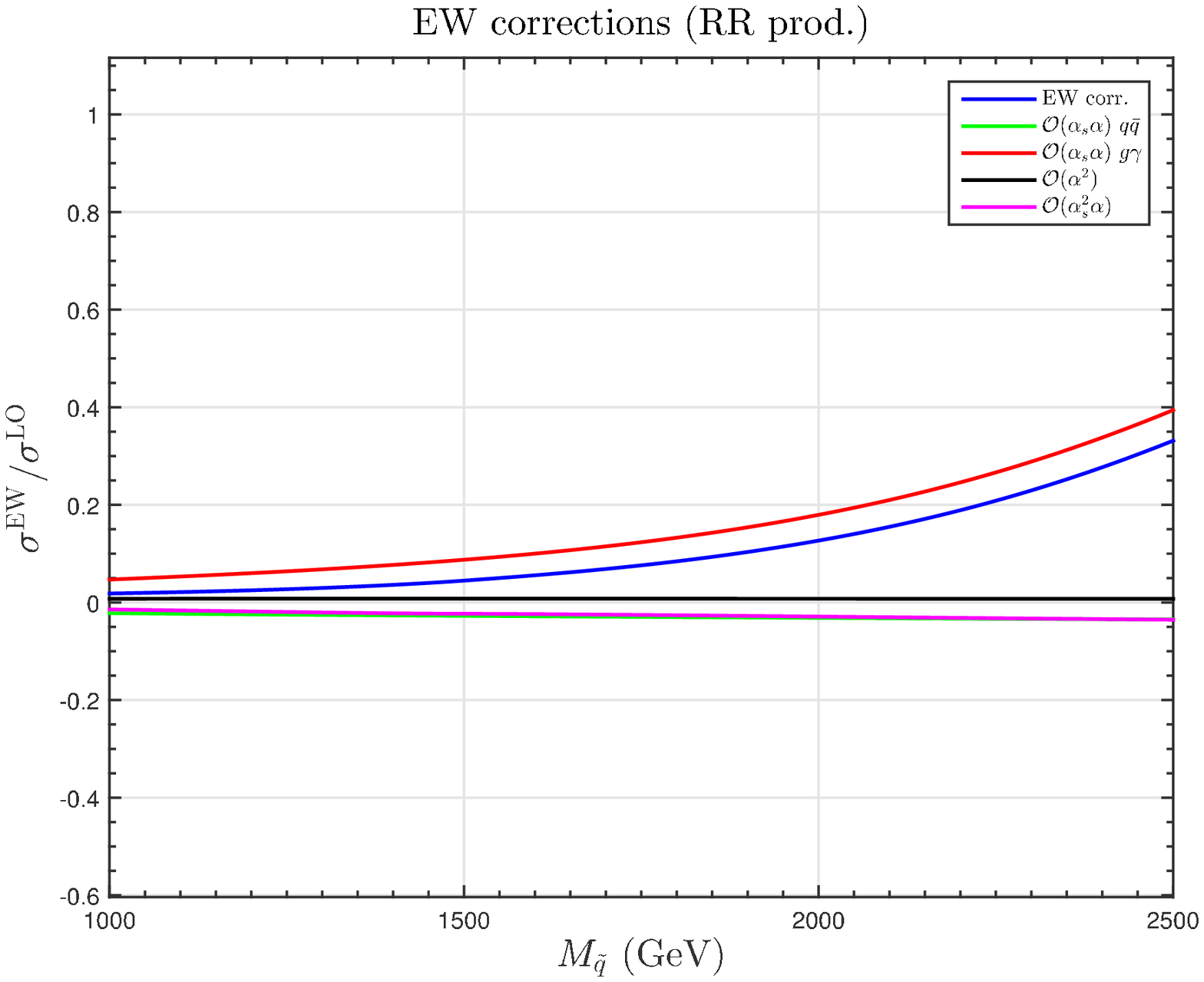}
\caption{}
\end{subfigure}
\begin{subfigure}[b]{0.5\textwidth}
\includegraphics[width=7.2cm,height=6.3cm]{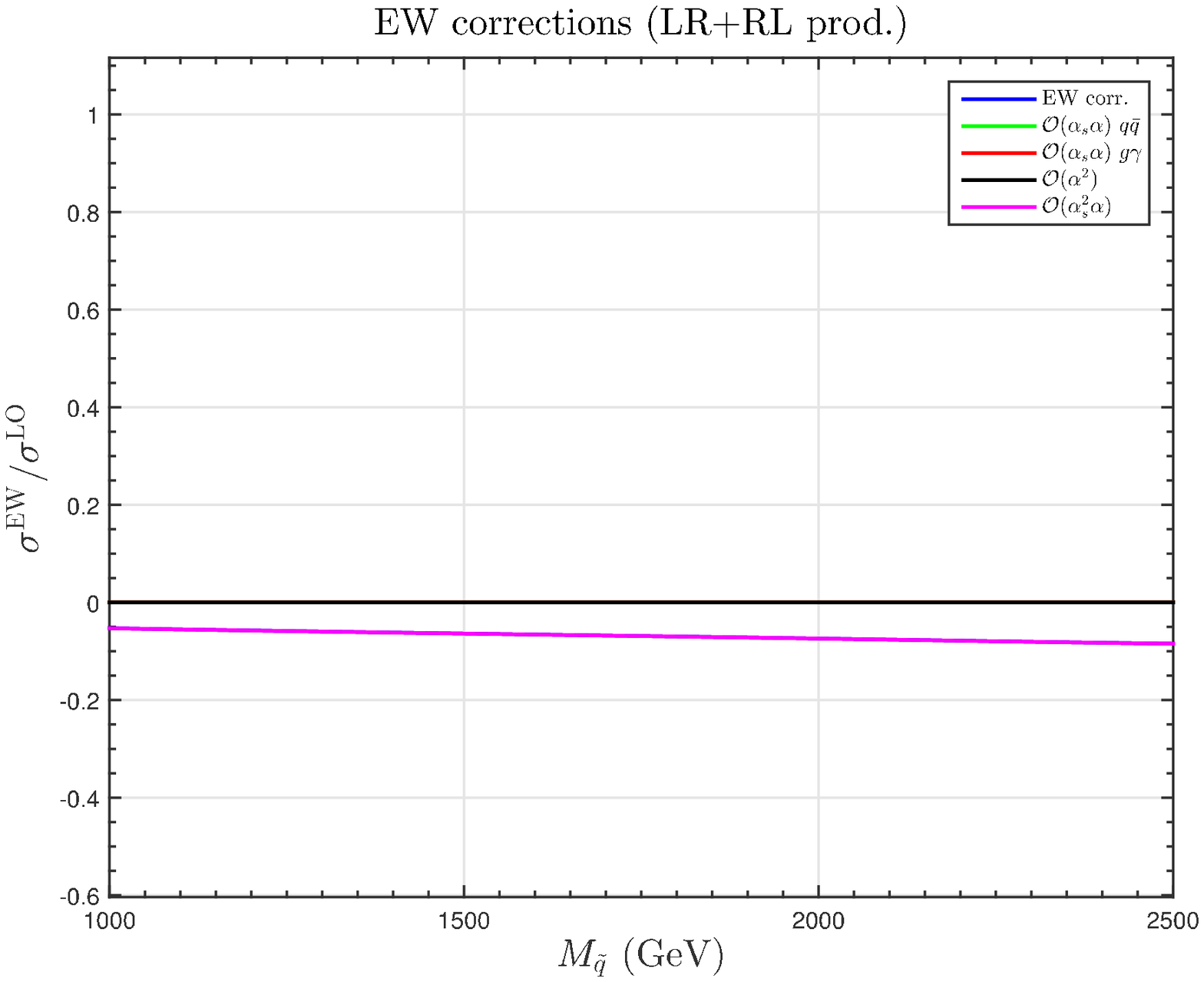}
\caption{}
\end{subfigure}
\caption[.]{Slope $S_3$: scan over $M_{\tilde q}$.   The value of the parameters not involved in the scan are collected in Table~\ref{Tab:Bench}. }
\label{Fig:LSS3_MSQ12}
 \end{figure}

\textbf{Slope $\mathbf{S_3}$}: 
In this scenarios (see Fig.~\ref{Fig:LSS3_MSQ12})
with $m_{\tilde q _{L}}=2m_{\tilde q _{R}}$, both LO predictions and higher-order corrections are mostly determined from RR production. The summed EW corrections are positive and increase as $M_{\tilde q}$ increases, reaching $10\%$  in correspondence to $M_{\tilde q} \simeq 2000$ GeV. The mass hierarchy is inverted wrt.  the one of the slope $S_2$, so  the qualitative discussion for the LL(RR) contribution in  $S_2$ applies here for the RR(LL) contribution. The main difference with $S_2$, as can be seen from Fig.~\ref{Fig:LSS3_MSQ12}(e), is  the fact that there are no large cancellations among the different EW contributions for RR(inclusive) production.

\begin{figure}[t]
\begin{subfigure}[b]{0.5\textwidth}
\includegraphics[width=7.2cm,height=6.3cm]{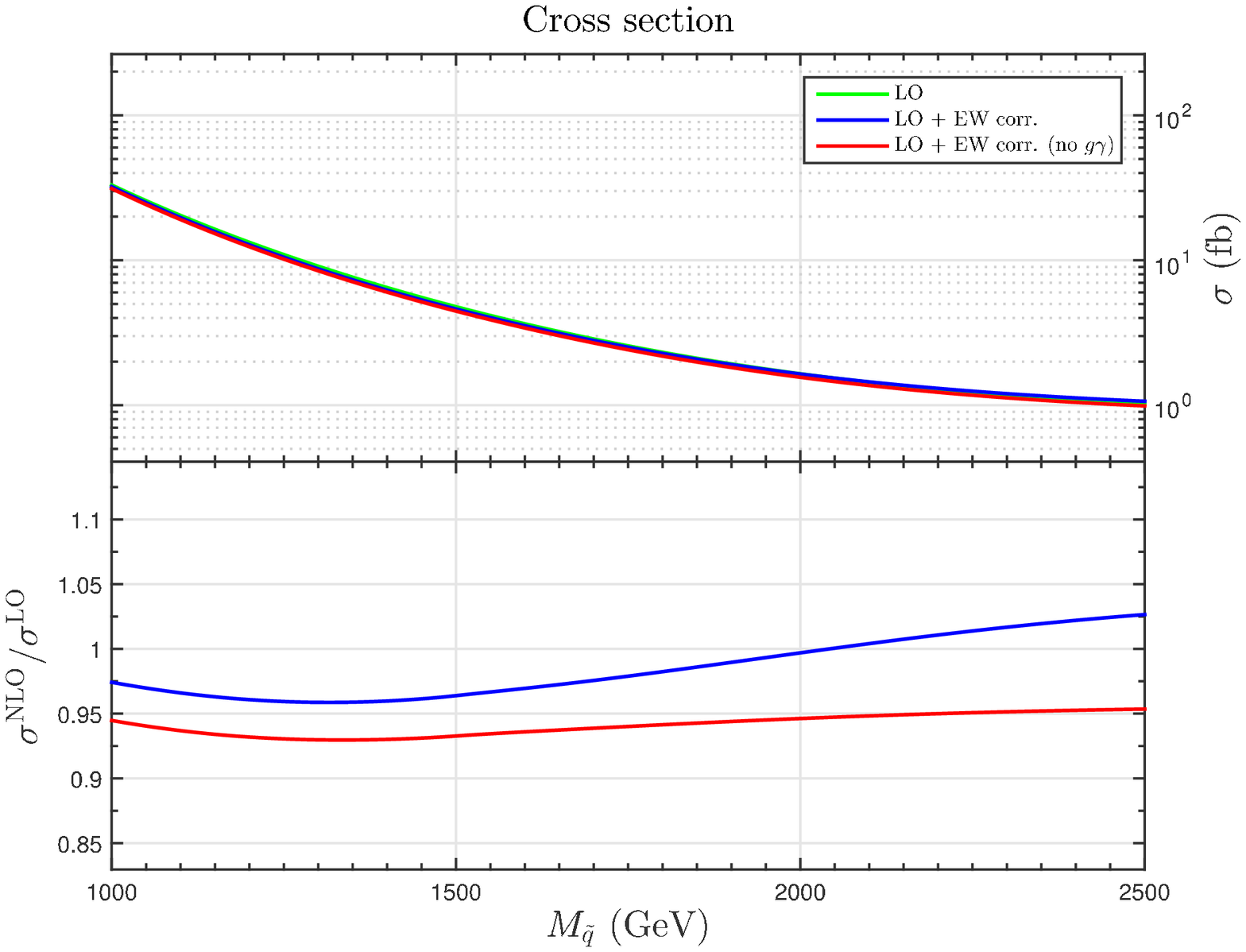}
\caption{}
\end{subfigure}
\begin{subfigure}[b]{0.5\textwidth}
\includegraphics[width=7.2cm,height=6.3cm]{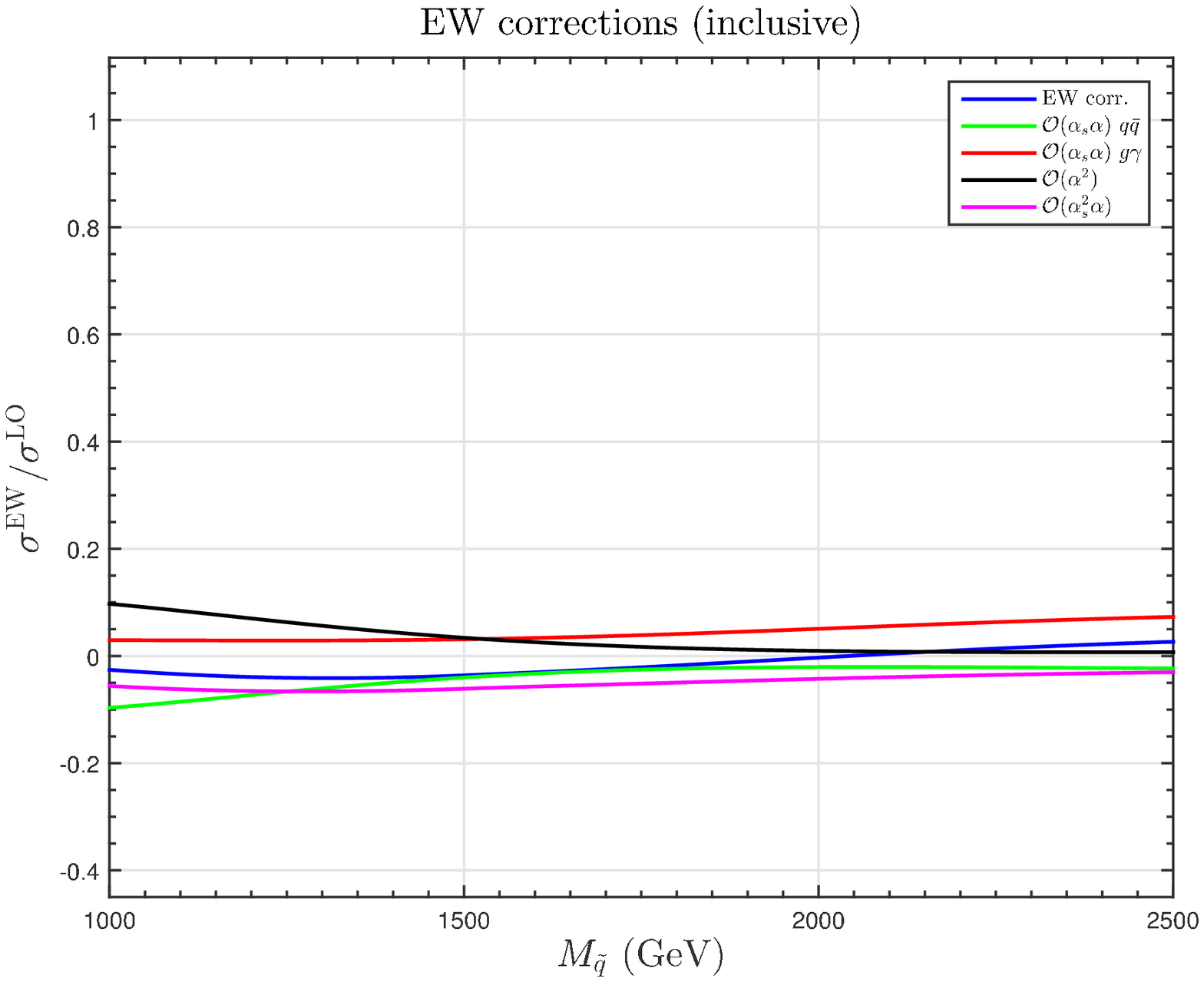}
\caption{}
\end{subfigure}
\phantom{pic}  \\
\begin{subfigure}[b]{0.5\textwidth}
\includegraphics[width=7.2cm,height=6.3cm]{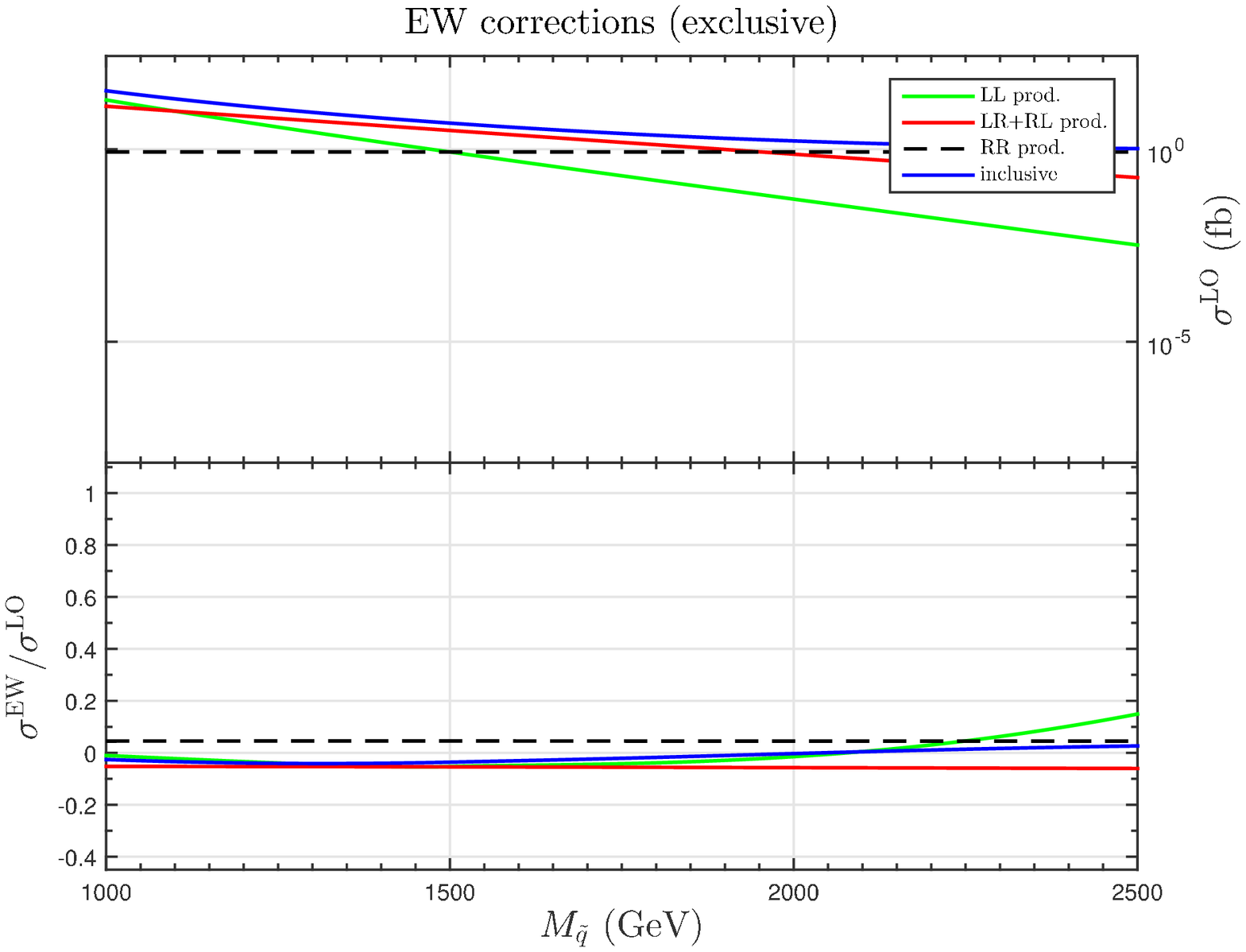}
\caption{}
\end{subfigure}
\begin{subfigure}[b]{0.5\textwidth}
\includegraphics[width=7.2cm,height=6.3cm]{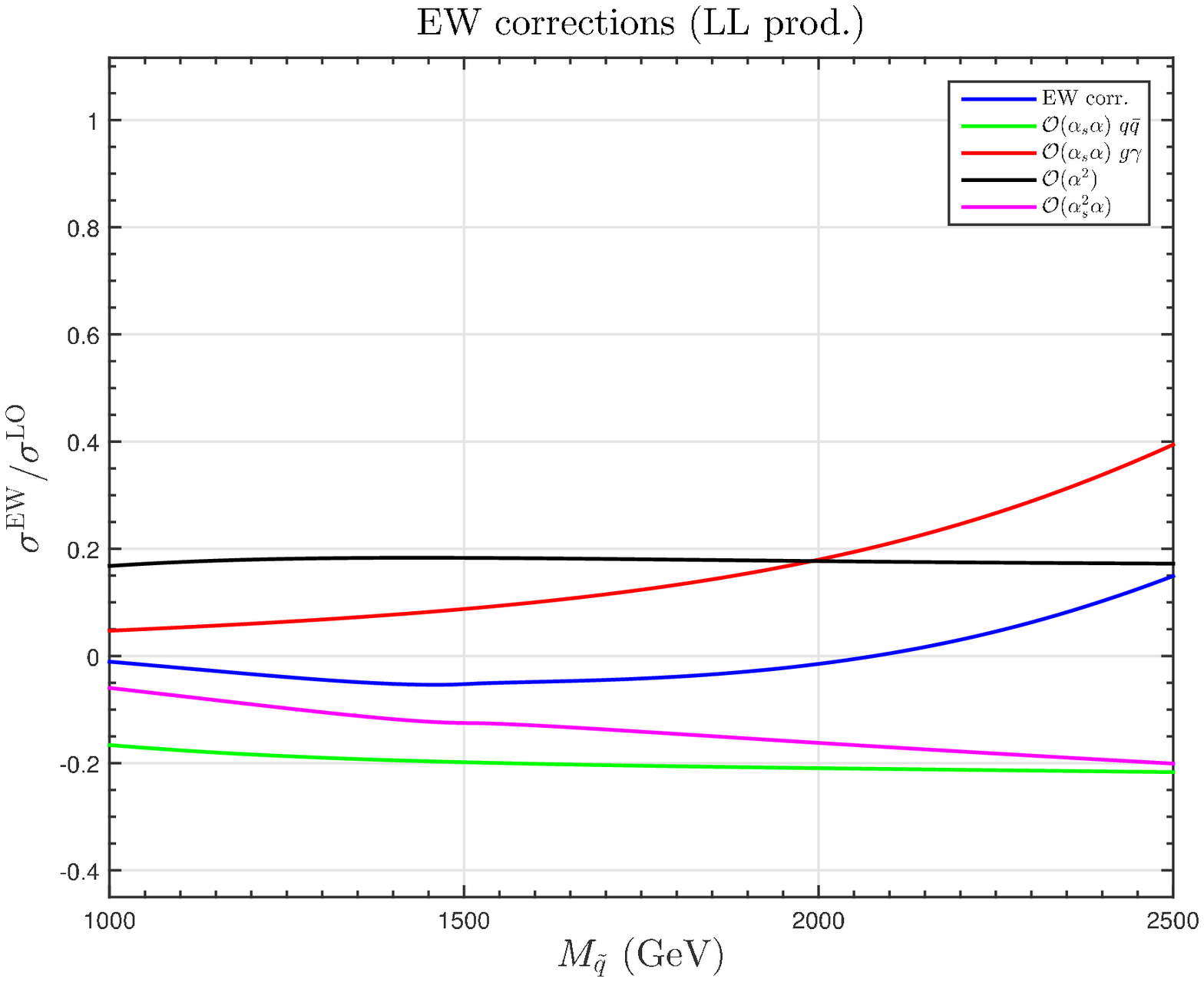}
\caption{}
\end{subfigure}
\phantom{pic}  \\
\begin{subfigure}[b]{0.5\textwidth}
\includegraphics[width=7.2cm,height=6.3cm]{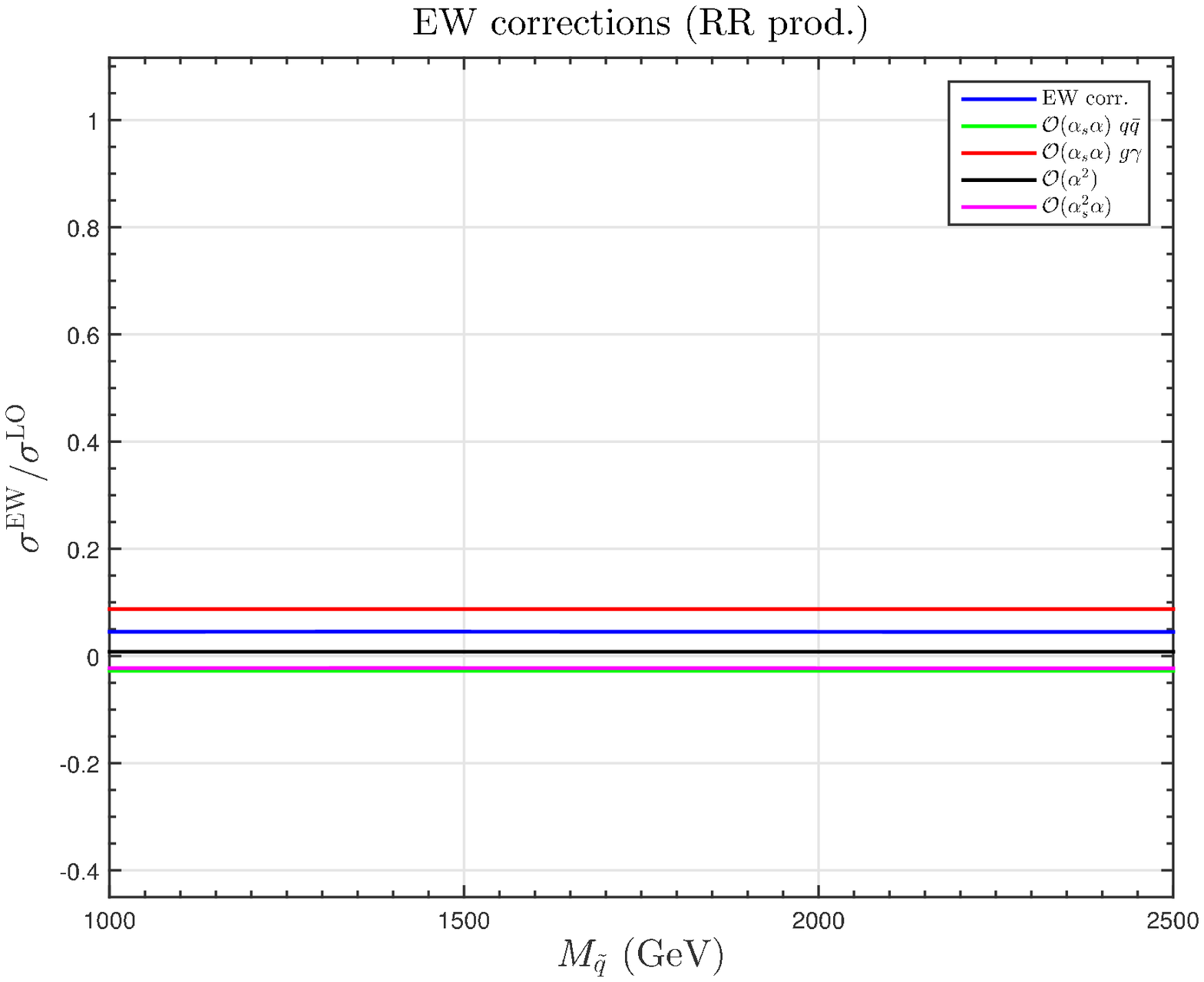}
\caption{}
\end{subfigure}
\begin{subfigure}[b]{0.5\textwidth}
\includegraphics[width=7.2cm,height=6.3cm]{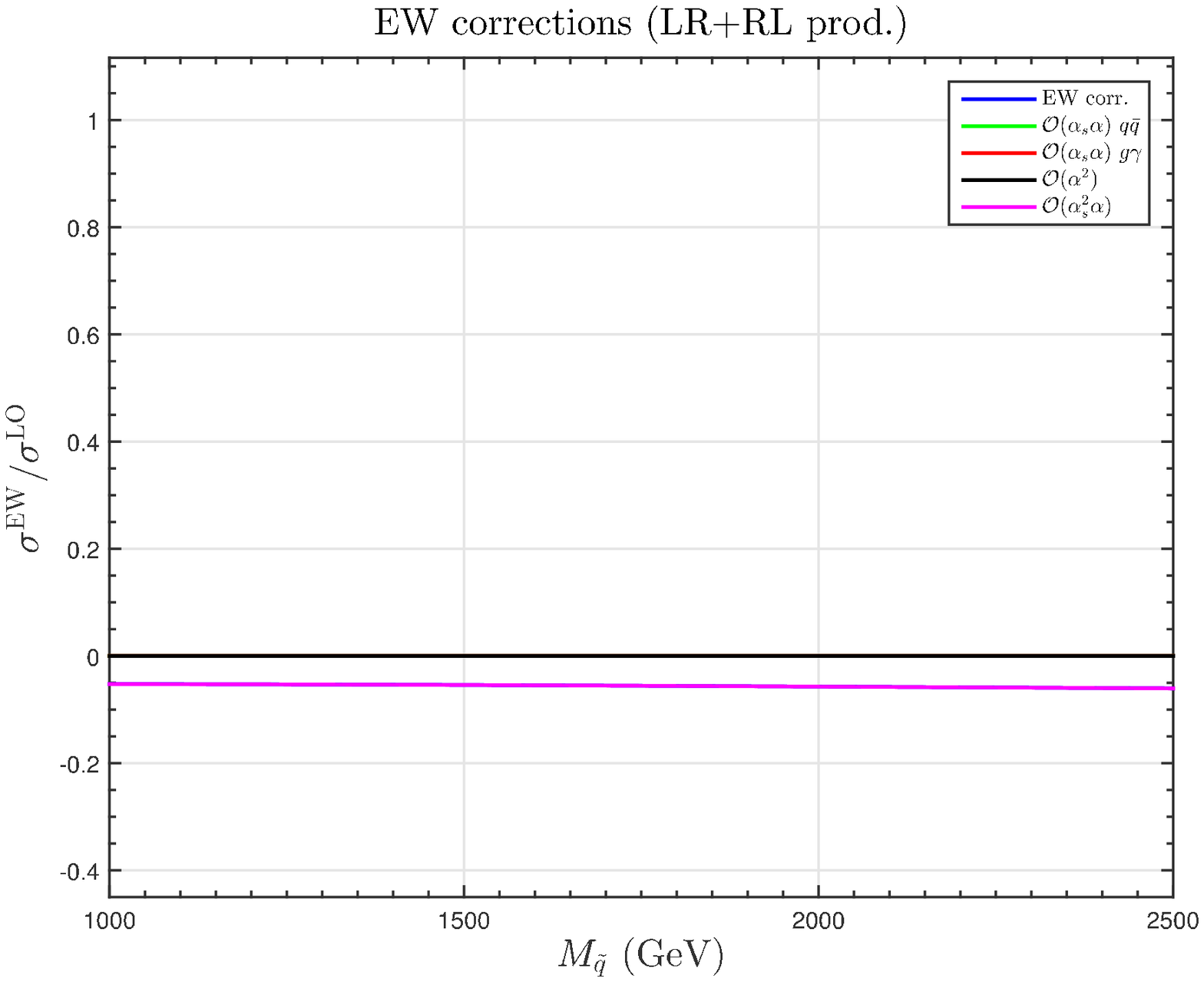}
\caption{}
\end{subfigure}
\caption[.]{Slope $S_4$: scan over $M_{\tilde q}$.   The value of the parameters not involved in the scan are collected in Table~\ref{Tab:Bench}. }
\label{Fig:LSS4_MSQ12}
 \end{figure}

\textbf{Slope $\mathbf{S_4}$}: In this scan (Fig.~\ref{Fig:LSS4_MSQ12}) the mass 
of the right-handed squark $m_{\tilde q _{R}}$ is kept fix at $m_{\tilde q _{R}}=1.5$~TeV, while $m_{\tilde q _{L}}=M_{\tilde q}$ is varied. As can be seen in the lower panel of Fig.~\ref{Fig:LSS4_MSQ12}(c), the hierarchy of the LO predictions for LL, RR and LR+RL depends on $M_{\tilde q}$. Consequentially, also the EW corrections and the individual perturbative orders receive the dominant contributions from LL, RR or LR+RL depending on the value of  $M_{\tilde q}$. In RR production, Fig.~\ref{Fig:LSS4_MSQ12}(e), the EW corrections are constant by construction and corresponds to those of the spectrum in Table~\ref{Tab:Bench}.  Also the summed LR+RL production, Fig.~\ref{Fig:LSS4_MSQ12}(f), does not show a visible dependence on $M_{\tilde q}$. LL production  shows a very similar behavior as observed for the slopes $S_1$ and $S_2$.

\begin{figure}[t]
\begin{subfigure}[b]{0.5\textwidth}
\includegraphics[width=7.2cm,height=6.3cm]{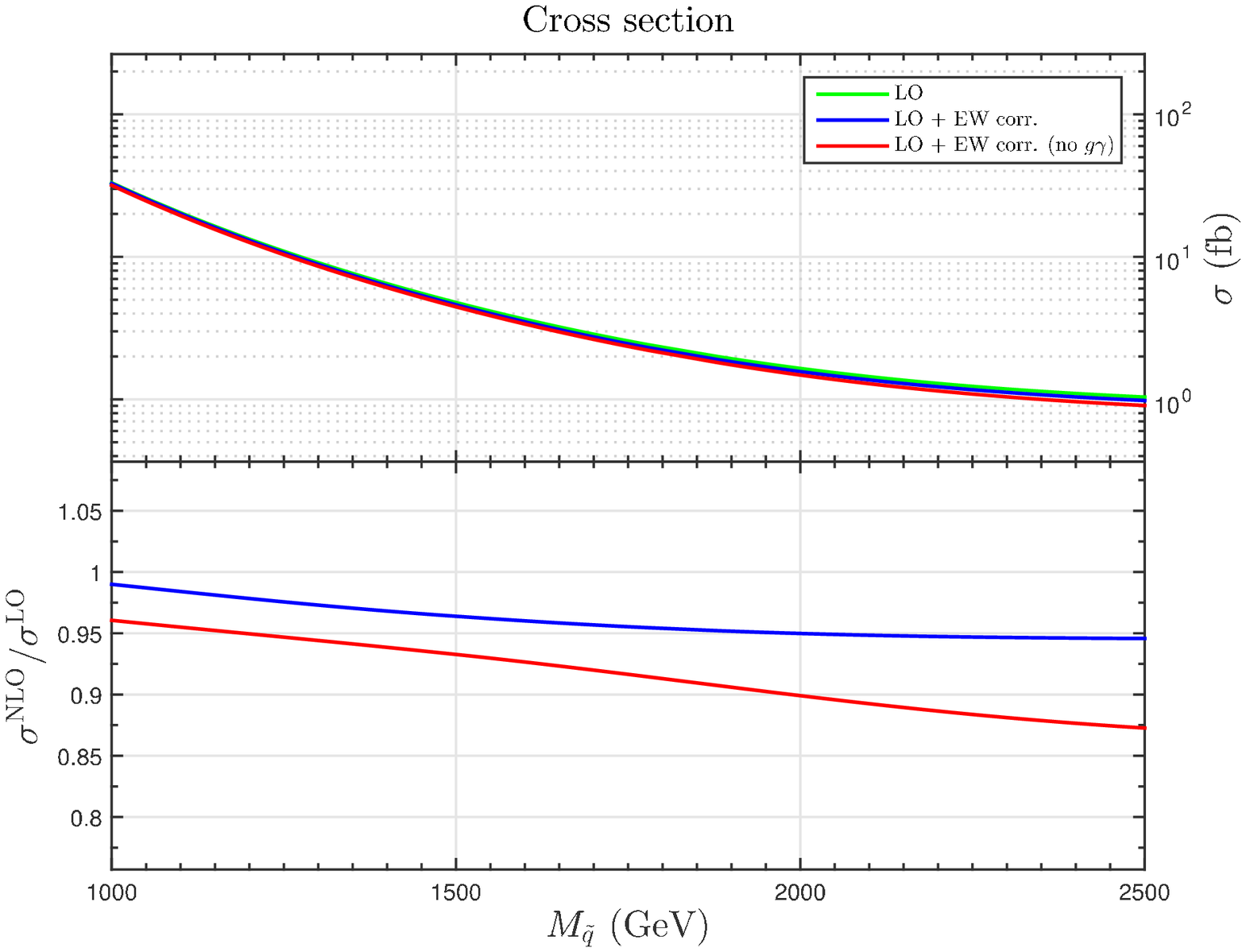}
\caption{}
\end{subfigure}
\begin{subfigure}[b]{0.5\textwidth}
\includegraphics[width=7.2cm,height=6.3cm]{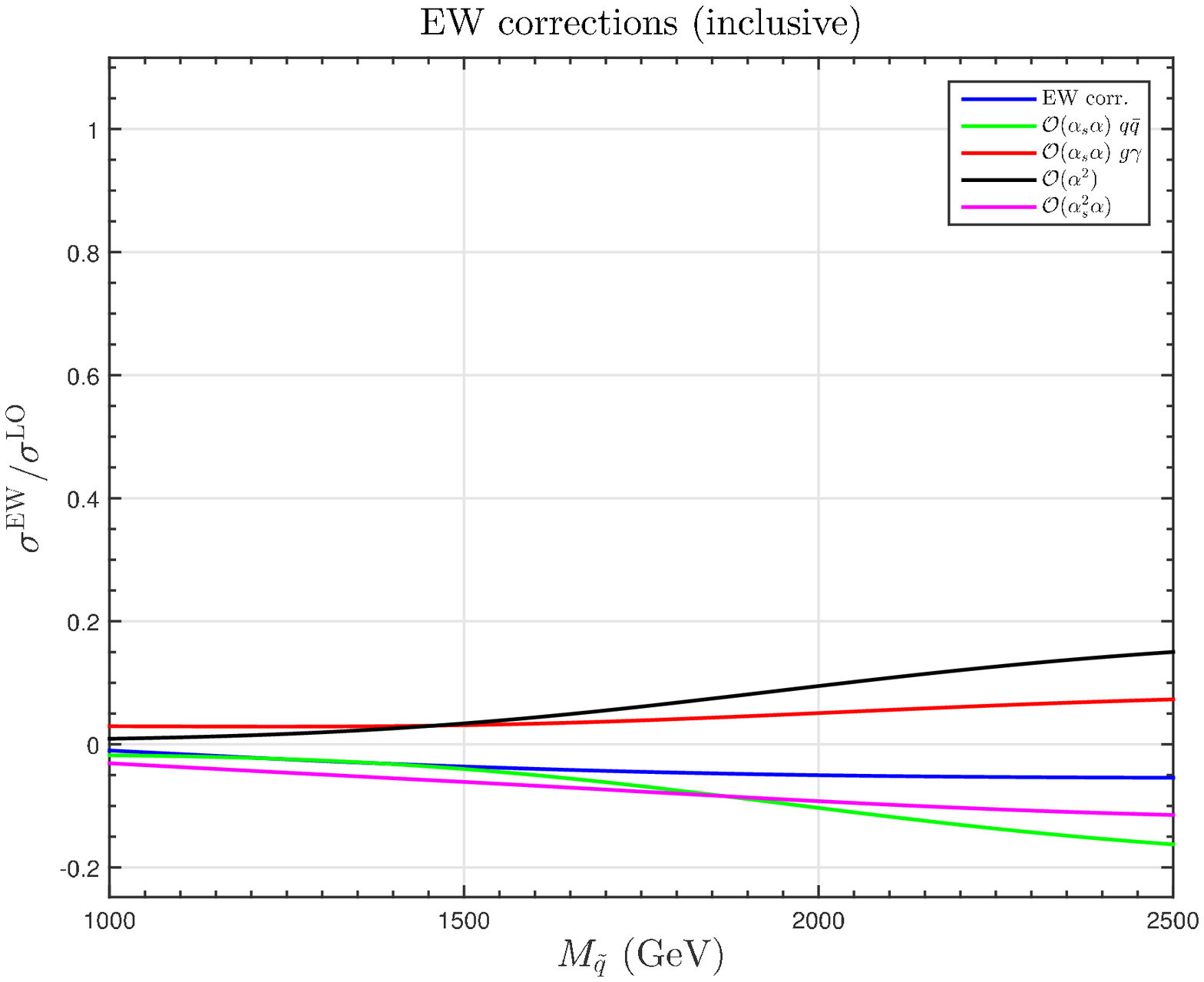}
\caption{}
\end{subfigure}
\phantom{pic}  \\
\begin{subfigure}[b]{0.5\textwidth}
\includegraphics[width=7.2cm,height=6.3cm]{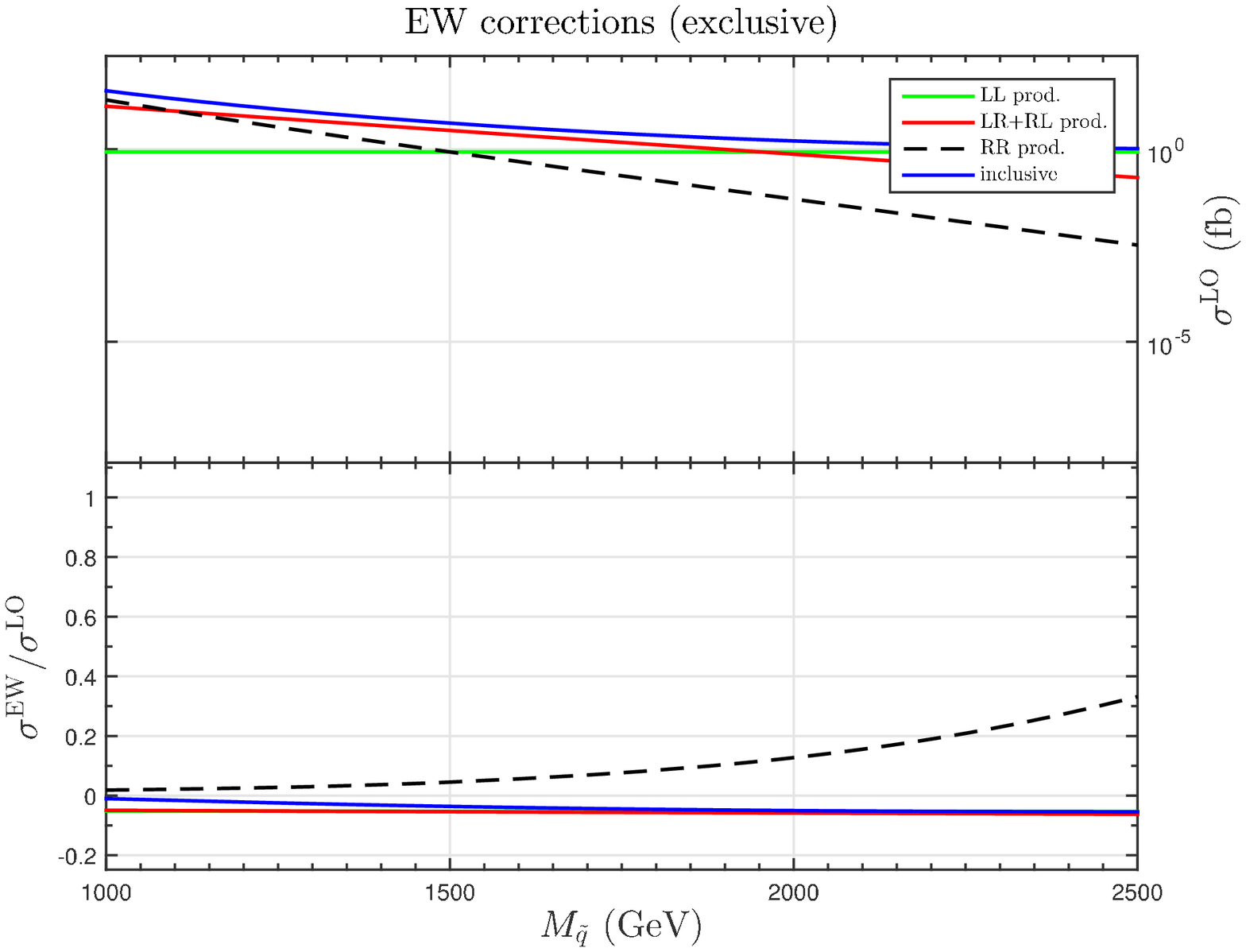}
\caption{}
\end{subfigure}
\begin{subfigure}[b]{0.5\textwidth}
\includegraphics[width=7.2cm,height=6.3cm]{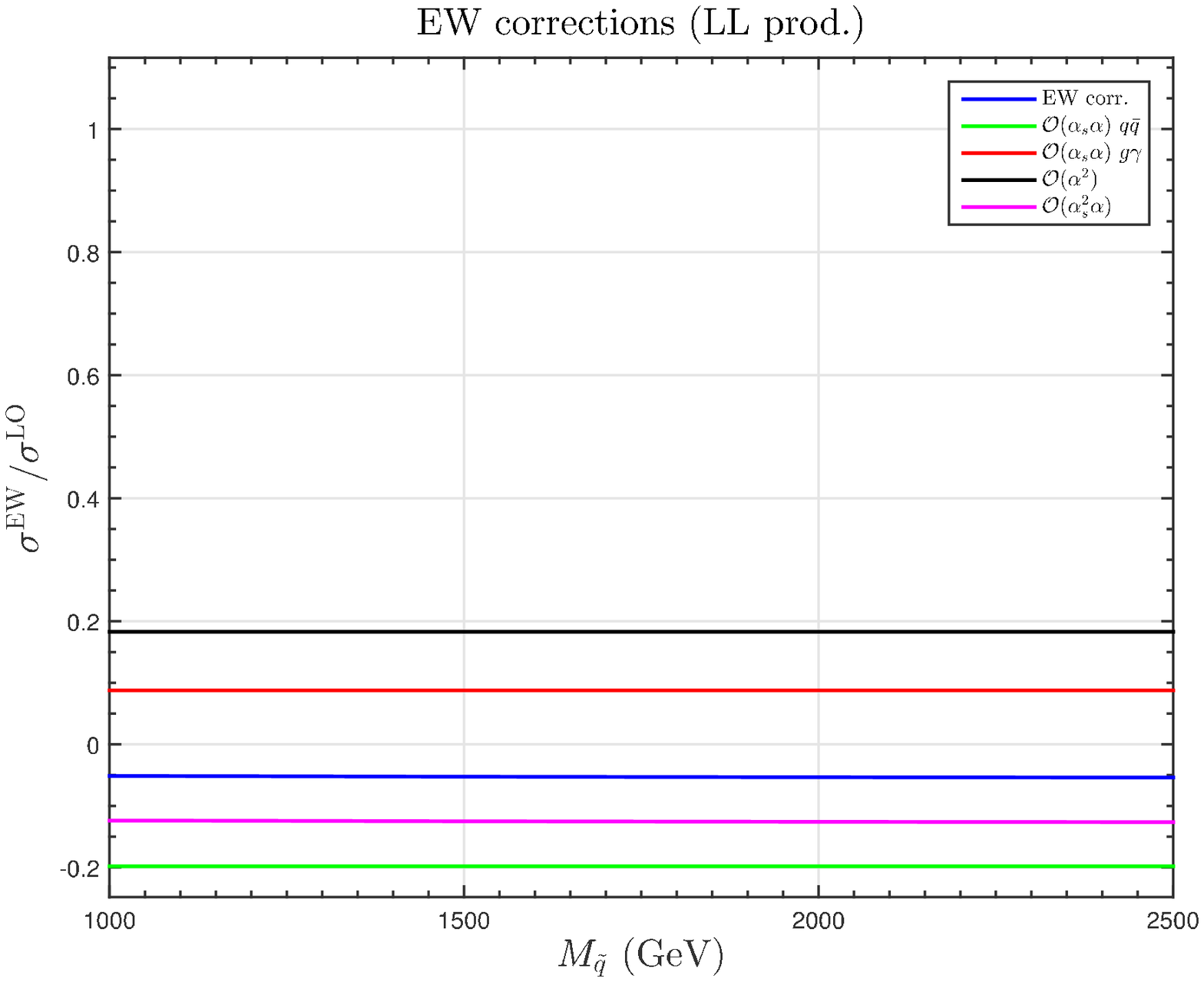}
\caption{}
\end{subfigure}
\phantom{pic}  \\
\begin{subfigure}[b]{0.5\textwidth}
\includegraphics[width=7.2cm,height=6.3cm]{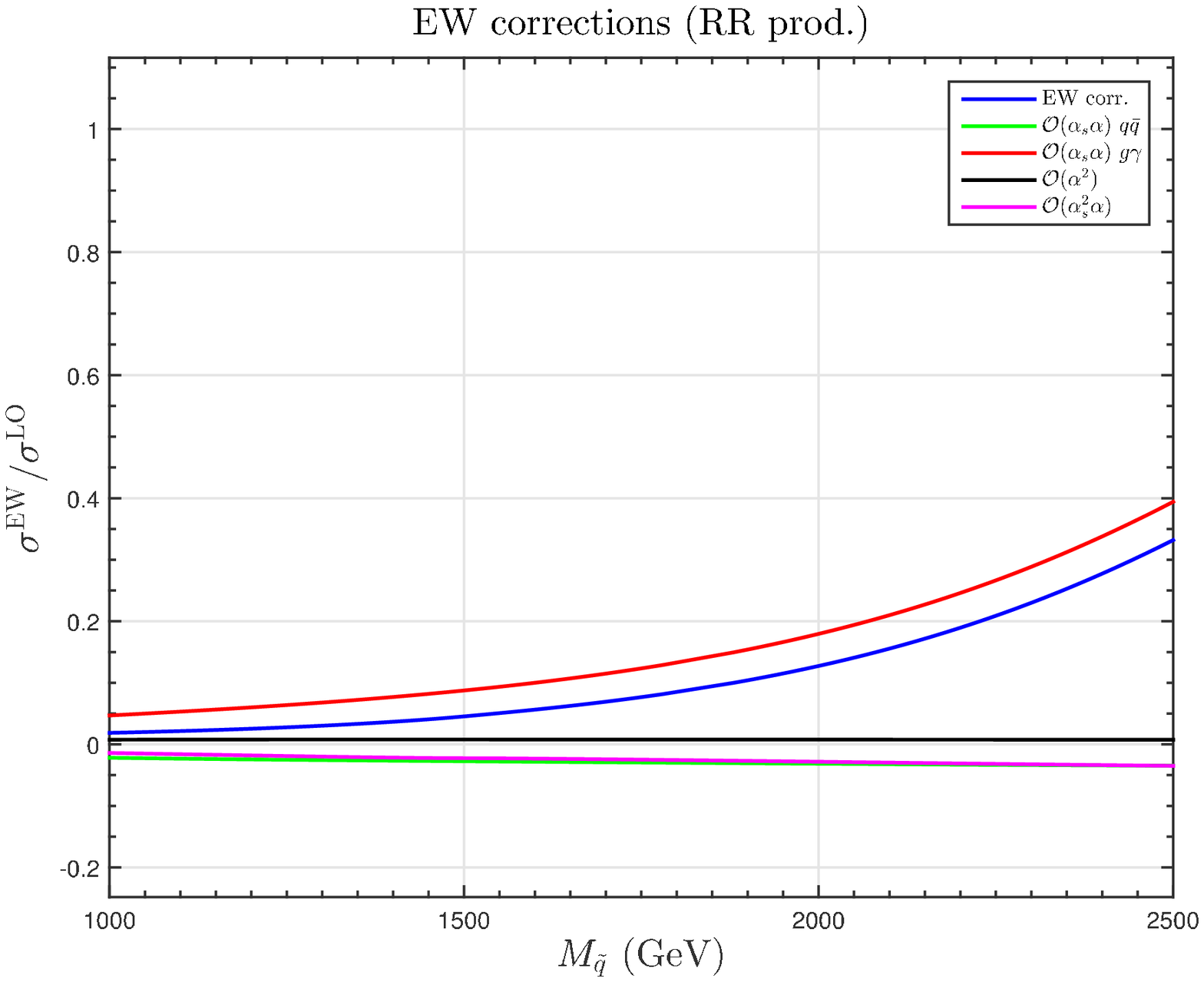}
\caption{}
\end{subfigure}
\begin{subfigure}[b]{0.5\textwidth}
\includegraphics[width=7.2cm,height=6.3cm]{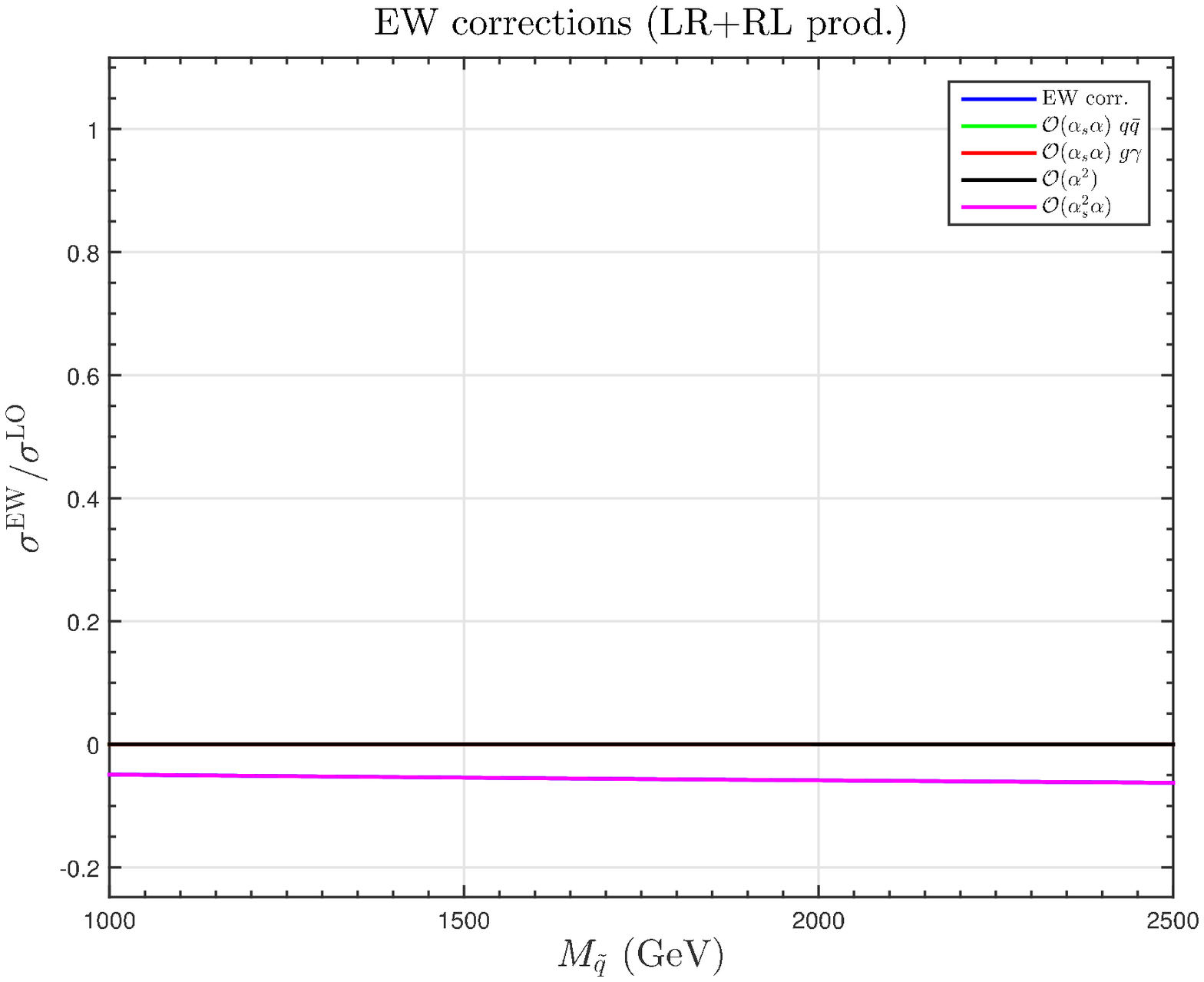}
\caption{}
\end{subfigure}
\caption[.]{Slope $S_5$: scan over $M_{\tilde q}$.   The value of the parameters not involved in the scan are collected in Table~\ref{Tab:Bench}.}
\label{Fig:LSS5_MSQ12}
 \end{figure}

\textbf{Slope $\mathbf{S_5}$}: In this scan, referring to Fig.~\ref{Fig:LSS5_MSQ12},
the mass of the right-handed squark $m_{\tilde q _{L}}$ is kept fix at $m_{\tilde q _{L}}=1.5$~TeV, while $m_{\tilde q _{R}}=M_{\tilde q}$ is varied. Thus the LO cross sections are those of Slope $S_4$ with LL and RR exchanged, cfr.~Fig.~\ref{Fig:LSS4_MSQ12}(c) and Fig.~\ref{Fig:LSS5_MSQ12}(c). The qualitative behavior of the  electroweak corrections to the different production channels can be understood as an interchange of LL with RR with respect to Slope $S_4$.

\subsection{Differential distributions}

In the previous subsection we studied the numerical impact of the  EW corrections to the total cross section for squark--antisquark production at the LHC.  
It is well known that, at high energies, Sudakov-type logarithms can enhance the EW contributions. 
Therefore, in the following we study the impact of the EW corrections on three kinematic distributions:  the transverse momentum of the produced squark, the maximal pseudo-rapidity of the squark and the anti-squark, and the invariant mass of the squark--antisquark system, defined as
 \begin{align}
 p_T  & \equiv  p_{T \, \tilde q_{\alpha}} \, ,   & \hspace{0.4cm} 
 \eta &\equiv  \left \{ \begin{array}{ll}
 \eta_{\tilde q_{\alpha}}  \quad \mbox{if} \quad    |\eta_{\tilde q_{\alpha}^{\phantom{\ast}}} | \ge |\eta_{ \tilde q_{\beta}^\ast}|   \\ [1.0ex]
\eta_{ \tilde q_{\beta}^\ast}  \quad \mbox{if} \quad    |\eta_{\tilde q_{\alpha}^{\phantom{\ast}}} | < |\eta_{ \tilde q_{\beta}^\ast}|   \end{array}  \right. \, , & & 
%
 M_{\rm inv} &\equiv \sqrt{\left ( p_{\tilde q_{\alpha}^{\phantom{\ast}}} + p_{\tilde q_{\beta}^\ast}\right)^2 }\, , \hspace{0.4cm} 
 \label{eq:diff_distributions}
 \end{align}
respectively. The quantities $p_j$, $p_{T\, j}$ and $\eta_j$ denote the four-momentum, the transverse momentum and the pseudo-rapidity  of the particle $j$. It is worth to remind that these distributions cannot be directly observed at the experimental level; a realistic phenomenological evaluation requires the combination of the production process with the decays of the squarks. Nevertheless, it is important and useful to identify the kinematic dependence of EW corrections at the production level, before performing a complete simulation including squark decays and acceptance cuts.

For illustration of the effects, we consider,
as a benchmark, the scenario defined in Table~\ref{Tab:Bench}, but now with $M_{\tilde q, \rm L}=M_{\tilde q, \rm R}=2000$~GeV, i.e. a scenario within the reach of Run-II of the LHC. The numerical results for distributions in $p_T$, $\eta$ and $M_{\rm inv}$ are collected in Fig.~\ref{Fig:dist_pt}, Fig.~\ref{Fig:dist_eta} and Fig.~\ref{Fig:dist_minv} respectively.
In the left plots we always show the absolute predictions at LO with
and without the EW corrections. The right plots contain the 
relative size of the  EW corrections with respect to the LO prediction,
and the breakdown to their various individual contributions.
From top to bottom, the plots are arranged to display the predictions for
inclusive, LL, RR, and LR+RL production.  
The overall behavior with respect to the different kinematic observables is strongly dependent on the chirality configuration of the produced squarks.

The EW corrections to the $p_T$ distribution, shown in
Fig.~\ref{Fig:dist_pt}, are largest for the LL squark configurations,
owing to their hypercharge. In particular they are negative and reach $20\%$ and more for large $p_T$. 
The LR+RL configurations, which dominate the inclusive cross section,
are influenced by  $-10\%$ for large $p_T$. 
In the RR configurations, the EW corrections always stay below  $5\%$.

The  $\ord(\alpha_s \alpha)$ and $\ord(\alpha^2)$ Born contributions,
relevant only for LL and RR production, have always opposite sign and
similar magnitude, resulting in mutual cancellations in most of the
$p_T$-spectrum. The photon-induced production channel, again only relevant for LL and RR production, can yield sizable contributions both at small and at very large $p_T$. In particular, in the case of  RR production it tends to overcompensate the negative NLO EW contribution.

Turning now to the EW corrections in the pseudo-rapidity $\eta$ of the produced squarks (Fig.~\ref{Fig:dist_eta}),
the origin of the large photon-induced contributions at small $p_T$ becomes evident. The photon induced production modes show a different angular dependence as compared to the LO prediction or the NLO EW, i.e.~they are strongly enhanced in the forward region at large $|\eta|$, while the NLO EW corrections are mildly enhanced for small $|\eta|$. Due to this large dependence on $\eta$, the relative corrections induced by photon-induced contributions on experimental rates can strongly be affected by the details of the analysis cuts on the squark decay products.

Finally we turn to the distributions in the invariant mass of the produced squark--antisquark pair (Fig.~\ref{Fig:dist_minv}).
Again we observe a partial mutual cancellation between the different contributions to the EW corrections. Moreover, the NLO EW corrections are not significantly enhanced at large invariant masses.

\section{Conclusions}
\label{sec:conclusions}

In this paper we have presented the first phenomenological study of the NLO EW corrections to squark--antisquark production at the LHC including all production channels and chirality combinations of the produced squarks.


Our analysis has shown that electroweak contributions to $\tilde q_\alpha \tilde q'^\ast_\beta$ production are not negligible even at
the inclusive level. The $\ord(\alpha_s^2\alpha)$ NLO EW contributions
are negative, in general sizable and increase in relative size with
the mass of the produced squarks. For the production of a left-handed
squark and antisquark pair they reach $-20\%$ for squark masses of 2 TeV.
For the production of two left- or two right-handed squarks these
negative contributions are typically (over-)compensated by  large
tree-level contributions from the photon--gluon initial state,
which also increase in relative size for large masses. 
However, these contributions are accompanied by very large intrinsic PDF uncertainties, which may substantially alter the size of the electroweak corrections and, for large masses, the accuracy of the total predictions.
The dependence on the remaining MSSM parameters is found to be very weak for the NLO EW $\ord(\alpha_s^2\alpha)$ and the photon-induced contributions. The subleading Born contributions of $\ord(\alpha_s\alpha)$ and $\ord(\alpha^2)$ are not negligible and show a moderate dependence on $m_{\tilde g}$ and $M_2$.

Besides inclusive cross sections, the electroweak corrections to
squark--antisquark production are also investigated at the
differential level. The $\ord(\alpha_s^2\alpha)$ NLO EW corrections
show a typical Sudakov behavior, i.e.~increasing influence at large transverse momenta of the produced squarks. As on the inclusive level, these corrections are partly compensated by large contributions from photon-induced  production. The origin of these large contributions can clearly be attributed to a qualitatively different behavior of the photon-induced channels at large  pseudo-rapidities.

In general, in studies for squark production as well as in corresponding data analyses,    
when EW effects are usually neglected for reasons of simplicity, 
the size of the EW contributions calculated and visualized in this paper,
can serve as an estimate of the uncertainty of the theoretical predictions, 
on top of the uncertainty resulting from the QCD side.

\begin{acknowledgments} 
We are grateful to S. Carrazza and J. Rojo for clarifications concerning NNPDF2.3QED. JML was supported by the European Commission through the ``LHCPhenoNet'' Initial Training Network PITN-GA-2010-264564. DP was supported by the the ERC grant 291377 ``LHCtheory: Theoretical predictions 
and analyses of LHC physics: advancing the precision frontier".
This research was supported by the Munich Institute for Astro- and Particle Physics (MIAPP) of the DFG cluster of excellence ``Origin and Structure of the Universe".
\end{acknowledgments}

\bibliographystyle{JHEP}

\providecommand{\href}[2]{#2}\begingroup\raggedright\endgroup


\newpage



%


\def\reldistplotwidth{0.94}
\def\reldistplotwidthratio{0.94}
\begin{figure}[t]
\begin{subfigure}[b]{0.5\textwidth}
\includegraphics[width=\reldistplotwidth\textwidth]{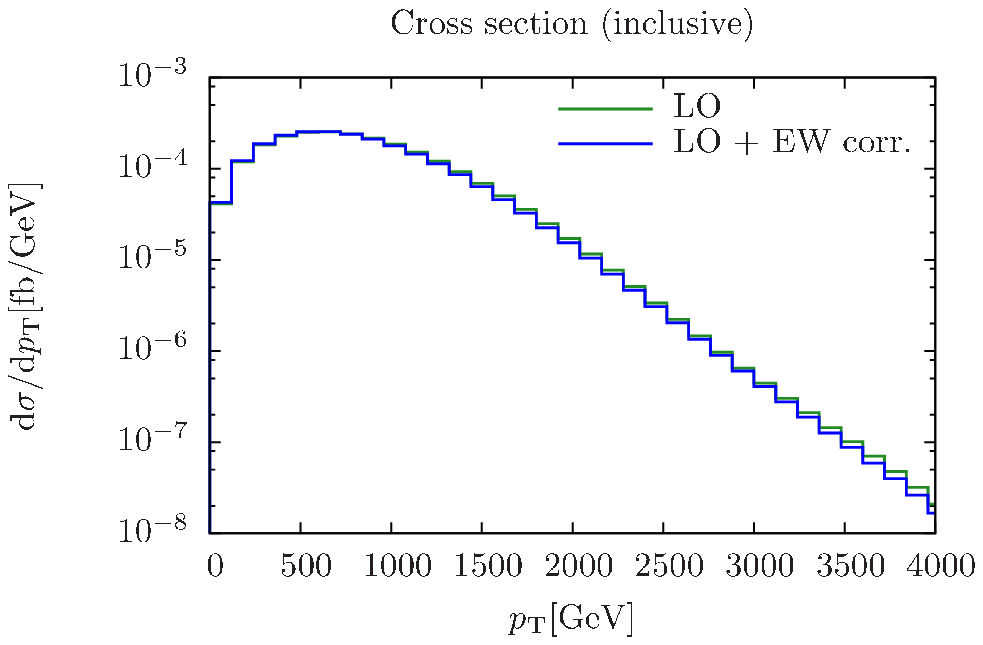}
\caption{}
\end{subfigure}
\begin{subfigure}[b]{0.5\textwidth}
\includegraphics[width=\reldistplotwidthratio\textwidth]{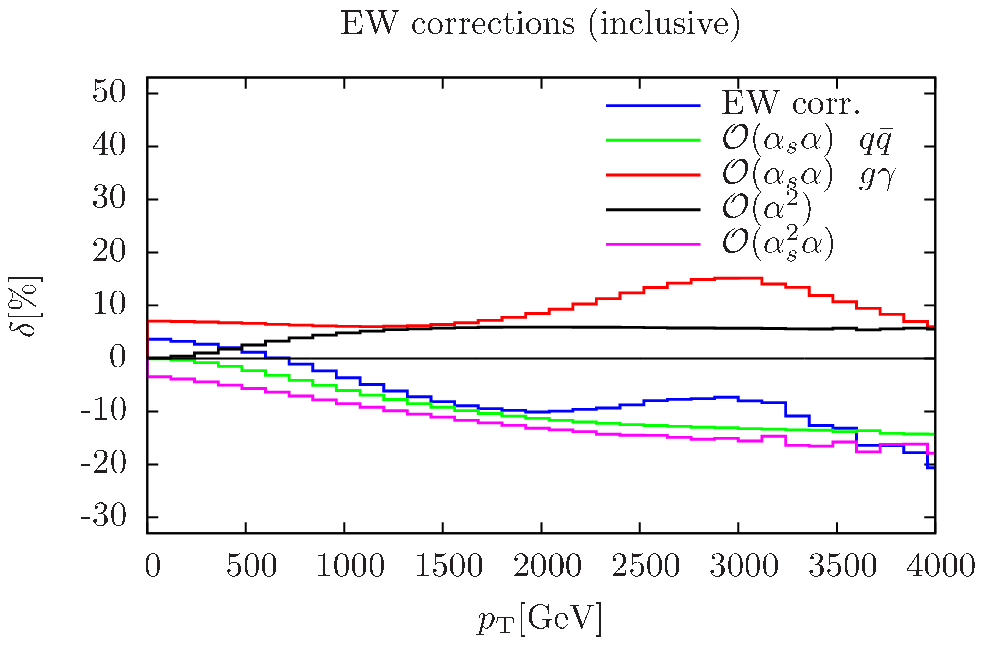}
\caption{}
\end{subfigure}
\phantom{pic}  \\
\begin{subfigure}[b]{0.5\textwidth}
\includegraphics[width=\reldistplotwidth\textwidth]{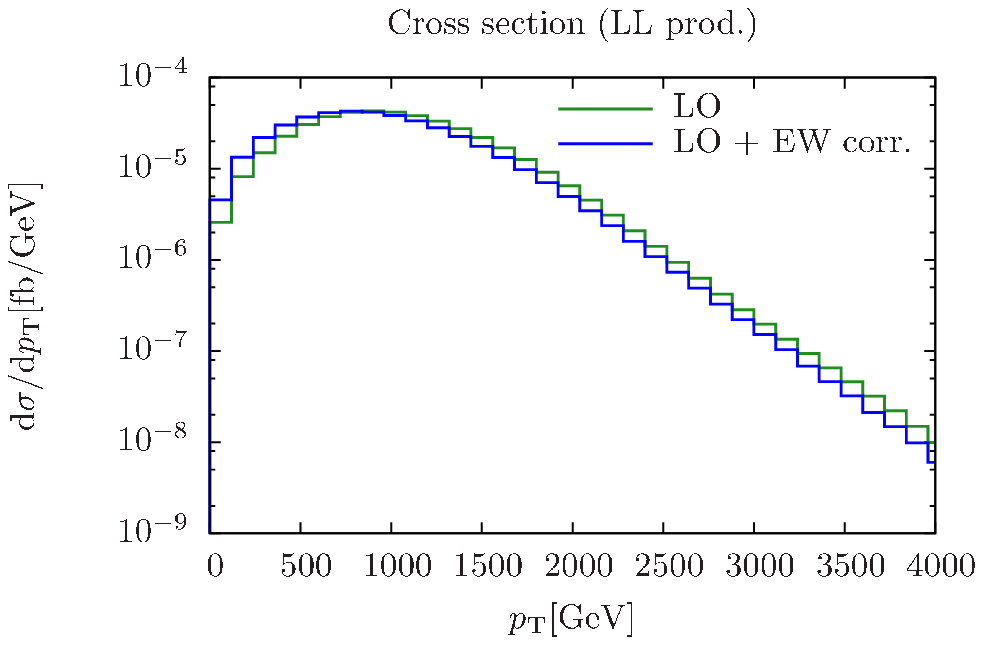}
\caption{}
\end{subfigure}
\begin{subfigure}[b]{0.5\textwidth}
\includegraphics[width=\reldistplotwidthratio\textwidth]{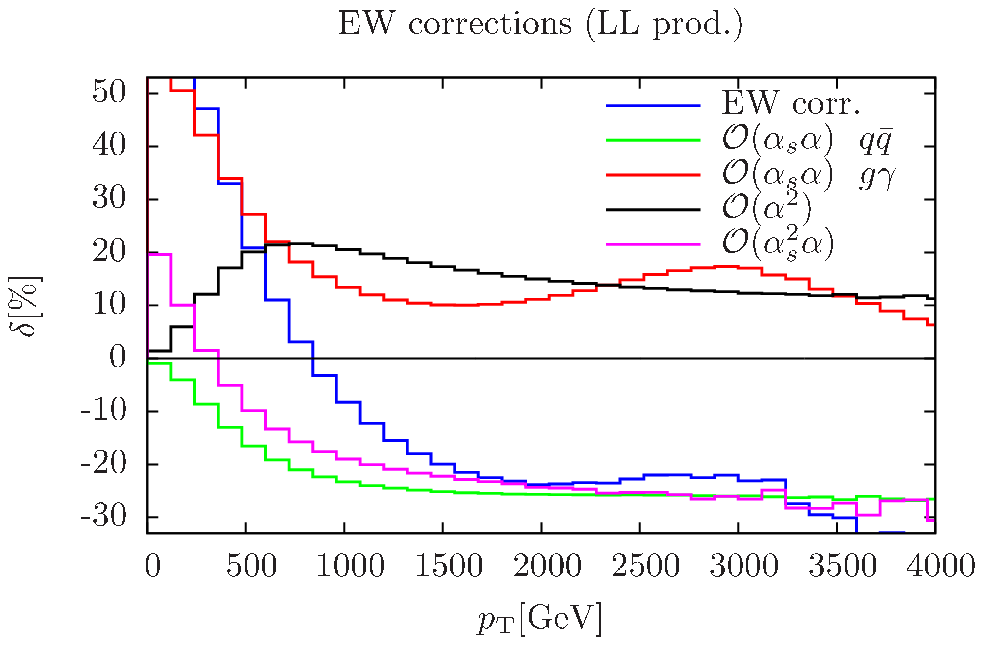}
\caption{}
\end{subfigure}
\phantom{pic}  \\
\begin{subfigure}[b]{0.5\textwidth}
\includegraphics[width=\reldistplotwidth\textwidth]{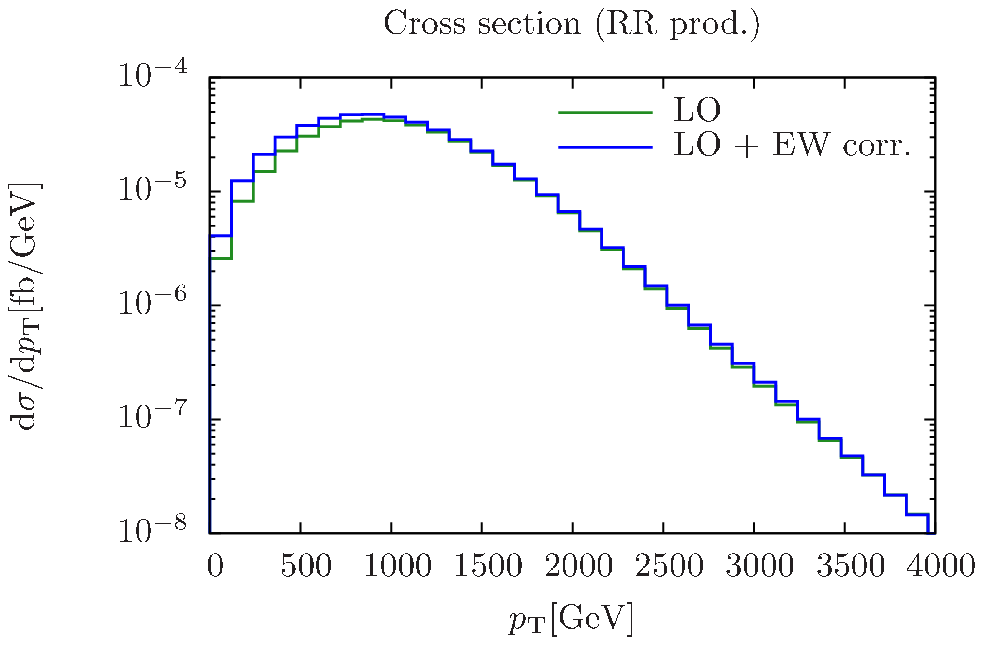}
\caption{}
\end{subfigure}
\begin{subfigure}[b]{0.5\textwidth}
\includegraphics[width=\reldistplotwidthratio\textwidth]{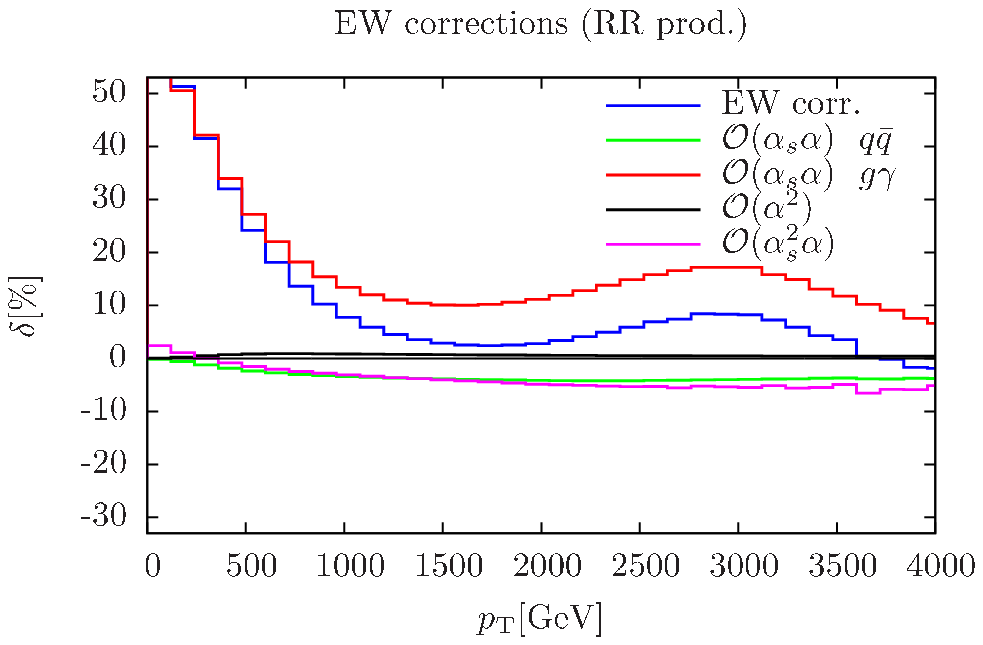}
\caption{}
\end{subfigure}
\phantom{pic}  \\
\begin{subfigure}[b]{0.5\textwidth}
\includegraphics[width=\reldistplotwidth\textwidth]{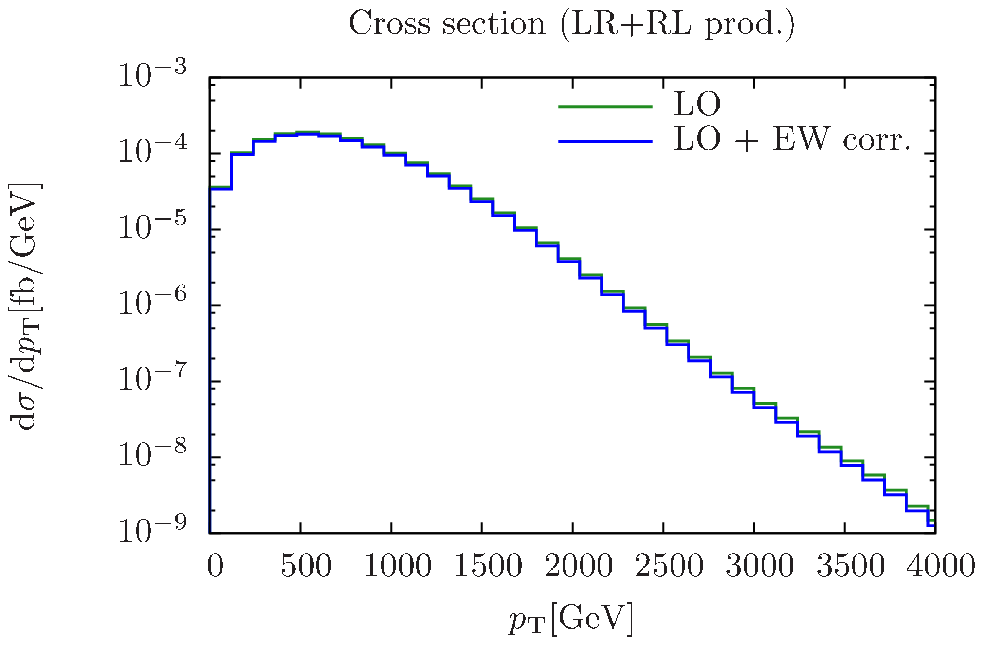}
\caption{}
\end{subfigure}
\begin{subfigure}[b]{0.5\textwidth}
\includegraphics[width=\reldistplotwidthratio\textwidth]{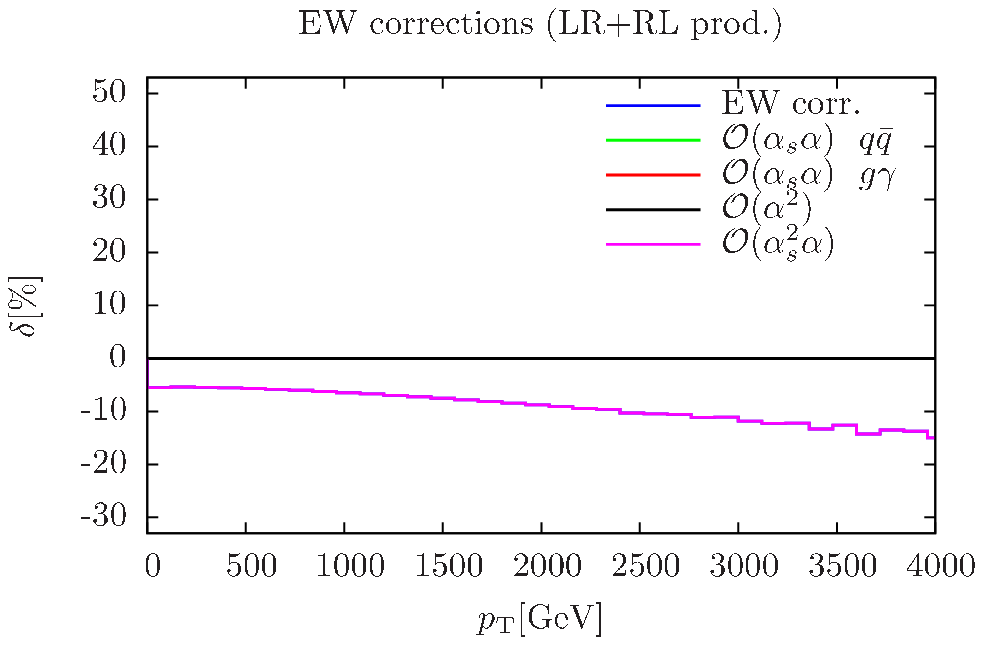}
\caption{}
\end{subfigure}
\caption[.]{Differential distributions in the transverse momentum $p_T$ of the produced squark.}
\label{Fig:dist_pt}
 \end{figure}

\def\reldistplotwidth{0.94}
\begin{figure}[t]
\begin{subfigure}[b]{0.5\textwidth}
\includegraphics[width=\reldistplotwidth\textwidth]{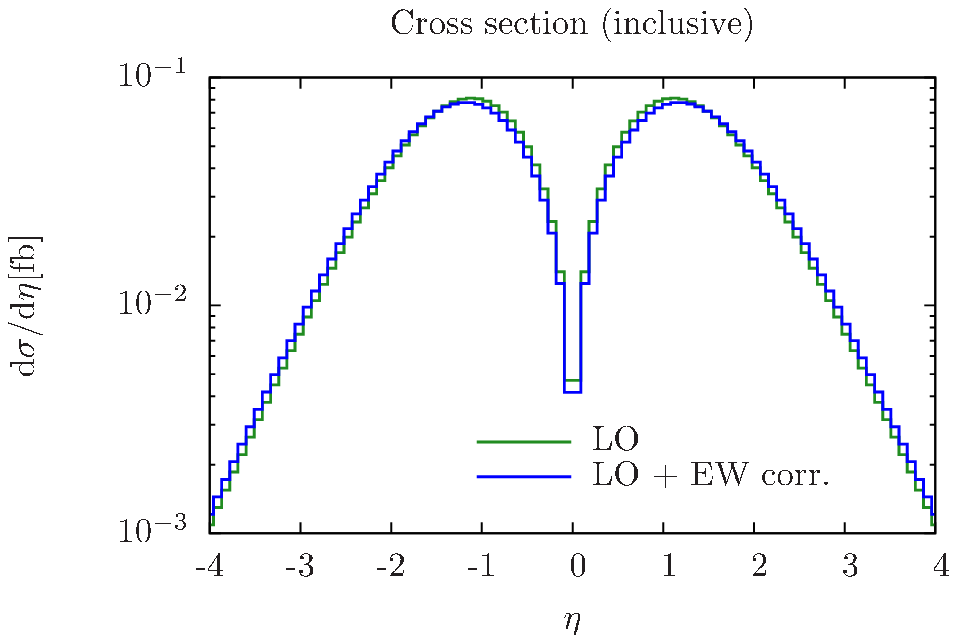}
\caption{}
\end{subfigure}
\begin{subfigure}[b]{0.5\textwidth}
\includegraphics[width=\reldistplotwidthratio\textwidth]{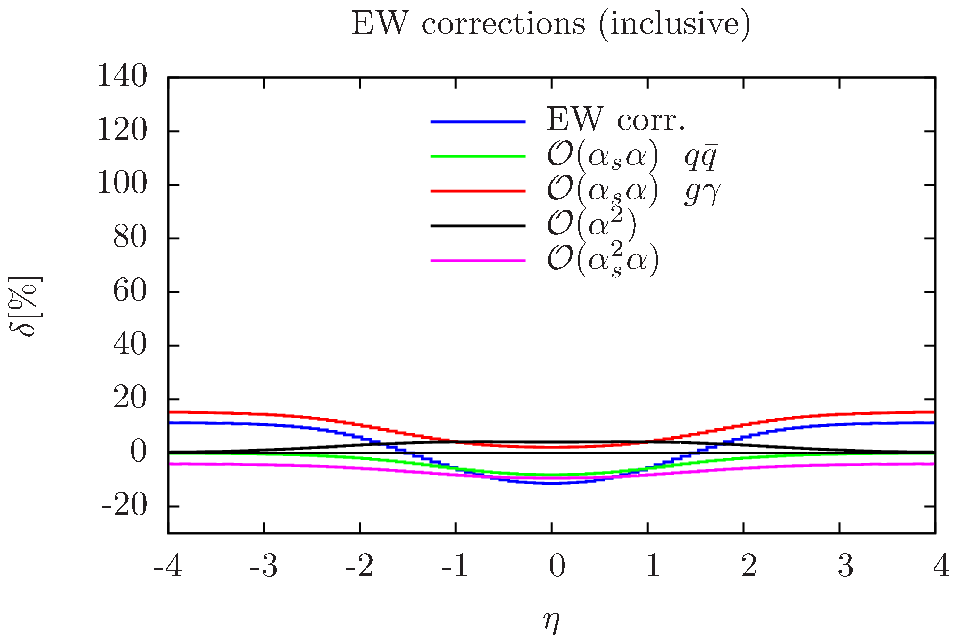}
\caption{}
\end{subfigure}
\phantom{pic}  
\\
\begin{subfigure}[b]{0.5\textwidth}
\includegraphics[width=\reldistplotwidth\textwidth]{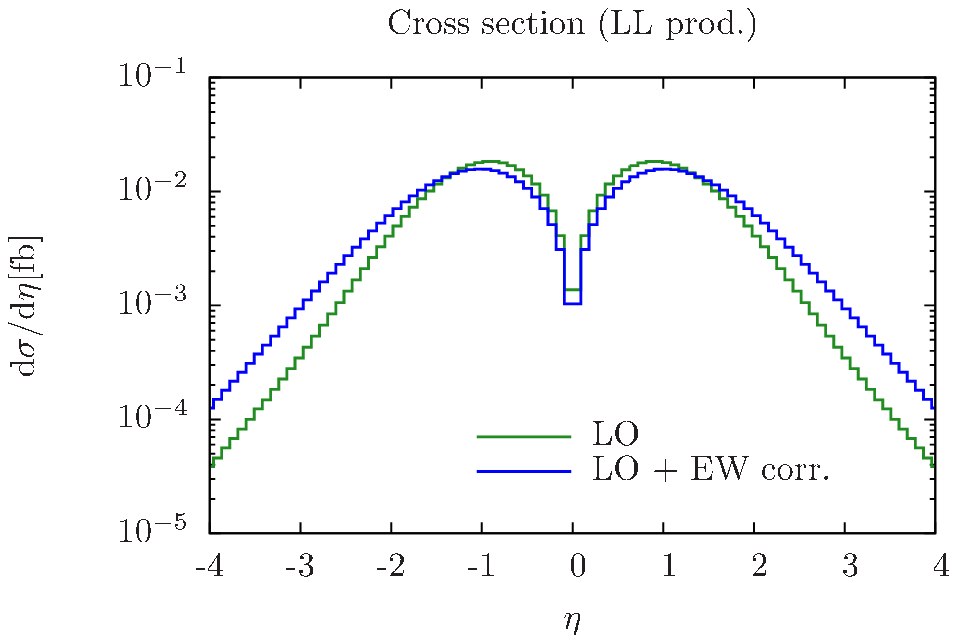}
\caption{}
\end{subfigure}
\begin{subfigure}[b]{0.5\textwidth}
\includegraphics[width=\reldistplotwidthratio\textwidth]{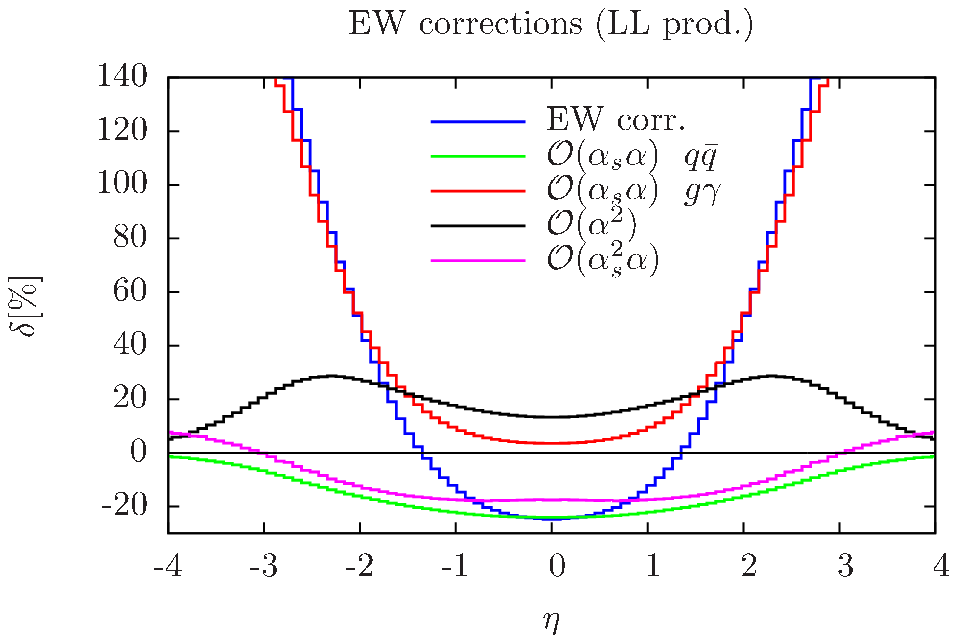}
\caption{}
\end{subfigure}
\phantom{pic}  \\
\begin{subfigure}[b]{0.5\textwidth}
\includegraphics[width=\reldistplotwidthratio\textwidth]{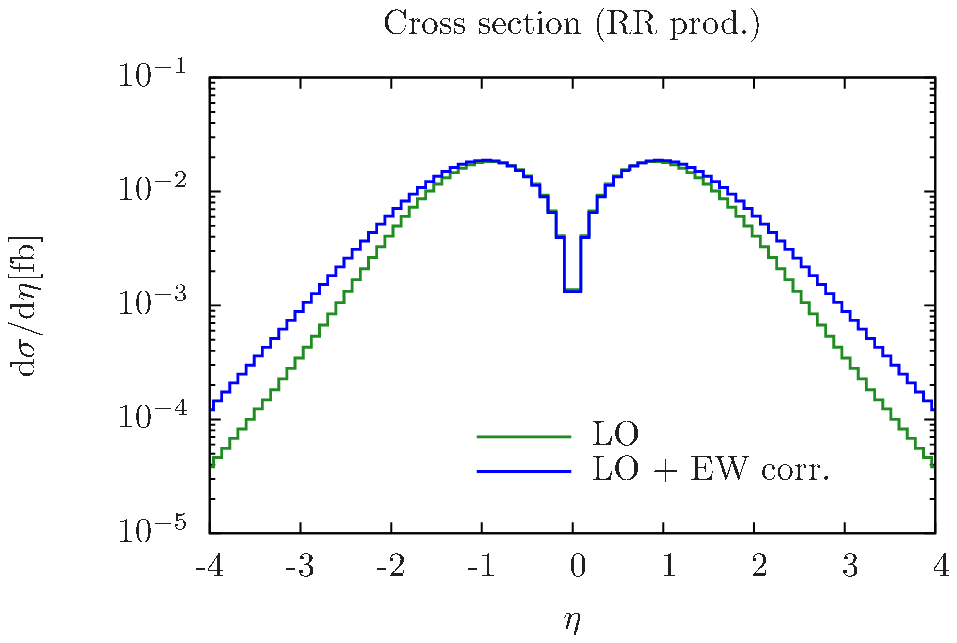}
\caption{}
\end{subfigure}
\begin{subfigure}[b]{0.5\textwidth}
\includegraphics[width=\reldistplotwidthratio\textwidth]{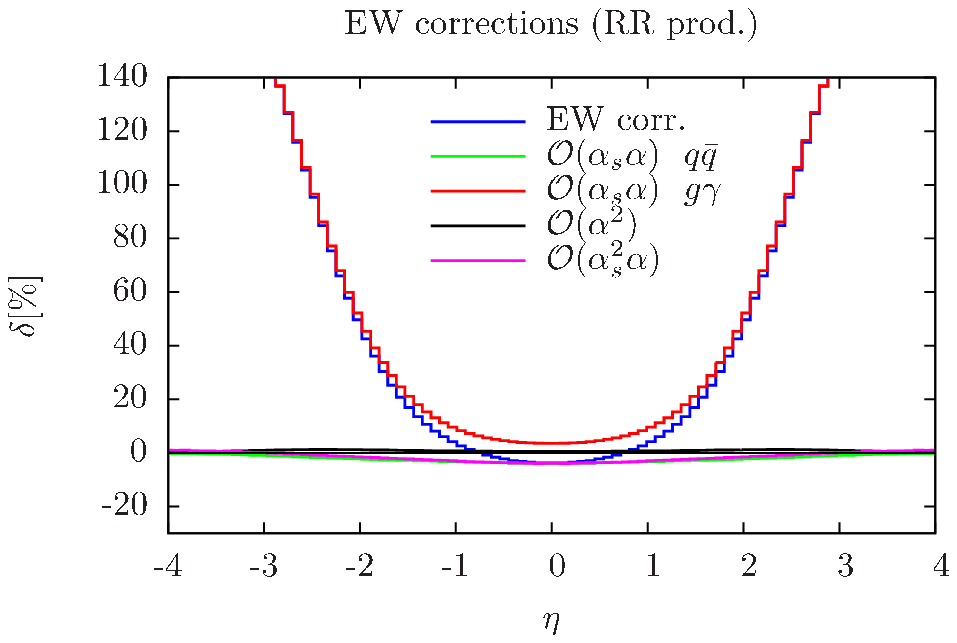}
\caption{}
\end{subfigure}
\phantom{pic}  \\
\begin{subfigure}[b]{0.5\textwidth}
\includegraphics[width=\reldistplotwidth\textwidth]{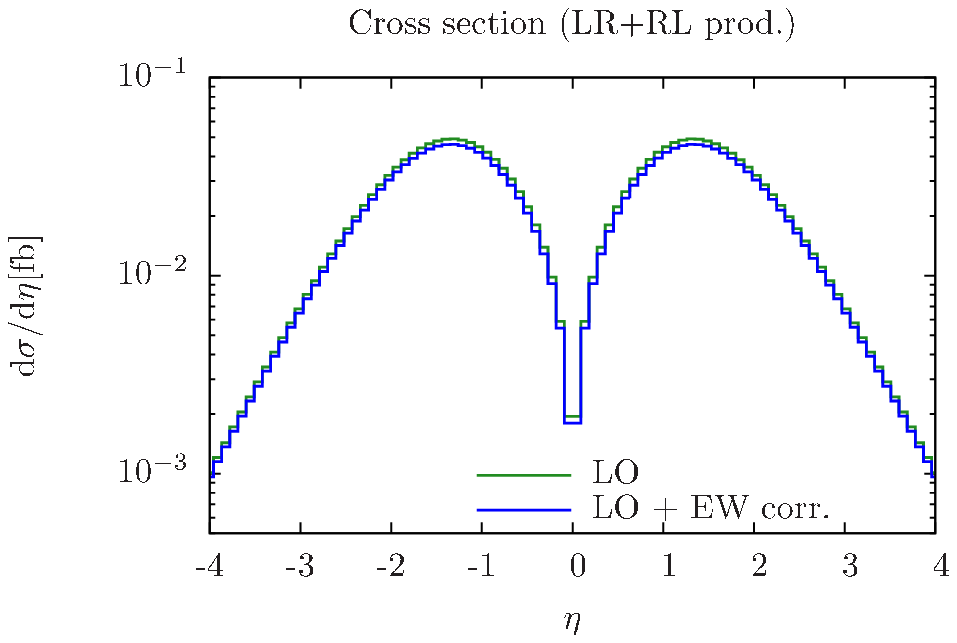}
\caption{}
\end{subfigure}
\begin{subfigure}[b]{0.5\textwidth}
\includegraphics[width=\reldistplotwidthratio\textwidth]{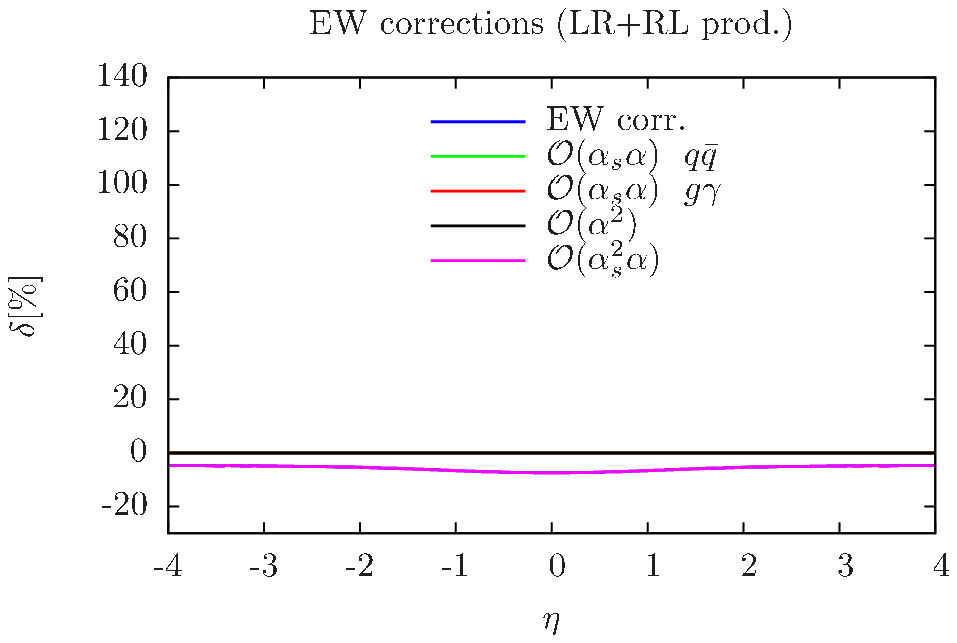}
\caption{}
\end{subfigure}
\caption[.]{Differential distributions in the maximal pseudorapidity $\eta$ of the produced squarks, as defined in eq.~\eqref{eq:diff_distributions}.}
\label{Fig:dist_eta}
 \end{figure}

\def\reldistplotwidth{0.94}
\begin{figure}[t]
\begin{subfigure}[b]{0.5\textwidth}
\includegraphics[width=\reldistplotwidth\textwidth]{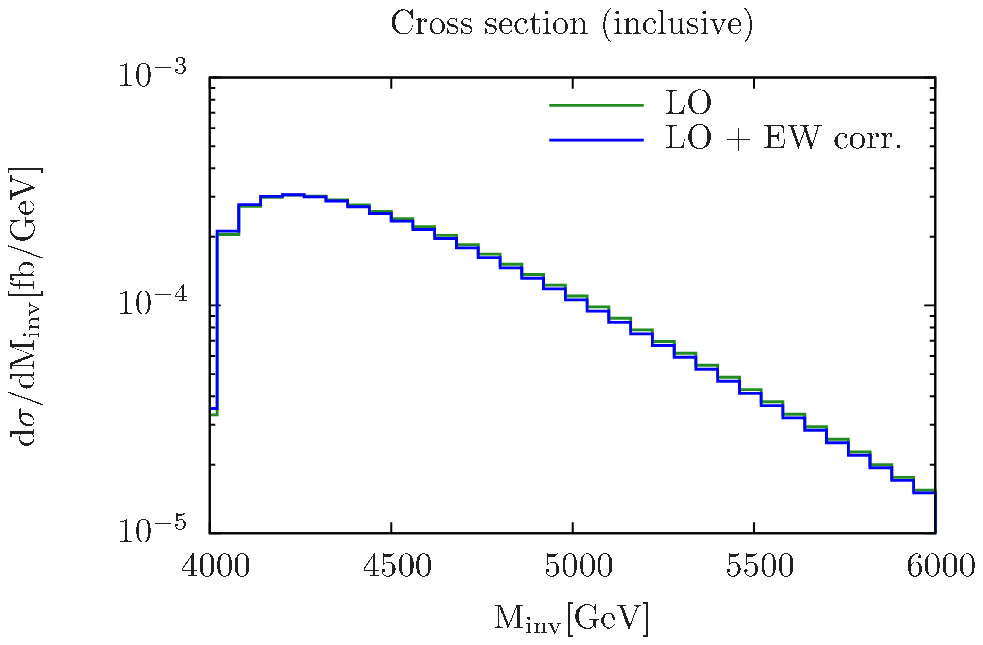}
\caption{}
\end{subfigure}
\begin{subfigure}[b]{0.5\textwidth}
\includegraphics[width=\reldistplotwidthratio\textwidth]{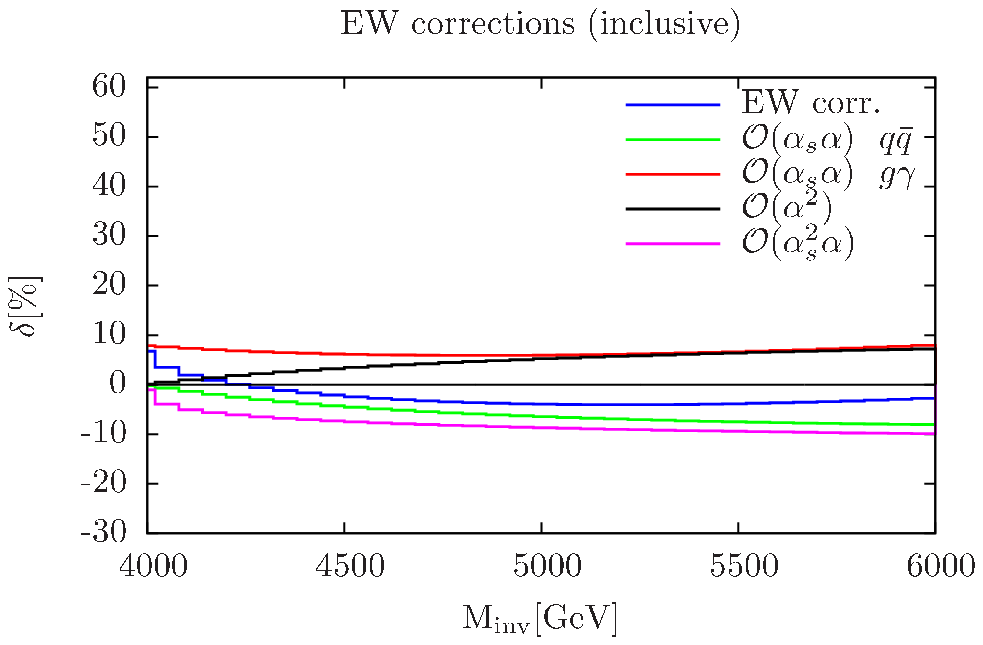}
\caption{}
\end{subfigure}
\phantom{pic}  \\
\begin{subfigure}[b]{0.5\textwidth}
\includegraphics[width=\reldistplotwidth\textwidth]{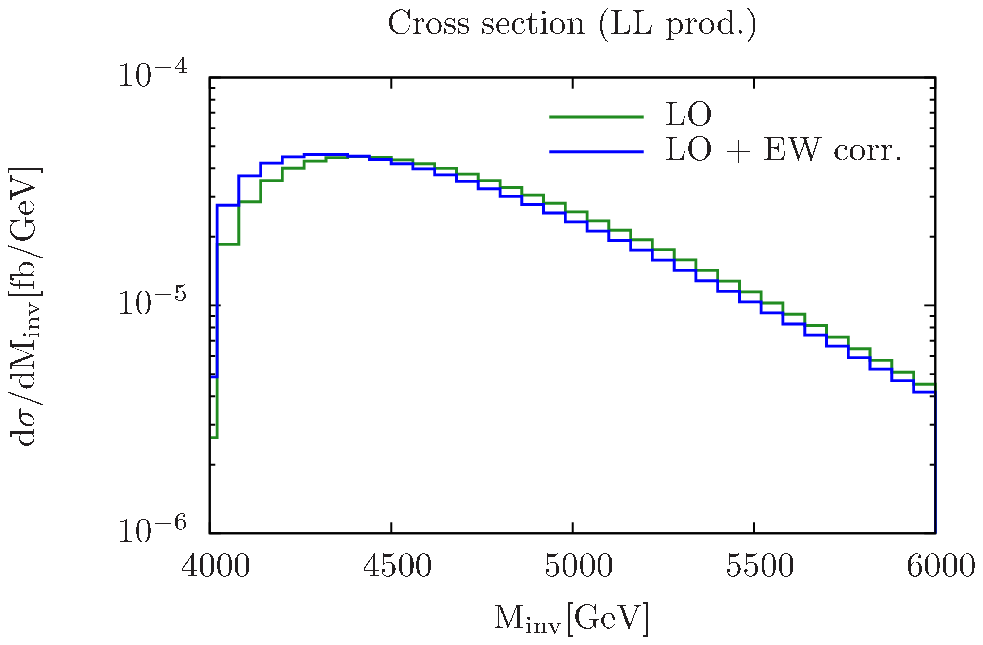}
\caption{}
\end{subfigure}
\begin{subfigure}[b]{0.5\textwidth}
\includegraphics[width=\reldistplotwidthratio\textwidth]{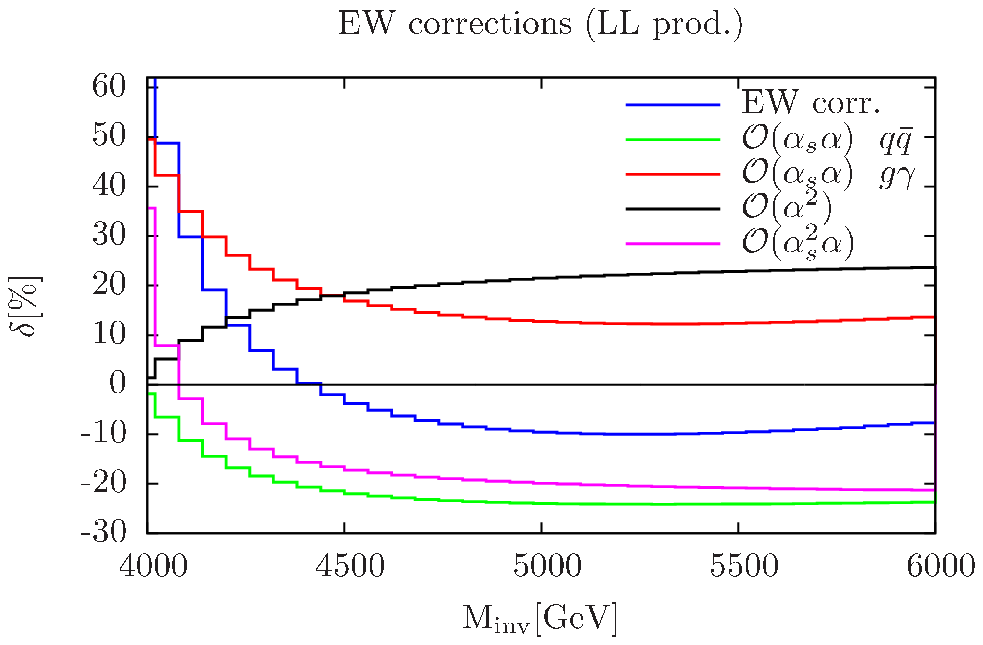}
\caption{}
\end{subfigure}
\phantom{pic}  \\
\begin{subfigure}[b]{0.5\textwidth}
\includegraphics[width=\reldistplotwidth\textwidth]{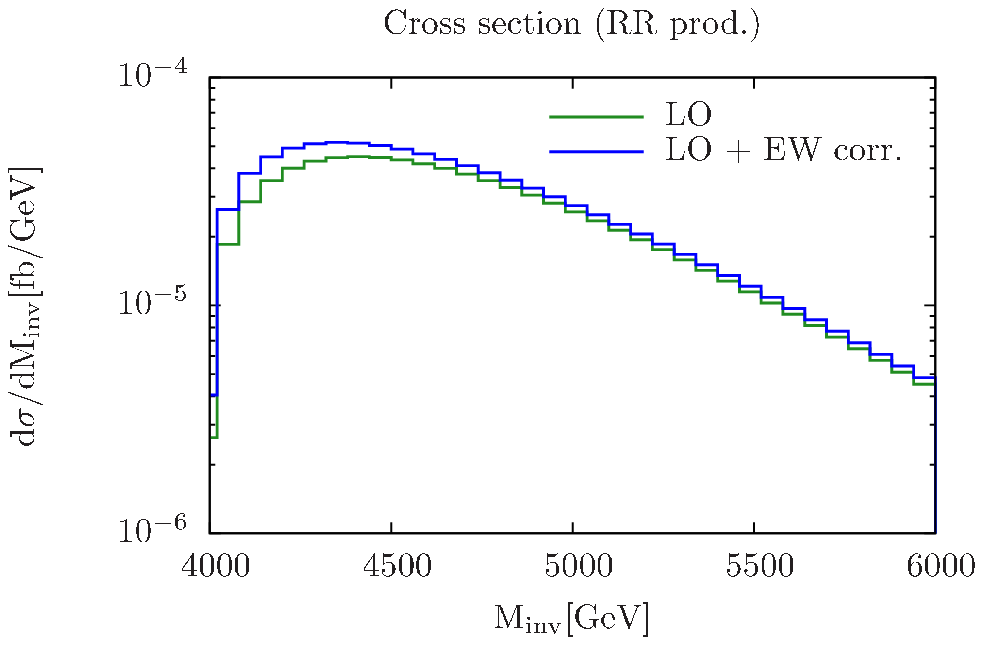}
\caption{}
\end{subfigure}
\begin{subfigure}[b]{0.5\textwidth}
\includegraphics[width=\reldistplotwidthratio\textwidth]{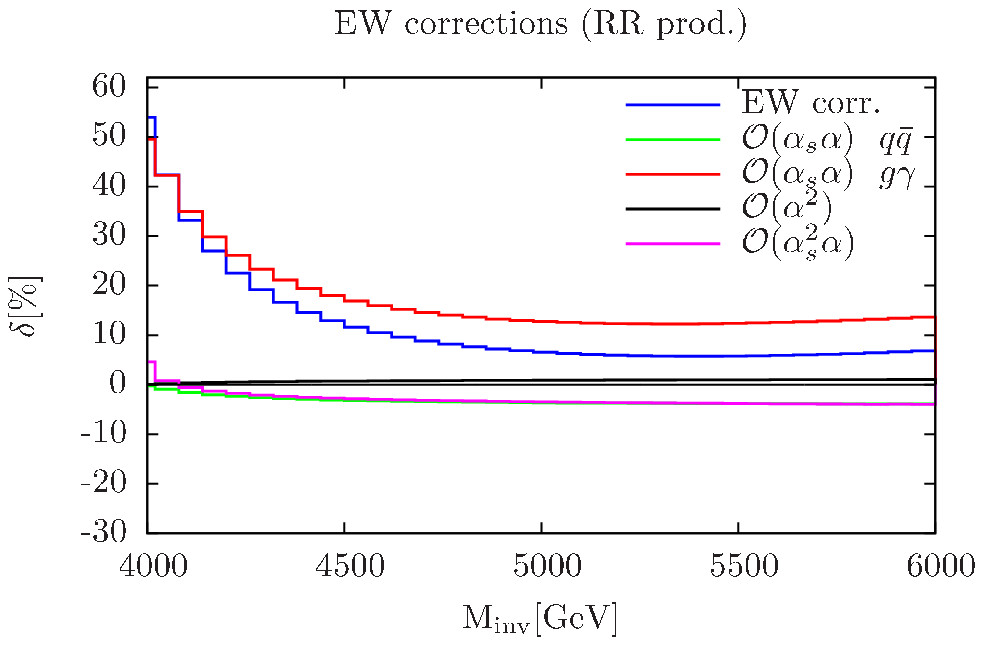}
\caption{}
\end{subfigure}
\phantom{pic}  \\
\begin{subfigure}[b]{0.5\textwidth}
\includegraphics[width=\reldistplotwidth\textwidth]{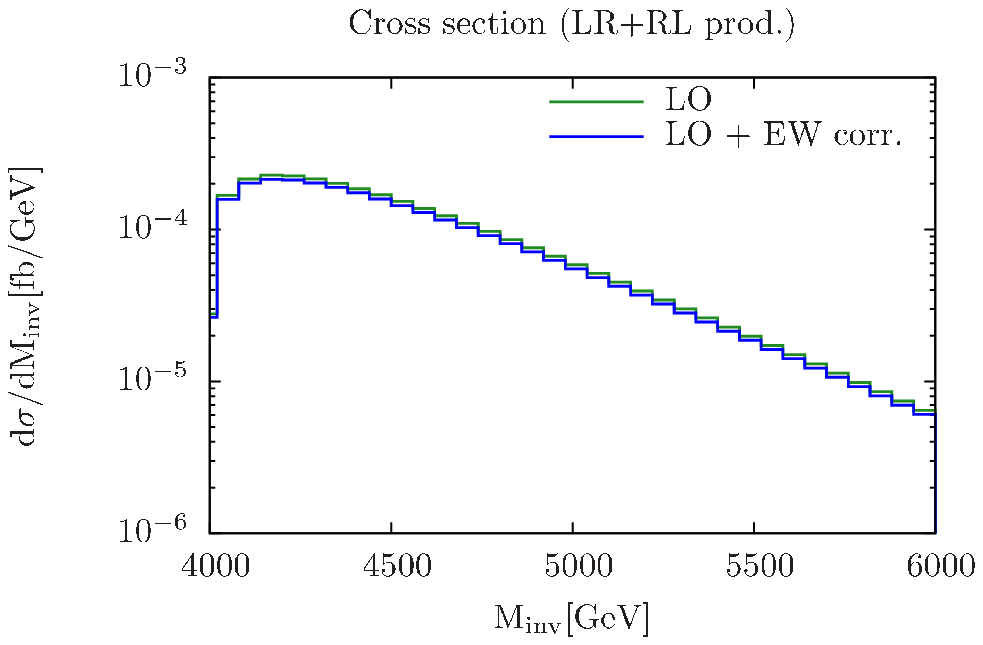}
\caption{}
\end{subfigure}
\begin{subfigure}[b]{0.5\textwidth}
\includegraphics[width=\reldistplotwidthratio\textwidth]{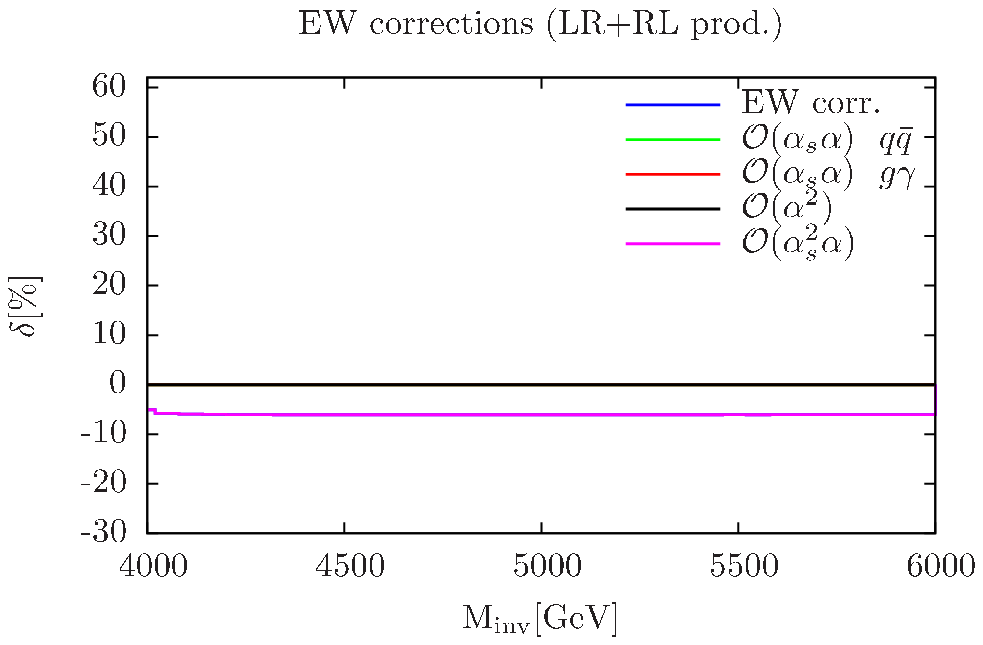}
\caption{}
\end{subfigure}
\caption[.]{Differential distributions in the invariant mass $M_{\text{inv}}$ of the produced squark--anti-squark pair.}
\label{Fig:dist_minv}
 \end{figure}

\end{document}